\documentclass[reqno,11pt]{amsart}
\usepackage[utf8]{inputenc}
\usepackage{amscd}
\usepackage{slashed}
\usepackage{amssymb}
\usepackage{esint}
\usepackage[usenames,dvipsnames]{pstricks}
\usepackage{epsfig}
\usepackage{pst-grad} 
\usepackage{pst-plot} 
\usepackage[mathscr]{eucal}
\numberwithin{equation}{section}
\allowdisplaybreaks[1]

\newcommand{\SetFigFont}[3]{}

\title[Lorentzian Spectral Geometry]{Lorentzian Spectral Geometry for \\ Globally Hyperbolic
Surfaces}

\author[F.\ Finster]{Felix Finster}
\address{Fakult\"at f\"ur Mathematik \\ Universit\"at Regensburg \\ D-93040 Regensburg \\ Germany}
\email{finster@ur.de, olaf.mueller@ur.de}
\author[O.\ M\"uller]{Olaf M\"uller \\ \\ November 2014}

\newtheorem{Def}{Definition}[section]
\newtheorem{Thm}[Def]{Theorem}
\newtheorem{Prp}[Def]{Proposition}
\newtheorem{Lemma}[Def]{Lemma}

\newtheorem{Corollary}[Def]{Corollary}
\newtheorem{Example}[Def]{Example}

\newcommand{\Thanks}{\vspace*{.5em} \noindent \thanks}
\newcommand{\beq}{\begin{equation}}
\newcommand{\eeq}{\end{equation}}
\newcommand{\Proof}{\begin{proof}}
\newcommand{\QED}{\end{proof} \noindent}
\newcommand{\QEDrem}{\hspace*{0.1em} \ \hfill $\Diamond$}

\newcommand{\la}{\langle}
\newcommand{\ra}{\rangle}
\newcommand{\bra}{\mathopen{<}}
\newcommand{\ket}{\mathclose{>}}
\newcommand{\Sl}{\mathopen{\prec}}
\newcommand{\Sr}{\mathclose{\succ}}

\newcommand{\C}{\mathbb{C}}
\newcommand{\R}{\mathbb{R}}
\newcommand{\1}{\mbox{\rm 1 \hspace{-1.05 em} 1}}
\newcommand{\Z}{\mathbb{Z}}
\newcommand{\N}{\mathbb{N}}

\newcommand{\nuslsh}{\slashed{\nu}}
\renewcommand{\H}{\mathscr{H}}

\newcommand{\SO}{{\rm{SO}}}

\newcommand{\e}{{\mathfrak{e}}}

\newcommand{\uslsh}{\slashed{u}}

\newcommand{\bep}{\begin{pmatrix}}
\newcommand{\enp}{\end{pmatrix}}

\renewcommand{\O}{\mathscr{O}}

\newcommand{\F}{{\mathscr{F}}}
\newcommand{\Dir}{{\mathcal{D}}}
\newcommand{\D}{{\mathscr{D}}}

\renewcommand{\O}{{\mathscr{O}}}

\newcommand{\s}{{\mathfrak{s}}}

\newcommand{\Lin}{\text{\rm{L}}}
\newcommand{\TV}{\text{\rm{TV}}}
\newcommand{\BV}{\text{\rm{BV}}}
\newcommand{\pu}{\partial_{\rm{u}}}
\newcommand{\pv}{\partial_{\rm{v}}}
\newcommand{\puv}{\partial_{\rm{uv}}}
\newcommand{\gu}{\gamma^{\rm{u}}}
\newcommand{\gv}{\gamma^{\rm{v}}}
\newcommand{\Cisc}{C^\infty_{\text{sc}}}

\newcommand{\ci}{\circ}

\newcommand{\bean}{\begin{eqnarray*}}
\newcommand{\eean}{\end{eqnarray*}}
\newcommand{\benu}{\begin{enumerate}}
\newcommand{\eenu}{\end{enumerate}}
\newcommand{\eea}{\end{eqnarray}}
\newcommand{\bea}{\begin{eqnarray}}
\newcommand{\pseudo}{\Gamma}
\newcommand{\SSS}{\mathbb{S}}

\DeclareMathOperator{\re}{Re}

\DeclareMathOperator{\Tr}{Tr}
\DeclareMathOperator{\tr}{tr}

\DeclareMathOperator{\diag}{diag}

\newcommand{\Sig}{\mathscr{S}}
\newcommand{\funk}{\text{$\Sig|_N$}}
\DeclareMathOperator{\ind}{ind}
\newcommand{\scrM}{\mycal M}
\newcommand{\scrN}{\mycal N}

\DeclareFontFamily{OT1}{rsfso}{}
\DeclareFontShape{OT1}{rsfso}{m}{n}{ <-7> rsfso5 <7-10> rsfso7 <10-> rsfso10}{}
\DeclareMathAlphabet{\mycal}{OT1}{rsfso}{m}{n}

\begin{document}
\maketitle

\begin{abstract}
The fermionic signature operator is analyzed on globally hyperbolic Lorentzian surfaces.
The connection between the spectrum of the fermionic signature operator and geometric
properties of the surface is studied. The findings are illustrated by simple examples and
counterexamples.
\end{abstract}

\tableofcontents

\section{Introduction}
It is a renowned mathematical problem if one can hear the shape of a drum, i.e.\ whether
the spectrum of the Laplace operator on a domain in~$\R^2$ with Dirichlet
boundary conditions determines the shape of the domain (see~\cite{kac, gordon1, gordon2}).
More generally, the mathematical area of spectral geometry is devoted to
studying the connection between the geometry of a Riemannian manifold~$(\scrM,g)$ and 
spectral properties of certain geometric operators on~$\scrM$ (see~\cite{gilkey} for a survey).
In the present paper, we propose a setting in which the objectives of spectral geometry
can be extended to Lorentzian signature.
In a more analytic language, our setting makes ideas and methods developed for elliptic differential
operators applicable to hyperbolic operators.
Moreover, we study the resulting ``Lorentzian spectral geometry''
in the simplest possible situations: for subsets of the Minkowski plane
and for Lorentzian surfaces.

In order to make the paper accessible to a broad readership, we now introduce the
problem for subsets of the Minkowski plane
without assuming a knowledge of differential geometry or Dirac spinors
(Section~\ref{secminkowski}).
Then we give a summary of the obtained results (Section~\ref{secsummary})
and outline our generalizations to Lorentzian surfaces with curvature (Section~\ref{seccurv}).
Finally, Section~\ref{secoutlook} puts our constructions and results into a more general context.

\subsection{The Minkowski Drum} \label{secminkowski}
Recall that in the classical drum problem one studies the eigenvalue problem
for the Laplacian on a bounded domain~$\Omega \subset \R^2$ with Dirichlet boundary values,
\[ -\Delta \phi = \lambda \phi \quad \text{in~$\Omega$} \:,\qquad
\phi|_{\partial \Omega} = 0 \:. \]
The naive approach to translate this problem to the Lorentzian setting is to replace the
Laplacian by the scalar wave operator. Denoting the variables in the plane by
by~$(t,x) \in \R^2$, we obtain the boundary value problem
\beq \label{naive}
\big(\partial_t^2 - \partial_x^2 \big) \phi(t,x) = \lambda\, \phi(t,x) \quad \forall\, (t,x) \in \Omega \subset \R^{1,1}
\:,\qquad \phi|_{\partial \Omega} = 0 \:.
\eeq
This is not a good problem to study, as we now explain. As a consequence of the minus sign, the
wave operator can be factorized into a product of two first order operators,
\beq \label{factor}
\big(\partial_t^2 - \partial_x^2 \big) = (\partial_t + \partial_x)(\partial_t - \partial_x) \:.
\eeq
This changes the analytic behavior of the solutions completely.
Namely, in the case of the scalar wave equation~$(\partial_t^2 - \partial_x^2 \big) \phi = 0$,
the factorization~\eqref{factor} implies that the general solution can be written as
\beq \label{phiLR}
\phi(t,x) = \phi_L(t+x) + \phi_R(t-x)
\eeq
with arbitrary real-valued functions~$\phi_L$ and~$\phi_R$. Thus, thinking of~$x$ as a spatial variable
and~$t$ as time, the solution can be decomposed into components~$\phi_L$ 
and~$\phi_R$ which propagate to the left respectively right, both with the characteristic speed one
(which can be thought of as the speed of light, which for convenience we set equal to one).
As a consequence, a boundary value problem makes no sense.
Namely, for Dirichlet boundary conditions, there is only the trivial solution~$\phi \equiv 0$.
Prescribing non-zero boundary values would give rise to consistency conditions for the
boundary values, which can only be satisfied for special boundary values.
If the ``eigenvalue'' $\lambda$ in~\eqref{naive} is non-zero, the structure of the equation
becomes more complicated because the components~$\phi_L$ and~$\phi_R$ are coupled
to each other. But again, boundary conditions give rise to consistency
conditions, which cannot in general be satisfied. Unless in very special cases, these
consistency condition cannot be met even for a countable set of values of~$\lambda \in \C$,
making it impossible to distinguish a discrete set of ``eigenvalues'' of the wave operator.

These mathematical problems reflect the fact that seeking for solutions of boundary value problems for the
scalar wave equation is not the correct question to ask. Instead, one should pose the problem in the way
it usually arises in the applications, namely as an {\em{initial-value problem}}.
Thus, instead of imposing boundary values, we should prescribe initial values up to first order $(\phi \vert_\scrN, \partial_t \phi \vert_\scrN)$ on a curve~$\scrN$ and seek for solutions
which satisfy the initial conditions. In order to avoid consistency conditions for the initial data,
the curve~$\scrN$ should be spacelike.
Moreover, this curve should be chosen in such a way that the initial conditions determine a unique global
solution in our ``space-time''. Such a curve is referred to as a {\em{Cauchy surface}},
and the existence of a Cauchy surface is subsumed in the notion that space-time should be
{\em{globally hyperbolic}}. We postpone the general definition of these notions to Section~\ref{secglobhyp},
and now simply explain what these notions mean for open subsets of the plane.

First of all, the {\em{Minkowski plane}}~$\R^{1,1}$ is the plane~$\R^2$
endowed with the inner product
\[ g \::\: \R^{1,1} \times \R^{1,1} \rightarrow \R\:,\qquad g\big((t,x), (t',x')\big) = t t' - x x' \:. \]
If the minus sign in the last equation were replaced by a plus sign, the inner product~$g$
would go over to the usual Euclidean scalar product on~$\R^2$. This minus sign accounts for
the Lorentzian signature. The inner product~$g$ is referred to as the {\em{Minkowski metric}}.
As already mentioned above, we regard the coordinates~$t$ and~$x$ of~$\R^{1,1}$
as time and space, respectively. The speed of light is set to one.
A regular smooth curve~$c(s)$ in the Minkowski plane parametrized by~$s \in (a,b) \subset \R$ 
with components~$(c(s)^0, c(s)^1)$ is said to be {\em{causal}} if it describes a motion
at most with the speed of light, i.e.
\begin{align*}
\text{causal curve:}\qquad \big|c'(s)^0\big| &\geq \big|c'(s)^1\big| \qquad \text{for all~$s \in (a,b)$} \:.
\intertext{The physical principle of {\em{causality}} states that information can be transmitted
only along causal curves. Similarly, the curve is said to be {\em{spacelike}} if its speed is
always faster than the speed of light,}
\text{spacelike curve:}\qquad \big|c'(s)^0\big| &< \big|c'(s)^1\big| \qquad \text{for all~$s \in (a,b)$} \:.
\end{align*}
Geometrically, for a spacelike curve the angle between the horizontal line and the tangent 
to the curve is less than $45^\circ$.
The causal structure can also be expressed in terms of the Minkowski metric. Namely, the curve
\[ \text{$c(s)$ is } \left\{ \begin{array}{cc} \text{causal} & \text{if }
g \big( c'(s), c'(s) \big) \geq 0 \\[0.3em]
\text{spacelike} & \text{if }
g \big( c'(s), c'(s) \big) < 0 \end{array} \right\} \text{ for all~$s \in (a,b)$} \:. \]
The {\em{domain of dependence}}~$D$ of a spacelike curve~$c$ is the
set of all points of~$\R^{1,1}$ such that every inextendible causal curve
through the point intersects~$c$.
It can also be characterized as the smallest rectangle
enclosing~$c$ whose sides have an angle of $45^\circ$ to the horizontal line.
More generally, we refer to a rectangle whose sides have an angle of $45^\circ$ as a {\em{causal diamond}}.

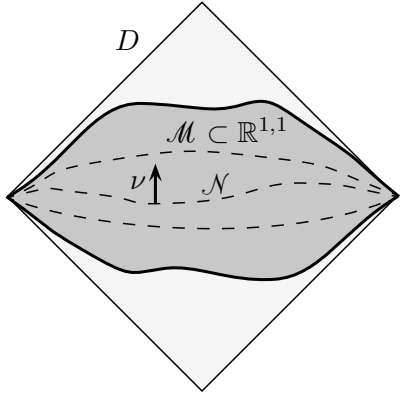
\begin{figure}
\scalebox{1} 
{
\begin{pspicture}(0,-2.5958583)(5.24,2.5958579)
\definecolor{color499b}{rgb}{0.9607843137254902,0.9607843137254902,0.9607843137254902}
\definecolor{color500g}{rgb}{0.7843137254901961,0.7843137254901961,0.7843137254901961}
\psdiamond[linewidth=0.02,dimen=outer,fillstyle=solid,fillcolor=color499b](2.6,0.0)(2.6,2.6)
\psbezier[linewidth=0.04,fillstyle=gradient,gradlines=2000,gradbegin=color500g,gradend=color500g,gradmidpoint=1.0](5.2,0.0)(4.48,-0.4270522)(4.18,-1.06)(3.56,-1.1)(2.94,-1.14)(2.530406,-0.89084834)(2.08,-0.9584512)(1.629594,-1.0260541)(1.66,-1.0864102)(1.04,-0.7242735)(0.42,-0.36213675)(0.7,-0.4972247)(0.0,0.0)
\psbezier[linewidth=0.04,fillstyle=gradient,gradlines=2000,gradbegin=color500g,gradend=color500g,gradmidpoint=1.0](0.02,0.0)(0.82,0.44)(1.12,1.26)(1.8,1.24)(2.48,1.22)(2.709594,1.0981123)(3.12,1.22)(3.530406,1.3418876)(3.56,1.2416117)(4.18,0.8277411)(4.8,0.41387054)(4.52,0.5682568)(5.22,0.0)
\psbezier[linewidth=0.02,linestyle=dashed,dash=0.16cm 0.16cm](0.02,0.0)(0.7874615,0.15166275)(0.3,0.12861046)(1.12,0.04)(1.94,-0.04861046)(1.4356606,-0.12686062)(2.36,-0.08)(3.2843394,-0.033139378)(3.32,0.2)(3.96,0.18)(4.6,0.16)(4.36,0.16)(5.2,0.0)
\usefont{T1}{ptm}{m}{n}
\rput(2.79,0.165){$\scrN$}
\usefont{T1}{ptm}{m}{n}
\rput(2.94,0.885){$\scrM \subset \R^{1,1}$}
\psline[linewidth=0.04cm,arrowsize=0.05291667cm 3.0,arrowlength=1.4,arrowinset=0.4]{->}(1.98,-0.1)(1.98,0.44)
\usefont{T1}{ptm}{m}{n}
\rput(1.75,0.185){$\nu$}
\psbezier[linewidth=0.02,linestyle=dashed,dash=0.16cm 0.16cm](0.02,-0.016585365)(1.1,-0.44)(3.7,-0.68)(5.2,0.0)
\psbezier[linewidth=0.02,linestyle=dashed,dash=0.16cm 0.16cm](0.02,-0.02)(0.76,0.3)(0.3635044,0.32446256)(1.34,0.48)(2.3164957,0.63553745)(2.46,0.64)(3.48,0.52)(4.5,0.4)(4.14,0.4)(5.2,0.02)
\usefont{T1}{ptm}{m}{n}
\rput(1.61,2.085){$D$}
\end{pspicture} 
}
\caption{A Minkowski drum.}
\label{figintro}
\end{figure}

We consider a bounded open subset~$\scrM \subset \R^{1,1}$ of the Minkowski plane.
The assumption of global hyperbolicity of~$\scrM$ implies that
there is a space-like curve~$\scrN$ such that its causal diamond~$D$ contains~$\scrM$
(see Figure~\ref{figintro}).
We refer to~$\scrM$ as a {\em{Minkowski drum}}.
It turns out that the curve~$\scrN$ is a Cauchy surface of~$\scrM$.
For the scalar wave equation, this can be verified by constructing the
solution of the Cauchy problem for initial data on~$\scrN$ explicitly by determining the
functions~$\phi_L$ and~$\phi_R$ in the general solution~\eqref{phiLR}.
If the parameter~$\lambda$ in~\eqref{naive} is present or if other linear geometric
equations are considered, the unique solvability of the Cauchy problem
follows for example from the theory of linear symmetric hyperbolic systems~\cite{john, taylor1}.

Clearly, the choice of the Cauchy surface is not unique, because any other spacelike
curve with the same endpoints is also a Cauchy surface (see the other dashed
lines in Figure~\ref{figintro}). Since we are interested in the geometry of space-time, our
constructions should not depend on the choice of the Cauchy surface~$\scrN$.

The basic question is which Hilbert space and which operator thereon
should be chosen as the basic objects of a Lorentzian spectral geometry.
An answer to this question is proposed in the recent paper~\cite{finite}, where
the so-called fermionic signature operator is introduced for space-times of finite lifetime
of general dimension. We here explain the idea and construction in the
simple setting of the Minkowski drum.
A main ingredient is that, instead of the scalar wave equation, we work with the
Dirac equation. We now explain how to get from the scalar wave equation to the Dirac
equation and why the Dirac equation is preferable for our purposes.
The first step for getting to the Dirac equation
is to write separate equations for the left- and right-moving components in~\eqref{phiLR},
\beq \label{Dirm0}
\begin{pmatrix} 0 & \partial_t + \partial_x \\ \partial_t - \partial_x & 0 \end{pmatrix} 
\begin{pmatrix} \phi_L \\ \phi_R \end{pmatrix} = 0 \:.
\eeq
Next, we replace~$\phi_L$ and~$\phi_R$ by complex-valued functions~$\psi_L, \psi_R$
and combine them to the so-called Dirac spinor~$\psi = (\psi_L, \psi_R) \in \C^2$
(for the formulation with vector bundles over a manifold see Section~\ref{seccurv}). Moreover, we
insert a parameter~$m \in \R$, the so-called rest mass, on the diagonal.
We thus obtain the {\em{Dirac equation}}
\[ \begin{pmatrix} im & \partial_t + \partial_x \\ \partial_t - \partial_x & im \end{pmatrix} 
\psi(t,x) = 0 \:. \]

Multiplying the differential operator from the left by the same differential operator with~$im$
replaced by~$-im$, we obtain
\[ \begin{pmatrix} -im & \partial_t + \partial_x \\ \partial_t - \partial_x & -im \end{pmatrix}
\begin{pmatrix} im & \partial_t + \partial_x \\ \partial_t - \partial_x & im \end{pmatrix} = \big(\partial_t^2-\partial_x^2
+m^2\big) \1_{\C^2} \:. \]
This shows that every component of a solution of the Dirac equation
 is also a solution of the wave equation in~\eqref{naive} with~$\lambda=-m^2$. Conversely,
 to a solution~$\phi$ of~\eqref{naive} we can associate a solution~$\psi$
of the Dirac equation by setting
\[ \psi = \begin{pmatrix} -im & \partial_t + \partial_x \\ \partial_t - \partial_x & -im \end{pmatrix} \phi\:. \]
We point out that this association is not one-to-one, because the mapping~$\phi \mapsto \psi$
is in general not injective (as is obvious in the example~$\phi \equiv 1$ in the massless case).
For this reason, we cannot identify the solutions of the Dirac equation with the solutions of
a scalar wave equation. In more general terms, the above consideration merely motivates
the Dirac equation, but it cannot be regarded as some kind of ``derivation'' of the Dirac equation.
Indeed, the Dirac equation is a different type of equation which cannot be derived mathematically
from a scalar equation. This is clear from the fact that the Dirac equation describes the
particle spin, a physical effect which is not taken into account by any scalar wave equation.

The Dirac equation gives rise to additional mathematical structures, which will be crucial for our
constructions. In preparation, it is useful to write the Dirac equation as the eigenvalue equation
for the {\em{Dirac operator}}~$\Dir$,
\beq \label{Dir}
\Dir \psi = m \psi  \qquad \text{with} \qquad
\Dir = i \gamma^0 \partial_t + i \gamma^1 \partial_x \:,
\eeq
where the Dirac matrices~$\gamma^0$ and~$\gamma^1$ are given by
\beq \label{gamma}
\gamma^0 = \begin{pmatrix} 0 & 1 \\ 1 & 0 \end{pmatrix} \:,\qquad
\gamma^1 = \begin{pmatrix} 0 & 1 \\ -1 & 0 \end{pmatrix} \:.
\eeq
The Dirac matrices can also be characterized in a basis-independent way
in terms of the anti-commutation relations
\[ \gamma^i \gamma^j + \gamma^j \gamma^i = 2 \, g^{ij}\,\1_{\C^2} \:, \]
where~$g^{ij} = \diag(1,-1)$ is again the Minkowski metric.
Next, it is useful to introduce an inner product~$\Sl .|. \Sr$ on the spinors such that
the Dirac matrices are symmetric with respect to it.
Clearly, the Dirac matrices are not symmetric with respect to the canonical scalar product on~$\C^2$,
because~$\gamma^1$ is anti-Hermitian. But they become symmetric if we introduce the
inner product~$\Sl .|. \Sr$ by
\beq \label{ssprod}
\Sl \psi | \phi \Sr = \la \psi | \begin{pmatrix} 0 & 1 \\ 1 & 0 \end{pmatrix} \phi \ra_{\C^2}\:.
\eeq
We denote~$\C^2$ with this inner product by~$(V \simeq \C^2, \Sl .|. \Sr)$
and refer to it as the {\em{spinor space}}.
To any two solutions~$\psi, \phi$ of the Dirac equation we can associate the vector field
\[ J^k(t,x) = \Sl \phi(t,x) \,|\, \gamma^k \,\psi(t,x) \Sr \:. \]
This vector field is divergence-free, as is verified by the following computation,
\begin{align*}
\partial_k J^k &= \partial_k \:\Sl \phi \,|\, \gamma^k \psi \Sr
= \Sl \partial_k \phi \,|\, \gamma^k \psi \Sr +
\Sl \phi \,|\, \gamma^k \partial_k \psi \Sr \\
&= \Sl \gamma^k \partial_k \phi \,|\, \psi \Sr + \Sl \phi \,|\, \gamma^k \partial_k \psi \Sr 
= \Sl (-im) \phi \,|\, \psi \Sr + \Sl \phi \,|\, (-i m) \psi \Sr = 0
\end{align*}
(here it is essential that~$m$ is real). Integrating this divergence over a region~$\Omega$ of~$\scrM$ and
applying the Gau{\ss} divergence theorem, one concludes that the flux of the vector field~$J$
through the boundary~$\partial \Omega$ vanishes
(the Gau{\ss} divergence theorem in Minkowski space is the same as
in Euclidean space, except that one must work with the normal with respect to the
Minkowski metric).
Choosing~$\Omega$ as the region between two Cauchy surfaces~$\scrN_1$ and~$\scrN_2$,
one sees that the following integral is independent of the choice of the Cauchy surface,
\beq \label{print}
( \psi | \phi) = 2 \pi \int_\scrN \Sl \psi | \gamma^j \phi \Sr\: \nu_j\, d\mu_\scrN \:.
\eeq
Here~$d\mu_\scrN$ the volume form of
the induced Riemannian metric on~$\scrN$, and~$\nu$ is the future-directed normal on~$\scrN$
(see Figure~\ref{figintro}).
Moreover, a direct computation shows that the inner product~$\Sl .| \gamma^j . \Sr\, \nu_j$
is positive definite, so that~\eqref{print} defines a scalar product on the solutions
of the Dirac equation. Forming the completion of the smooth solutions,
we obtain a Hilbert space~$(\H_m, (.|.))$. We remark that in physics, the vector
field~$\Sl \psi | \gamma^j \psi \Sr$
is called {\em{Dirac current}}, and the fact that it is divergence-free is referred to as
{\em{current conservation}}. Thus the conservation of the Dirac current is essential
for introducing the Hilbert space~$\H_m$ of solutions, independent of the choice of the Cauchy
surface~$\scrN$. We also remark that the function~$\Sl \psi | \gamma^j \psi \Sr\: \nu_j$
has the physical interpretation as the probability density for the Dirac particle described by
the wave function~$\psi$ to be
at a certain position on the space-like hypersurface~$\scrN$.
In this context, current conservation corresponds to the fact that the probability
of the particle to be anywhere in space must be preserved in time.
Therefore, current conservation is intimately connected with the probabilistic interpretation
of the wave function in quantum mechanics.

In order to encode the global behavior of the solutions in space-time in an operator, we
introduce on~$\H_m$ the inner product
\beq \label{stip}
\bra \psi|\phi \ket := \int_\scrM \Sl \psi | \phi \Sr\: d\mu\:,
\eeq
where~$d\mu = dt\, dx$ is the Lebesgue measure.
Using that~$\scrM$ is a bounded set, it follows that this inner product is bounded, i.e.\
there is a constant~$c>0$ such that
\beq \label{bounded}
|\bra \psi | \phi \ket| \leq c \:\|\psi\|\: \|\phi\| \qquad \text{for all~$\psi, \phi \in \H_m$}
\eeq
(for details see~\cite[Section~3]{finite}).
Thus for any~$\phi \in \H_m$, the anti-linear form~$\bra \,.\,|\, \phi \ket : \H_m \rightarrow \C$
is continuous. By the Fr{\'e}chet-Riesz theorem (see for example~\cite[Section~6.3]{lax}),
there is a unique vector~$u \in \H_m$ such that
\[ \bra \psi | \phi \ket = (\psi \,|\, u) \qquad \text{for all~$\psi \in \H_m$}\:. \]
The mapping~$\phi \mapsto u$ is linear and bounded. We thus obtain a bounded linear
operator~$\Sig \in \Lin(\H_m)$ such that
\beq \label{Sdef}
\bra \phi | \psi \ket = ( \phi \,|\, \Sig\, \psi) \qquad \forall \, \phi, \psi \in \H_m \:.
\eeq
Moreover, taking the complex conjugate of~\eqref{Sdef} and exchanging~$\phi$ and~$\psi$,
one sees that the operator~$\Sig$ is symmetric.
The operator~$\Sig$ is referred to as the {\em{fermionic signature operator}}.

With the fermionic signature operator~$\Sig$ on the Hilbert space~$(\H_m, (.|.))$,
we have introduced the objects of our spectral geometry.
We are interested in the question if the geometry
of~$(\scrM,g)$ is encoded in the operator~$\Sig$ and how this geometric information can be retrieved.

\subsection{Summary of Results for the Minkowski Drum} \label{secsummary}
In short, our analysis reveals that the spectrum of the fermionic signature operator encodes many
geometric properties of~$\scrM$. However, it does not determine the geometry of~$\scrM$ completely.

The {\em{massless}} case~$m=0$ is easier to analyze because, similar as explained
in~\eqref{phiLR} for the scalar wave equation, the Dirac equation has a simple explicit solution.
For this reason, in Section~\ref{secmassless} we begin with the massless case.
We first consider so-called {\em{simple domains}} for which the fermionic signature operator
can be represented by an explicit finite-dimensional matrix (see Definition~\ref{defsimple}
and Lemma~\ref{lemmasimple}).
We construct one-parameter families of simple domains which are {\em{isospectral
but not isometric}}, showing that the spectrum of~$\Sig$ does not determine the geometry completely
(Example~\ref{excounter}).
Then we represent the fermionic signature operator for general Minkowski drums
as an integral operator and show that it is Hilbert-Schmidt (Proposition~\ref{prpintrep}).
Moreover, the spectrum of the fermionic signature operator is shown to be
symmetric with respect to the origin (Proposition~\ref{prpsymm}).
We proceed by computing the trace of powers of~$\Sig$.
The trace of~$\Sig^2$ encodes the total volume of space-time (Proposition~\ref{prpS2}),
\beq \label{volrep}
\tr \big( \Sig^2 \big) = \frac{\mu(\scrM)}{4 \pi^2} \:.
\eeq
The traces of higher powers give additional geometric information, as is explained
in the example of~$\tr(\Sig^4)$ in Proposition~\ref{prp71}.
Then we explore the connection between
the spectrum of~$\Sig$ and the lengths of curves.
We prove that length~$\ell$ of any {\em{timelike}} curve is bounded from above
by the largest eigenvalue~$\lambda$ of~$\Sig$ by (Proposition~\ref{prp81})
\[ \lambda \geq \frac{\ell}{4 \pi} \:. \]
Moreover, it is shown that the length~$\ell$ of any
{\em{spacelike}} curve is bounded from above by the trace over the positive spectral subspace of~$\Sig$
(Proposition~\ref{prp91}),
\[ \tr \Big( \chi_{(0, \infty)}(\Sig) \,\Sig \Big)  \geq \frac{\ell}{4 \pi} \:.\]
Finally, it is shown that the geometry of~$\scrM$ is completely determined
if~$\Sig$ is given as an integral operator acting
on the initial data set of any Cauchy hypersurface (Theorem~\ref{thmreconstruct}).

In Section~\ref{secmassive} we turn attention to the {\em{massive}} case.
A general solution of the Cauchy problem is constructed using the Green's function
which is given in terms of Bessel functions (Lemma~\ref{lemmabessel}).
Next, we analyze how the regularity of the image of~$\Sig$
depends on the smoothness of the boundary (see Propositions~\ref{prpholder} and~\ref{prpBV}).
This also makes it possible to estimate the asymptotics
of the eigenvalues near the origin in terms of the total variation of the boundary
curve (Theorem~\ref{thmasy}).
In Proposition~\ref{prpsymmmass} it is shown that the spectrum of~$\Sig$ is again symmetric
with respect to the origin, but with a different method than in the massless case.
We proceed by computing the trace of powers of the fermionic signature operator.
The dependence on the mass parameter gives additional geometric information,
as is explained in Proposition~\ref{prpS2massive} for the trace of~$\Sig^2$.

\subsection{Lorentzian Surfaces in the Massless Case} \label{seccurv}
In generalization of the Minkowski drum, in this paper we also consider Lorentzian surfaces with curvature.
The point of interest is to analyze how the curvature of the surface affects the spectrum of
the fermionic signature operator. For technical simplicity, we restrict attention to the massless
case. This case is mathematically appealing because we can make use of the {\em{conformal invariance}}
of the massless Dirac equation, making the tools of conformal geometry available.

The structures introduced in Section~\ref{secminkowski} for the Minkowski drum
all generalize to the setting with curvature:
We let~$(\scrM,g)$ be a two-dimensional globally hyperbolic Lorentzian manifold.
Globally hyperbolic means that there is a space-like curve~$\scrN$ being a Cauchy surface
(for the precise definition of global hyperbolicity and a Cauchy surface see Section~\ref{secglobhyp}).
The Cauchy surface can be either compact or non-compact. In the first case, it is diffeomorphic to a sphere,
whereas in the latter case, it is diffeomorphic to an open interval. For simplicity, we here restrict attention
to the latter case. 
We let~$S\scrM$ be the spinor bundle on~$\scrM$.
The fibers~$S_x\scrM$ are isomorphic to~$\C^2$. They are endowed with an inner product of signature~$(1,1)$,
which we denote by~$\Sl .|. \Sr_x$.
The smooth sections of the spinor bundle are denoted by~$C^\infty(\scrM, S\scrM)$.
The Lorentzian metric induces a Levi-Civita connection
and a spin connection, which we both denote by~$\nabla$.
Every vector of the tangent space acts on the corresponding spinor space by Clifford multiplication.
We denote the corresponding map from the
tangent space to the linear operators on the spinor space by~$\gamma
\::\: T_x\scrM \rightarrow \Lin(S_x\scrM)$.
Clifford multiplication is related to the Lorentzian metric via the anti-commutation relations
\[ \gamma(u) \,\gamma(v) + \gamma(v) \,\gamma(u) = 2 \, g(u,v)\,\1_{S_x(\scrM)} \:. \]
We also write Clifford multiplication in components with the Dirac matrices~$\gamma^j$
and use the short notation with the Feynman dagger, $\gamma(u) \equiv u^j \gamma_j \equiv \uslsh$.
The connections, inner products and Clifford multiplication satisfy Leibniz rules and compatibility
conditions; we refer to~\cite{baum, lawson+michelsohn} for details.
Combining the spin connection with Clifford multiplication gives the geometric Dirac operator~$\Dir = i \gamma^j
\nabla_j$. The massless Dirac equation reads
\[ \Dir \psi = 0 \:. \]
We remark for clarity that in the case with curvature, the square of the Dirac operator
no longer coincides with the wave operator. Indeed, by the
Schr\"odinger-Lichnerowicz-Weitzenb\"ock formula $\Dir^2 = -\nabla_j \nabla^j + \frac{R}{4}$
these operators differ by a multiple of scalar curvature~$R$.

In the Cauchy problem, one seeks for a solution of the Dirac equation with
initial data~$\psi_\scrN$ prescribed on a given Cauchy surface~$\scrN$. Thus in the smooth setting,
\[ \D \psi = 0 \:,\qquad \psi|_{\scrN} = \psi_\scrN \in C^\infty(\scrN, S\scrM) \:. \]
This Cauchy problem has a unique solution~$\psi \in C^\infty(\scrM, S\scrM)$.
This can be seen either by considering energy estimates for symmetric hyperbolic systems
(see for example~\cite{john}) or alternatively by constructing the Green's kernel (see for
example~\cite{baer+ginoux}). These methods also show that the Dirac equation is causal,
meaning that the solution of the Cauchy problem only depends on the initial data in the causal
past or future. In particular, if~$\psi_\scrN$ has compact support, the solution~$\psi$ will also have compact
support on any other Cauchy hypersurface. This leads us to consider solutions~$\psi$
in the class~$\Cisc(\scrM, S\scrM)$ of smooth sections with spatially compact support. On solutions in this class,
one again introduces the scalar product~\eqref{print},
where~$\nuslsh$ denotes Clifford multiplication by the future-directed normal~$\nu$
(we always adopt the convention that the inner product~$\Sl . | \nuslsh . \Sr_x$ is
{\em{positive}} definite). Using current conservation~$\nabla_j \Sl \psi | \phi \Sr = 0$,
the scalar product~\eqref{print} is independent of the choice of the Cauchy surface
(similar as explained in Section~\ref{secminkowski} for the Minkowski drum).
Now the fermionic signature operator~$\Sig$ is defined exactly as for the Minkowski
drum by expressing the space-time inner product~\eqref{stip} in terms of the scalar product
in the form~\eqref{Sdef}.

For globally hyperbolic Lorentzian surfaces of finite lifetime and finite volume having a
non-compact Cauchy surface, we show that the fermionic signature operator encodes
the volume and the curvature
in the following way. First, the Hilbert-Schmidt norm of~$\Sig$ again encodes the volume~\eqref{volrep},
where~$\mu$ now is the volume measure corresponding to the Lorentzian metric.
The formula for~$\tr(\Sig^4)$ involves integrals of curvature:
\[ \tr \big(\Sig^4\big) = 
\frac{1}{8 \pi^4} \int_\scrM d\mu(\zeta) \int_{J(\zeta)} 
\exp \bigg( \frac{1}{4} \int_{D(\zeta, \zeta')} R\: d\mu \bigg) \:d\mu(\zeta')\:, \]
where~$J(\zeta)$ denotes all space-time points which can be connected to~$\zeta$
by a causal curve. Moreover,
$D(\zeta, \zeta')$ is the causal diamond of the space-time points~$\zeta$ and~$\zeta'$, i.e.\
\beq \label{causaldiamond}
D(\zeta, \zeta') = \big( J^\vee(\zeta) \cap J^\wedge(\zeta') \big) \cup  
\big( J^\vee(\zeta') \cap J^\wedge(\zeta) \big) \:,
\eeq
where~$J^\vee(\zeta)$ and~$J^\wedge(\zeta)$ denotes the points which can be connected to~$\zeta$
via a future- and past-directed causal curve, respectively.

Finally, we show that the geometry of~$\scrM$ can be reconstructed if~$\Sig$ is given as an integral
operator acting on the initial data set of any Cauchy hypersurface (Theorem~\ref{thmreconstruct2}).

\subsection{Outlook: The Chiral Index and Causal Fermion Systems} \label{secoutlook}
We now put the ideas and constructions given in this paper into a more general context,
also indicating possible directions of future research.

We first point out that for simplicity, we here restrict attention to two-dimensional
space-times and mainly the massless Dirac equation. But most constructions could be
generalized to globally hyperbolic Lorentzian manifolds of arbitrary dimension.
The continuity of the space-time inner product~\eqref{bounded} can be subsumed
in the notion that the space-time should be $m$-finite, which means qualitatively that
the space-time must have {\em{finite lifetime}} (for details see~\cite[Section~3.2]{finite}).
However, many interesting space-times like asymptotically flat Lorentzian manifolds or
Lorentzian manifolds with asymptotic ends have {\em{infinite life-time}}, implying that
the continuity condition~\eqref{bounded} fails. In such space-times, one must use
a different construction which relies on the so-called {\em{mass oscillation property}}
introduced in~\cite{infinite}.
In all these situations, the connection between
the spectrum of the fermionic signature operator and the geometry of the Lorentzian manifold
is largely unknown, leaving many interesting mathematical questions open.

We next remark that it is possible to associate an index to the fermionic signature operator,
which takes integer values.
In~\cite{index} simple examples of space-times with a non-trivial index are constructed, and the stability
of the index under homotopies is studied. But it is unknown if and how this
index is related to the geometry or the topology of the space-time.
In order to introduce this index, one needs as an additional structure
a {\em{chiral grading operator}}~$\pseudo$ which acts on the spinor spaces
and has for all~$u \in T_x\scrM$ the properties
\beq \label{gammaprop}
\pseudo^* = -\pseudo\:,\qquad \pseudo^2 = \1 \:,\qquad 
\pseudo\, \gamma(u) = -\gamma(u) \,\pseudo \:,\qquad \nabla \pseudo = 0 \:,
\eeq
where~$\gamma$ is Clifford multiplication and
the star denotes the adjoint with respect to the inner product~$\Sl .|. \Sr_x$.
More generally, the operator~$\pseudo$ is defined in any even space-time dimension
by Clifford multiplication with the volume form.
In physics, $\pseudo$ is called ``pseudoscalar operator'' and is usually
denoted by~$\gamma^5$.
The grading operator gives rise to the two idempotent operators
\[ \chi_L = \frac{1}{2} \big( \1 - \pseudo \big) \qquad \text{and} \qquad \chi_R = \frac{1}{2}
\big( \1 + \pseudo \big) \:, \]
referred to as the chiral projections (on the left respectively right handed component of the spinors).
The {\em{chiral signature operators}} $\Sig_L$ and~$\Sig_R$ are defined by
inserting the chiral projections into~\eqref{Sdef},
\[ \bra \phi \,|\, \chi_{L\!/\!R} \,\psi \ket = ( \phi \,|\, \Sig_{L\!/\!R}\, \psi) \:. \]
The first relation in~\eqref{gammaprop} implies that~$\Sig_L^*=\Sig_R$.
We thus define the {\em{chiral index}} as the Noether index of~$\Sig_L$
(sometimes called Fredholm index; for basics see for example~\cite[\S27.1]{lax}),
\begin{align*}
\ind \Sig &:= \dim \ker \Sig_L - \dim \text{coker} \,\Sig_L \\
&\;= \dim \ker \Sig_L - \dim \ker \Sig_R \:.
\end{align*}

The fermionic signature operator is a technical tool in the fermionic projector approach
to quantum field theory and is also used for constructing examples of causal fermion systems.
We now outline these connections.
The {\em{fermionic projector}}~$P$ is obtained by composing the causal fundamental solution~$k_m$
with the projection operator on the negative spectral subspace of~$\Sig$,
\beq \label{Pdef}
P = -\chi_{(-\infty, 0)}(\Sig)\, k_m \:.
\eeq
This distribution implements the physical concept of the ``Dirac sea''.
Next, particles and anti-particles are introduced to the system by adding to~\eqref{Pdef}
additional occupied states or by creating ``holes''.
We refer the interested reader to the constructions in~\cite[Section~3]{finite} and the
survey article~\cite{srev}.

It is a general idea behind the fermionic projector approach that the geometry of space-time
as well as all the objects therein should be described purely in terms of the physical wave functions
of the system. This idea is made mathematically precise in the notion of a {\em{causal fermion system}}
as introduced in~\cite{rrev}. In order to get into this framework, one chooses~$\H_\text{\tiny{particle}}$
as a subspace of the solution space of the Dirac operator.
A typical example is to choose~$\H_\text{\tiny{particle}} = \chi_{(-\infty, 0)}(\Sig)$ as the
image of the fermionic projector~\eqref{Pdef}.
By introducing an ultraviolet regularization, one arranges that the functions in~$\H_\text{\tiny{particle}}$
are continuous (for details see~\cite[Section]{finite}). Then for any space-time point~$x$,
one can introduce the so-called {\em{local correlation operator}}~$F(x) \in \Lin(\H_\text{\tiny{particle}})$
via the relations
\[ (\psi \,|\, F(x)\, \phi) = -\Sl \psi(x) | \phi(x) \Sr_x \qquad \text{for all~$\psi, \phi
\in \H_\text{\tiny{particle}}$} \:. \]
Denoting the signature of the spin scalar product by~$(n,n)$,
the local correlation operator is a symmetric operator in~$\Lin(\H_\text{\tiny{particle}})$
of rank at most~$2n$,
which has at most $n$ positive and at most $n$ negative eigenvalues.
Finally, we introduce the {\em{universal measure}}~$d\rho= F_* \,d\mu$ as the push-forward
of the volume measure on~$\scrM$ under the mapping~$F$
(thus~$\rho(\Omega) := \mu((F)^{-1}(\Omega))$).
Omitting the subscript ``particle'', we thus obtain
a causal fermion system as defined in~\cite[Section~1.2]{rrev}:

\begin{Def} 
Given a complex Hilbert space~$(\H, \la .|. \ra_\H)$
and a parameter~$n \in \N$ (the {\bf{``spin dimension''}}), we let~$\F \subset \Lin(\H)$ be the set of all
self-adjoint operators on~$\H$ of finite rank, which (counting with multiplicities) have
at most~$n$ positive and at most~$n$ negative eigenvalues. On~$\F$ we are given
a positive measure~$\rho$ (defined on a $\sigma$-algebra of subsets of~$\F$), the so-called
{\bf{universal measure}}. We refer to~$(\H, \F, \rho)$ as a {\bf{causal fermion system}}.
\end{Def}

Causal fermion systems provide a general mathematical framework
in which there are many inherent analytic, geometric and topological structures.
This concept makes it possible to generalize notions of differential geometry to the non-smooth setting.
From the physical point of view, it is a proposal for quantum geometry and an approach to quantum gravity.
We refer the interested mathematical reader to the research papers~\cite{lqg, topology}
or the textbooks~\cite{cfs, intro}.
In the setting of causal fermion systems, the physical equations are formulated
in terms of the {\em{causal action principle}} (see~\cite{continuum} or~\cite[Section~1.1]{cfs}).

In the setting of causal fermion systems, the fermionic signature operator is simply defined by the integral
\[ \Sig = -\int_\F x \: d\rho(x) \]
(this operator can be viewed as the restriction of the fermionic signature operator defined by~\eqref{Sdef}
to~$\H_\text{\tiny{particle}}$; for details see~\cite[Section~4]{index}).
In this context, the objectives of our ``Lorentzian spectral geometry'' generalize to
the question of how the spectrum of~$\Sig$ is related to the objects
of the Lorentzian quantum geometry as introduced in~\cite{lqg}.

\section{The Massless Case} \label{secmassless}
\subsection{Simple Domains} \label{secsimple}
\begin{figure}
\scalebox{1} 
{
\begin{pspicture}(0,-1.61)(13.26,1.61)
\definecolor{color88b}{rgb}{0.9607843137254902,0.9607843137254902,0.9607843137254902}
\definecolor{color89g}{rgb}{0.7843137254901961,0.7843137254901961,0.7843137254901961}
\definecolor{color105b}{rgb}{0.9254901960784314,0.9254901960784314,0.9254901960784314}
\pspolygon[linewidth=0.02,linecolor=white,fillstyle=solid,fillcolor=color88b](7.22,-0.8)(9.62,1.6)(12.42,-0.8)(11.82,-1.4)(7.82,-1.4)
\psbezier[linewidth=0.04,fillstyle=gradient,gradlines=2000,gradbegin=color89g,gradend=color89g,gradmidpoint=1.0](7.22,-0.8)(7.94,-0.37294778)(8.613909,0.32)(9.32,0.09487893)(10.026091,-0.13024215)(9.889594,0.090848334)(10.34,0.15845121)(10.790406,0.22605409)(10.76,0.2864102)(11.38,-0.075726524)(12.0,-0.43786326)(11.72,-0.30277532)(12.42,-0.8)
\usefont{T1}{ptm}{m}{n}
\rput(3.04,0.545){$\scrM \subset \R^{1,1}$}
\psbezier[linewidth=0.04,fillstyle=gradient,gradlines=2000,gradbegin=color89g,gradend=color89g,gradmidpoint=1.0](0.0,-0.8)(0.72,-0.37294778)(1.3939091,0.32)(2.1,0.09487893)(2.8060908,-0.13024215)(2.669594,0.090848334)(3.12,0.15845121)(3.570406,0.22605409)(3.54,0.2864102)(4.16,-0.075726524)(4.78,-0.43786326)(4.5,-0.30277532)(5.2,-0.8)
\psbezier[linewidth=0.04,fillstyle=gradient,gradlines=2000,gradbegin=white,gradend=white,gradmidpoint=1.0](0.0,-0.8)(0.84,-0.668)(1.22,-0.437)(1.88,-0.558)(2.54,-0.679)(2.2,-0.448)(3.0,-0.437)(3.8,-0.426)(3.28,-0.36)(4.16,-0.547)(5.04,-0.734)(4.18,-0.547)(5.2,-0.8)
\psbezier[linewidth=0.02,linestyle=dashed,dash=0.16cm 0.16cm](0.02,-0.78)(0.82,-0.5222222)(0.0,-0.73703706)(0.88,-0.49)(1.76,-0.24296296)(1.560809,-0.3102668)(2.56,-0.27404255)(3.559191,-0.23781832)(3.46,-0.2)(4.26,-0.4148148)(5.06,-0.6296296)(4.36,-0.4344681)(5.2,-0.78)
\usefont{T1}{ptm}{m}{n}
\rput(3.47,-0.055){$\scrN$}
\psbezier[linewidth=0.04,fillstyle=solid,fillcolor=color105b](7.22,-0.8)(8.06,-0.668)(8.44,-0.437)(9.1,-0.558)(9.76,-0.679)(9.42,-0.448)(10.22,-0.437)(11.02,-0.426)(10.5,-0.36)(11.38,-0.547)(12.26,-0.734)(11.4,-0.547)(12.42,-0.8)
\psline[linewidth=0.02cm,linestyle=dashed,dash=0.16cm 0.16cm](7.24,-0.82)(12.38,-0.8)
\usefont{T1}{ptm}{m}{n}
\rput(9.83,-1.135){$\scrN$}
\usefont{T1}{ptm}{m}{n}
\rput(8.93,1.365){$D$}
\usefont{T1}{ptm}{m}{n}
\rput(6.99,-1.215){$(0,0)$}
\usefont{T1}{ptm}{m}{n}
\rput(12.69,-1.235){$(0,b)$}
\psline[linewidth=0.02cm,linecolor=white](7.62,-1.6)(7.62,-1.2)
\psline[linewidth=0.02](7.82,-1.4)(7.22,-0.8)(9.62,1.6)(12.42,-0.8)(11.82,-1.4)
\end{pspicture} 
}
\caption{Choosing the Cauchy surface at~$t=0$.}
\label{figcauchy}
\end{figure}
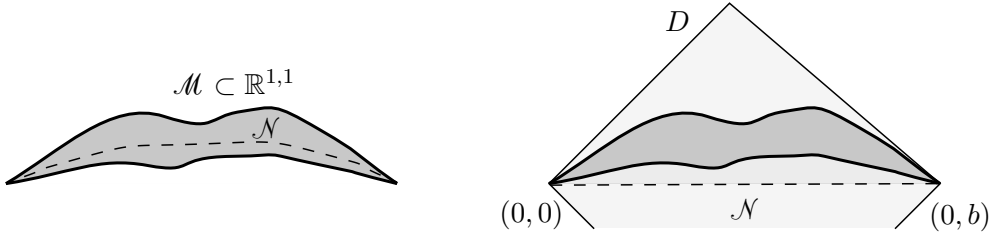

Let~$\scrM \subset \R^{1,1}$ be an open, bounded, globally hyperbolic subset
of the Minkowski plane~$\R^{1,1}$ (see the left of Figure~\ref{figcauchy}, where
also one Cauchy surface~$\scrN$ is shown).
The maximal solution for initial values on a Cauchy surface~$\scrN$ is defined
on a causal diamond~$D$ in the Minkowski plane.
The following constructions will depend only on the space of solutions in this causal diamond,
but the choice of the Cauchy surface will be irrelevant. In particular, the Cauchy surface
does not necessarily need to lie entirely in~$\scrM$.
With this in mind, we simply choose~$\scrN$ as the straight line joining the left and right corners
of the causal diamond. Moreover, by a Poincar{\'e} transformation we can arrange that the left
corner is the origin, whereas the right corner has the coordinates~$(0,b)$ with
a parameter~$b>0$. Then the scalar product~\eqref{print} simplifies to
\beq \label{print2}
( \psi | \phi) = 2 \pi \int_0^b \Sl \psi(0,x) \,|\, \gamma^0 \phi(0,x) \Sr\: dx
= 2 \pi \int_0^b \la \psi(0,x) , \phi(0,x) \ra_{\C^2}\: dx
\eeq
(where in the last step we used~\eqref{ssprod}).

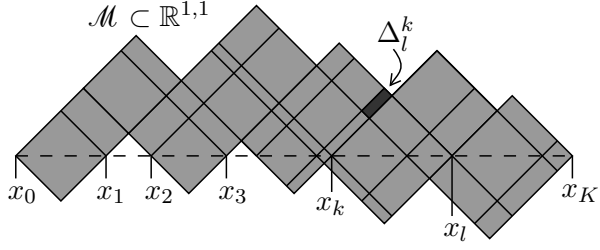
\begin{figure}
\scalebox{1} 
{
\begin{pspicture}(0,-1.66)(8.4,1.66)
\definecolor{color696b}{rgb}{0.6,0.6,0.6}
\definecolor{color743b}{rgb}{0.2,0.2,0.2}
\pspolygon[linewidth=0.02,fillstyle=solid,fillcolor=color696b](6.16,-0.42)(5.36,0.38)(5.96,0.98)(6.76,0.18)
\usefont{T1}{ptm}{m}{n}
\rput(2.14,1.465){$\scrM \subset \R^{1,1}$}
\pspolygon[linewidth=0.02,fillstyle=solid,fillcolor=color696b](0.36,-0.42)(1.26,0.48)(1.86,-0.12)(0.96,-1.02)
\pspolygon[linewidth=0.02,fillstyle=solid,fillcolor=color696b](1.86,-0.12)(3.56,1.58)(4.36,0.78)(2.66,-0.92)
\pspolygon[linewidth=0.02,fillstyle=solid,fillcolor=color696b](5.86,-0.72)(6.96,0.38)(7.76,-0.42)(6.66,-1.52)
\pspolygon[linewidth=0.02,fillstyle=solid,fillcolor=color696b](3.36,-0.22)(4.66,1.08)(5.36,0.38)(4.06,-0.92)
\pspolygon[linewidth=0.02,fillstyle=solid,fillcolor=color696b](4.46,-0.52)(5.36,0.38)(6.16,-0.42)(5.26,-1.32)
\psline[linewidth=0.02cm,linestyle=dashed,dash=0.16cm 0.16cm](0.36,-0.42)(7.66,-0.42)
\psline[linewidth=0.02cm](6.16,-0.42)(6.96,-1.22)
\psline[linewidth=0.02cm](4.46,-0.52)(2.96,0.98)
\psline[linewidth=0.02cm](4.36,0.78)(5.86,-0.72)
\psline[linewidth=0.02cm](3.56,-0.42)(2.56,0.58)
\psline[linewidth=0.02cm](5.26,0.48)(3.96,-0.82)
\psline[linewidth=0.02cm](5.06,0.08)(5.96,0.98)
\psline[linewidth=0.02cm](5.96,0.98)(7.56,-0.62)
\psline[linewidth=0.02cm](1.56,-0.42)(0.96,0.18)
\psline[linewidth=0.02cm](2.16,-0.42)(3.86,1.28)
\psline[linewidth=0.02cm](3.56,-0.42)(4.86,0.88)
\psline[linewidth=0.02cm](5.36,-1.22)(3.06,1.08)
\psline[linewidth=0.02cm](4.96,-1.02)(6.46,0.48)
\psline[linewidth=0.02cm](7.56,-0.22)(6.46,-1.32)
\usefont{T1}{ptm}{m}{n}
\rput(0.45,-0.935){$x_0$}
\usefont{T1}{ptm}{m}{n}
\rput(1.63,-0.935){$x_1$}
\usefont{T1}{ptm}{m}{n}
\rput(7.87,-0.915){$x_K$}
\usefont{T1}{ptm}{m}{n}
\rput(2.27,-0.935){$x_2$}
\usefont{T1}{ptm}{m}{n}
\rput(3.25,-0.935){$x_3$}
\psline[linewidth=0.02cm](7.76,-0.42)(7.76,-0.72)
\psline[linewidth=0.02cm](3.16,-0.42)(3.16,-0.72)
\psline[linewidth=0.02cm](2.16,-0.42)(2.16,-0.72)
\psline[linewidth=0.02cm](1.56,-0.42)(1.56,-0.72)
\psline[linewidth=0.02cm](0.36,-0.42)(0.36,-0.72)
\pspolygon[linewidth=0.02,fillstyle=solid,fillcolor=color696b](1.86,-0.12)(1.26,0.48)(1.96,1.18)(2.56,0.58)
\psline[linewidth=0.02cm](4.56,-0.42)(4.56,-0.9)
\usefont{T1}{ptm}{m}{n}
\rput(4.55,-1.095){$x_k$}
\usefont{T1}{ptm}{m}{n}
\rput(6.21,-1.435){$x_l$}
\psline[linewidth=0.02cm](6.16,-0.42)(6.16,-1.22)
\pspolygon[linewidth=0.02,fillstyle=solid,fillcolor=color743b](4.96,0.18)(5.26,0.48)(5.36,0.38)(5.06,0.08)
\usefont{T1}{ptm}{m}{n}
\rput(5.41,1.205){$\Delta^k_l$}
\psbezier[linewidth=0.02](5.38,1.0)(5.58,0.74)(5.44,0.68)(5.34,0.52)
\psline[linewidth=0.02](5.36,0.68)(5.34,0.52)(5.48,0.58)
\psline[linewidth=0.02cm](3.16,-0.42)(1.76,0.98)
\end{pspicture} 
}
\caption{A simple domain.}
\label{figsimple}
\end{figure}

For simplicity, we begin the analysis in the massless case.
Then the Dirac equation~\eqref{Dirm0} has the general solution
\beq \label{psiLR}
\psi(t,x) =  \begin{pmatrix} \psi_L(t+x) \\ \psi_R(t-x)  \end{pmatrix}
\eeq
with complex-valued functions~$\psi_L$ and~$\psi_R$.
In view of~\eqref{print2}, the left- and right-moving components are orthogonal.
The following assumption makes it possible to analyze~$\Sig$ explicitly.
\begin{Def} \label{defsimple}
$\scrM$ is a {\bf{simple domain}} if there are finitely many points
\[ 0 = x_0 < x_1 < \cdots < x_K = b \]
such that the boundary of~$\scrM$ is contained in the lightlike
curves through these points,
\[ \partial \scrM \subset \big\{x_0, \ldots, x_K \big\} + \R \,(1,1) + 
\R \,(1,-1) \:. \]
\end{Def} \noindent
The name ``simple domain'' is motivated by simple functions in measure theory which
take only a finite number of values. Figure~\ref{figsimple} shows an example.

We now introduce a basis of~$\H_0$ in which the operator $\Sig$ will turn out to have a particularly simple form.
To this end, for~$c \in \{L, R\}$, $k \in \{1, \ldots, K\}$ and~$n \in \Z$ we define functions
which are plane waves on the subintervals,
\beq \label{ONB}
\psi_c^{k,n}(x) = \frac{1}{\sqrt{2 \pi\,(x_k-x_{k-1})}}\:\chi_{(x_{k-1}, x_k]}(x)\: e^{\frac{2 \pi i}{x_k-x_{k-1}}\: n x}
\eeq
(where~$\chi$ denotes the characteristic function). As in~\eqref{psiLR} we regard the
functions~$\psi_R^{k,n}$ and~$\psi_L^{k,n}$ as spinors in the first and second component, respectively.
Solving the Cauchy problem, we obtain the corresponding Dirac solutions
\[ \begin{pmatrix} \psi_L^{k,n}(t+x) \\ 0 \end{pmatrix} \qquad \text{and} \qquad
\begin{pmatrix} 0 \\ \psi_R^{k,n}(t-x) \end{pmatrix} \:, \]
which with a slight abuse of notation we again denote by~$\psi_c^{k,n} \in \H_0$
(note that these are solutions only in the weak sense, as they are not continuous).
A short computation shows that these vectors are orthonormal,
\[ ( \psi_c^{k,n} | \psi_{c'}^{k',n'}) = \delta_{c,c'} \,\delta^{k,k'} \,\delta^{n,n'}\:. \]
Using that the plane waves~$e^{\frac{2 \pi i}{x_k-x_{k-1}}\: n x}$ form a Fourier basis
of~$L^2((x_{k-1}, x_k])$, one also sees that the vectors~$(\psi_c^{k,n})$ form a basis of~$\H_0$.
Hence~$(\psi_c^{k,n})$ is an orthonormal basis of~$\H_0$.
Moreover, a short computation shows that the space-time inner product~$\bra \psi_c^{k,n} | \psi_{c'}^{k',n'}
\ket$ vanishes if~$n$ or~$n'$ are non-zero (because one integrates over a full period of a plane wave)
or if~$c=c'$ (because the inner product~\eqref{ssprod} involves an off-diagonal matrix).
The remaining inner products are computed by
\begin{align*}
\bra &\psi_R^{k,n} | \psi_L^{l,n'} \ket = 
\bra \psi_L^{l,n'} | \psi_R^{k,n} \ket \notag \\
& =\frac{\delta^n_0\, \delta^{n'}_0}{2 \pi\, \sqrt{(x_k-x_{k-1}) (x_l-x_{l-1})}}
\: \mu(\Delta^k_l)
= \frac{\delta^n_0\, \delta^{n'}_0}{2 \pi \sqrt{2}} \: \sqrt{\mu(\Delta^k_l)} \:,
\end{align*}
where~$\mu$ is the Lebesgue measure on~$\R^{1,1}$ and
\[ \Delta^k_l = \Big( (x_{k-1}, x_k] + \R \,(1,1) \Big) \cap  \Big( (x_{l-1}, x_l] + \R \,(1,-1) \Big) \cap \scrM \]
(see Figure~\ref{figsimple}). We thus obtain the following result:
\begin{Lemma} \label{lemmasimple}
On a simple domain~$\scrM$ and for zero mass, the fermionic signature operator~$\Sig$
has finite rank. More precisely, choosing the orthonormal basis~\eqref{ONB},
\[ (\psi_c^{k,n}) \qquad \text{with} \qquad \text{$c \in \{L, R\}$, $k \in \{1, \ldots, K\}$, $n \in \Z$} \:, \]
it has rank at most~$2K$. It vanishes on all the vectors~$\psi_c^{k,n}$ with~$n \neq 0$.
On the subspace spanned by the basis vectors~$\psi_L^{1,0}, \ldots, \psi_L^{K,0}, \psi_R^{1,0}, \ldots,
\psi_R^{K,0}$, it has the block matrix representation
\[ \Sig = \frac{1}{2 \pi \sqrt{2}} \begin{pmatrix} 0 & T^* \\ T & 0 \end{pmatrix} \:, \]
where~$T$ is the matrix with components
\beq \label{Telements}
T^k_l = \sqrt{\mu(\Delta^k_l)}\:.
\eeq
\end{Lemma}

From this matrix representation one can read off a few general properties of the spectrum of
the fermionic signature operator:
\begin{Corollary} \label{corm0} On a simple domain~$\scrM$ and for zero mass, the following
statements hold.
\begin{itemize}
\item[(i)] The spectrum of~$\Sig$ is symmetric with respect to the origin and
\[ \sigma(\Sig) =  \sigma \big( \sqrt{T^* T} \big) \cup - \sigma \big( \sqrt{T^* T} \big)  \:. \]
For an eigenvector~$u$ of~$\sqrt{T^* T}$ corresponding to the non-zero eigenvalue~$\lambda$,
the eigenvectors of~$\Sig$ corresponding to the eigenvalues~$\pm \lambda$ have the form
\[ \psi = \begin{pmatrix} \lambda u \\ \pm T u \end{pmatrix} \:. \]
\item[(ii)] The eigenvector corresponding to the largest eigenvalue of~$\Sig$
is non-degenerate. Its components can be chosen to be non-negative.
\item[(iii)] $\displaystyle \Tr \big( \Sig^2 \big) = \frac{\mu(\scrM)}{4 \pi^2}\:. $
\end{itemize}
\end{Corollary}
\Proof Follows from a direct computation. Part~(ii) is a consequence of the
Perron-Frobenius theorem for matrices with positive entries (see~\cite[Chapter~5]{serreM}).
\QED

The spectrum of~$\Sig$ does not determine the geometry completely, as the following example
shows.
\begin{Example} {\bf{(Isospectral simple domains)}} \label{excounter} {\em{ Consider the matrices
\[ T = \begin{pmatrix} a & \sqrt{ab} & 0 \\ 0 & b & \sqrt{bc} \\ 0 & 0 & c \end{pmatrix} 
\qquad \text{and} \qquad
\tilde{T} = \begin{pmatrix} d & \sqrt{de} & \sqrt{df} \\ 0 & e & \sqrt{ef} \\ 0 & 0 & f \end{pmatrix} , \]
where~$a, \ldots, f$ are strictly positive parameters.
The form of the off-diagonal matrix elements ensures that these matrices 
can be realized in the form~\eqref{Telements} by simple domains. More precisely,
in order to realize the matrix~$T$, one chooses~$K=3$ and
\[ x_1-x_0 = \sqrt{2}\, a \:,\qquad x_2-x_1 = \sqrt{2}\, b\:,\qquad x_3-x_2 = \sqrt{2}\, c \:. \]
The simple domain is then chosen as all the squares in the future, except for the
square on top (see the Figure~\ref{figT} on the left).
\begin{figure}
\scalebox{1} 
{
\begin{pspicture}(0,-1.26)(9.34,1.26)
\definecolor{color81b}{rgb}{0.6,0.6,0.6}
\pspolygon[linewidth=0.02,fillstyle=solid,fillcolor=color81b](0.36,-0.45)(0.96,0.15)(1.56,-0.45)(0.96,-1.05)
\usefont{T1}{ptm}{m}{n}
\rput(0.45,-0.965){$x_0$}
\usefont{T1}{ptm}{m}{n}
\rput(1.63,-0.965){$x_1$}
\usefont{T1}{ptm}{m}{n}
\rput(2.57,-0.965){$x_2$}
\usefont{T1}{ptm}{m}{n}
\rput(4.05,-0.965){$x_3$}
\psline[linewidth=0.02cm](2.06,-0.45)(2.06,-0.75)
\psline[linewidth=0.02cm](1.56,-0.45)(1.56,-0.75)
\psline[linewidth=0.02cm](0.36,-0.45)(0.36,-0.75)
\pspolygon[linewidth=0.02,fillstyle=solid,fillcolor=color81b](2.46,-0.45)(3.26,0.35)(4.06,-0.45)(3.26,-1.25)
\pspolygon[linewidth=0.02,fillstyle=solid,fillcolor=color81b](0.96,0.15)(1.41,0.6)(2.01,0.0)(1.56,-0.45)
\pspolygon[linewidth=0.02,fillstyle=solid,fillcolor=color81b](1.56,-0.45)(2.01,0.0)(2.46,-0.45)(2.01,-0.9)
\pspolygon[linewidth=0.02,fillstyle=solid,fillcolor=color81b](2.01,0.0)(2.81,0.8)(3.26,0.35)(2.46,-0.45)
\psline[linewidth=0.02cm](2.46,-0.45)(2.46,-0.75)
\psline[linewidth=0.02cm,linestyle=dashed,dash=0.16cm 0.16cm](0.36,-0.45)(4.06,-0.45)
\psline[linewidth=0.02cm](4.06,-0.45)(4.06,-0.75)
\pspolygon[linewidth=0.02,fillstyle=solid,fillcolor=color81b](7.26,-0.45)(8.06,0.35)(8.86,-0.45)(8.06,-1.25)
\pspolygon[linewidth=0.02,fillstyle=solid,fillcolor=color81b](6.66,-0.45)(6.96,-0.15)(7.26,-0.45)(6.96,-0.75)
\pspolygon[linewidth=0.02,fillstyle=solid,fillcolor=color81b](5.46,-0.45)(6.06,0.15)(6.66,-0.45)(6.06,-1.05)
\pspolygon[linewidth=0.02,fillstyle=solid,fillcolor=color81b](6.06,0.15)(7.16,1.25)(8.06,0.35)(6.96,-0.75)
\psline[linewidth=0.02cm,linestyle=dashed,dash=0.16cm 0.16cm](5.46,-0.45)(8.86,-0.45)
\psline[linewidth=0.02cm](6.06,-1.05)(7.76,0.65)
\psline[linewidth=0.02cm](6.36,0.45)(8.06,-1.25)
\usefont{T1}{ptm}{m}{n}
\rput(5.45,-0.965){$x_0$}
\usefont{T1}{ptm}{m}{n}
\rput(6.63,-0.965){$x_1$}
\usefont{T1}{ptm}{m}{n}
\rput(7.27,-0.965){$x_2$}
\usefont{T1}{ptm}{m}{n}
\rput(8.85,-0.965){$x_3$}
\psline[linewidth=0.02cm](5.46,-0.45)(5.46,-0.75)
\psline[linewidth=0.02cm](6.66,-0.45)(6.66,-0.75)
\psline[linewidth=0.02cm](7.26,-0.45)(7.26,-0.75)
\psline[linewidth=0.02cm](8.86,-0.45)(8.86,-0.75)
\end{pspicture} 
}
\caption{Simple domains corresponding to the matrices~$T$ (left) and~$\tilde{T}$ (right).}
\label{figT}
\end{figure}
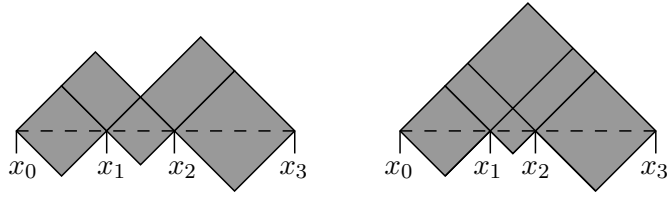
The matrix~$\tilde{T}$ is realized similarly by a simple domain with~$K=3$, but this
time without removing the square on top (see Figure~\ref{figT} on the right).
Obviously, the resulting simple domains are not isometric. We want to show that for suitable
values of the parameters, the matrices~$T^* T$ and~$\tilde{T}^* \tilde{T}$ are isospectral.
In view of Corollary~\ref{corm0}, this implies that the fermionic signature operators corresponding
to the two simple domains are isospectral.

We choose the parameters~$d=e=1$ and~$f = \delta$ with a small parameter~$\delta>0$, so that
\[ \tilde{T} = \begin{pmatrix} 1 & 1 & \sqrt{\delta} \\ 0 & 1 & \sqrt{\delta} \\ 0 & 0 & \delta \end{pmatrix} \:. \]
As a necessary condition for~$T^* T$ and~$\tilde{T}^* \tilde{T}$ to be isospectral,
the matrices~$T$ and~$\tilde{T}$ must have the same determinant
(note that these determinants are obviously positive). We satisfy this
condition by choosing
\[ c = \frac{\delta}{a b}\:. \]
Computing and comparing the characteristic polynomials of~$T^* T$ and~$\tilde{T}^* \tilde{T}$, 
one sees that it remains to satisfy the conditions
\begin{align}
-1+a^2 b^2-\delta +a \delta +b \delta -3 \delta ^2+\frac{\delta ^2}{a^2}+\frac{\delta ^2}{b^2}+\frac{\delta ^2}{a b}
&= 0 \label{eq1} \\
3-a^2-a b-b^2+2 \delta -\frac{\delta }{a}+\delta ^2-\frac{\delta ^2}{a^2 b^2} &= 0 \:.
\end{align}
Multiplying the second equation by~$a^2$ and adding the first equation, we obtain a quadratic
equation for~$b$ of positive discriminant. Choosing the explicit solution which for~$\delta=0$ and~$a=1$ gives~$b=1$
(so that we get back the matrix~$\tilde{T}$), and substituting this solution into~\eqref{eq1},
we obtain one equation for the remaining unknown~$a$. Expanding this equation
around~$\delta=0$ and~$a=1$, we obtain the condition
\[ 5 \delta - 8 (a-1)^2 + \O\big( \delta^2 \big) + \O\big( \delta\, (a-1) \big) + \O\big( (a-1)^3 \big) = 0 \:. \]
Thus for sufficiently small~$\delta>0$ there are solutions~$a$ of the form
\[ a = 1 \pm \sqrt{\frac{5 \delta}{8}} + \O(\delta) \:. \]
We thus obtain a one-parameter family of isospectral pairs of matrices~$T^* T$ and~$\tilde{T}^* \tilde{T}$.
}} \QEDrem
\end{Example}

\subsection{Representation of~$\Sig$ as an Integral Operator} \label{sec22}
Let~$\scrM \subset \R^{1,1}$ be a bounded, globally hyperbolic subset of
the Minkowski plane. As explained at the beginning of Section~\ref{secsimple}, we
again choose the Cauchy surface~$(0,b)$ at time~$t=0$.

\begin{Prp} \label{prpintrep}
The fermionic signature operator can be represented as an integral operator
\beq \label{Sig0intrep}
(\Sig \psi)(x) = \int_0^b \Sig(x,y)\, \psi(y)\, dy
\eeq
with a bounded integral kernel,
\beq \label{Sigbound}
\Sig(.,.) \in L^\infty \big( (0,b) \times (0,b), \Lin(V) \big) \:.
\eeq
\end{Prp}
\Proof
The Cauchy problem with initial data at~$t=0$ has the explicit solution
\beq \label{solCauchy}
\psi(t,x) = \begin{pmatrix} \psi_L(0, x+t) \\ \psi_R(0, x-t) \end{pmatrix} \:.
\eeq
Using this representation in~\eqref{stip}, multiplying out and estimating each term gives
\begin{align*}
\big| \bra \psi | \phi \ket \big| &\leq
\int_\scrM \Big( \|\psi_R(0, x-t)\| \, \|\phi_L(0,x+t)\| + \|\psi_L(0, x+t)\| \, \|\phi_R(0,x-t)\| \Big) \:dt \,dx \:.
\end{align*}
Estimating the integral by the integral over the whole causal diamond, we obtain
\begin{align*}
\int_\scrM &\|\psi_R(0, x-t)\| \, \|\phi_L(0,x+t)\| \: dt \,dx \\
&\leq \int_D \|\psi_R(0, x-t)\| \, \|\phi_L(0,x+t)\| \: dt \,dx \\
&= \frac{1}{2} \left( \int_0^b \|\psi_R(0, x)\| \, dx \right) \left( \int_0^b \|\phi_L(0, x')\| \, dx' \right) .
\end{align*}
We conclude that for all~$\psi, \phi \in \H_0$,
\[ \big| \bra \psi | \phi \ket \big| \leq \left( \int_0^b \|\psi(0, x)\| \, dx \right) \left( \int_0^b \|\phi(0,x')\| \, dx' \right) . \]
This means that the bilinear form~$\bra .|. \ket$ can be estimated
in terms of the $L^1$-norm of both arguments on the Cauchy surface~$t=0$.
In other words,
\[ \bra .|. \ket \in L^1 \big( (0,b) \times (0,b), dx\, dy \big)^* \:. \]
Since the dual space of~$L^1(dx\, dy)$ is the Banach space of $L^\infty$-functions
acting by weak evaluation, we conclude that there is a kernel~$\Sig(.,.) \in L^\infty(
(0,b) \times (0,b), dx\,dy)$ such that
\[  \bra \psi | \phi \ket = 2 \pi \int_0^b \int_0^b \la \psi(x), \Sig(x,y) \phi(y) \ra_{\C^2}\: dx\, dy \:. \]
Comparing with~\eqref{Sdef} and~\eqref{print2} gives the result.
\QED
We remark that the estimates used in the proof of this proposition will be generalized
and refined in Section~\ref{secreg}.

The fact that the kernel is pointwise bounded~\eqref{Sigbound} and the domain~$(0,b)$ has finite volume
implies that the trace of any even power of~$\Sig$ is finite and can be computed with the
standard formula:
\begin{Corollary} \label{cortrace}
The fermionic signature operator~$\Sig$ is Hilbert-Schmidt. Moreover, the
traces of even powers of~$\Sig^{2q}$, $q \in \N$, are given by the integrals
\beq \label{trS2q1}
\tr(\Sig^{2q}) = \int_0^b dx_1 \ldots \int_0^b dx_{2q}
\Tr \big(\Sig(x_1, x_2) \cdots \Sig(x_{2q}, x_1) \big) \:,
\eeq
where~$\tr$ denotes the trace of an operator on the Hilbert space,
and~$\Tr$ is the trace of a $(2 \times 2)$-matrix.
\end{Corollary}
\Proof According to~\eqref{Sigbound}, we know that~$\|\Sig(x,y) \| \leq C$ for almost all~$x,y \in (0,b)$
(where~$\| \cdot \|$ denotes the Hilbert-Schmidt norm of a $(2 \times 2)$-matrix). Hence
\beq \label{intes}
\int_0^b dx \int_0^b dy \:\big\| \Sig(x,y) \big\|^2 < C^2\, b^2 \:.
\eeq
We now choose on~$\H_0 = L^2((0,b), \C^2)$ the orthonormal basis~$(\psi^{n}_{c})$
with~$n \in \Z$, $c \in \{L, R\}$ given by plane waves
\[ \psi^{n}_{c} = \frac{1}{\sqrt{2 \pi b} }\:\psi_c \; e^{\frac{2 \pi i}{b}\:n x} \qquad \text{where} \qquad
\psi_L = \begin{pmatrix} 1 \\ 0 \end{pmatrix} \:,\quad
\psi_R = \begin{pmatrix} 0 \\ 1 \end{pmatrix} . \]
Then Parseval's identity for double Fourier series shows that the series
\[ \frac{1}{(2 \pi)^2} \sum_{c,c' \in \{L, R\}} \sum_{n,n' \in \Z} \big| ( \psi^n_c | \Sig \,\psi^{n'}_{c'} ) \big|^2 \]
coincides with the integral in~\eqref{intes}.
We conclude that the operator~$\Sig$ is Hilbert-Schmidt. Moreover, the trace of~$\Sig^2$ can be computed
by~\eqref{trS2q1} specialized to the case~$q=1$.

By iterating~\eqref{Sig0intrep} and using Fubini's theorem,
one obtains an integral representation of the operator~$\Sig^q$
again with a pointwise bounded kernel. Repeating the above argument with~$\Sig$ replaced by the
operator~$\Sig^q$, we conclude that also the operator~$\Sig^q$ is Hilbert-Schmidt,
and that its Hilbert-Schmidt norm can be computed by~\eqref{trS2q1}. This concludes the proof.
\QED
This corollary shows in particular that the operator~$\Sig$ is compact.
Thus~$\Sig$ has a pure point spectrum and finite-dimensional eigenspaces. Moreover, the eigenvalues can
accumulate only at the origin.

\subsection{Symmetry of the Spectrum} \label{sec23}
In Corollary~\ref{corm0}~(i) we saw that for a simple domain, the spectrum is symmetric with respect
to the origin. The next proposition shows why this is true even for general domains.

\begin{Prp} \label{prpsymm} The spectrum of~$\Sig$ is symmetric with respect to the origin.
\end{Prp}
\Proof The matrix
\beq \label{pseudodef}
\pseudo := \begin{pmatrix} -1 & 0 \\ 0 & 1 \end{pmatrix}
\eeq
obviously anti-commutes with the Dirac matrices~\eqref{gamma} and 
the Dirac operator~\eqref{Dir}, i.e.\ $\pseudo \Dir = - \Dir \pseudo$.
Hence if~$\psi$ is a solution the massless Dirac equation, the same is true for~$\pseudo \psi$.
In other words, $\pseudo$ maps the solution space of the Dirac equation to itself,
\[ \pseudo \::\: \H_0 \rightarrow \H_0 \:. \]
Using~\eqref{ssprod}, one sees that~$\Gamma$ is anti-symmetric with respect to the
spin scalar product~$\Sl \psi | \pseudo \phi \Sr = - \Sl \pseudo \psi | \phi \Sr$.
Moreover, using~\eqref{print} and~\eqref{stip}, we find that for all~$\psi, \phi \in \H_0$,
\begin{align}
\bra \psi | \pseudo \phi \ket &= \int_\scrM \Sl \psi | \pseudo \phi \Sr_x\: d\mu(x) 
= - \int_\scrM \Sl \pseudo \psi | \phi \Sr_x\: d\mu(x) = -\bra \pseudo \psi | \phi \ket \label{symm1} \\
(\psi | \pseudo \phi) &= 2 \pi \int_\scrN \Sl \psi | \nuslsh \pseudo \phi \Sr_x\: d\mu_\scrN(x)
=-2 \pi \int_\scrN \Sl \psi |  \pseudo \nuslsh \phi \Sr_x\: d\mu_\scrN(x) \notag \\
&=2 \pi \int_\scrN \Sl \pseudo \psi | \nuslsh \phi \Sr_x\: d\mu_\scrN(x) = (\pseudo \psi | \phi) \:. \label{symm2}
\end{align}

In view of~\eqref{Sdef}, the eigenvalue equation~$\Sig \psi = \lambda \psi$ can be written as
\[ \bra \phi | \psi \ket = \lambda\, (\phi | \psi) \qquad \text{for all~$\phi \in \H_0$}\:. \]
Using the symmetries~\eqref{symm1} and~\eqref{symm2}, one sees that
if~$\psi$ is an eigenvector corresponding to the eigenvalue~$\lambda$,
then~$\pseudo \psi$ is an eigenvector corresponding to the eigenvalue~$-\lambda$.
\QED

\subsection{Computation of~$\tr(\Sig^{2q})$, Recovering the Volume} \label{sectrace}
Proposition~\ref{prpintrep} also implies that the operators~$\Sig^p$ are trace class for any~$p \in \N$.
The symmetry of the spectrum shown Proposition~\ref{prpsymm} implies that the trace
of an odd power of~$\Sig$ vanishes.
We now want to compute the trace of even powers of~$\Sig$.
For computational purposes, it is most convenient to work with the causal fundamental
solution in light cone coordinates. In order to keep the setting as simple as possible,
we here introduce the causal fundamental solution~$k_0$ simply as a device
for expressing the solution of the Cauchy problem.

\begin{Lemma} The solution~$\psi$ of the Cauchy problem
\[ \Dir \psi = 0 \:,\qquad \psi|_{t=0} = \psi_0 \in C^0((0,b)) \]
has the representation
\beq \label{solCauchy2}
\psi(t,x) = 2 \pi \int_0^b k(t,x-y)\, \gamma^0\, \psi_0(y)\: dy\:,
\eeq
where~$k(t,x)$ is the distribution
\beq \label{k0}
k(t,x) = \frac{1}{4\pi} \,(\gamma^0+\gamma^1) \:\delta(t+x) +
\frac{1}{4\pi} \,(\gamma^0-\gamma^1) \:\delta(t-x) \:.
\eeq
\end{Lemma}
\Proof A direct computation using~\eqref{gamma} shows that~\eqref{solCauchy2}
indeed agrees with~\eqref{solCauchy}.
\QED
The distribution~$k(t,x)$ is referred to as the {\em{causal fundamental solution}}.
At first sight, the method of this lemma seems unnecessarily complicated,
because~\eqref{solCauchy} is much simpler than~\eqref{solCauchy2}.
The advantage of~\eqref{solCauchy2} is that this formula generalizes
to the massive case (see Section~\ref{secgenmass}) and even to globally hyperbolic
space-times in arbitrary dimension (see for example~\cite[Lemma~2.1]{finite}).

The integral in~\eqref{solCauchy2} can be regarded as a {\em{time evolution operator}}
which maps the solution at some initial time~$t=0$ to the solution at a final time~$t$.
Clearly, if one first takes the time evolution from time~$t_0$ to~$t_1$ and then
the time evolution from~$t_1$ to~$t_2$, one gets the same as if one takes the
time evolution directly from~$t_0$ to~$t_2$. This fact is often referred to as
a {\em{group property}} of the time evolution, where the group operation is the multiplication
of the time evolution operators and the inverse of the time evolution from~$t_0$ to~$t_1$
is the time evolution from~$t_1$ to~$t_0$. This group property is reflected in a property
of the causal fundamental solution. Namely, denoting a space-time point for simplicity
by~$\zeta=(t,x)$, we have for any~$\zeta, \tilde{\zeta} \in \scrM$,
\beq \label{semigroup0}
k(\zeta-\zeta') = 2 \pi \int_0^b k(\zeta^0, \zeta^1-x)\, \gamma^0\, k(-\tilde{\zeta}^0, x-\tilde{\zeta}^1)\: dx
\eeq
(this relation can also be verified by direct computation using~\eqref{k0}).
Moreover, one sees directly from~\eqref{k0} that the kernel~$k(t,x)$ is symmetric in the sense that
\beq \label{ksymm}
k(t,x)^* = k(-t,-x)
\eeq
(where the star denotes the adjoint with respect to the inner product~$\Sl .|. \Sr_x$).

We next express the integral kernel of~$\Sig$ in terms of the causal fundamental solution.
\begin{Lemma} \label{lemma0kernel} The kernel~$\Sig(x,y)$ of the fermionic signature operator
in~\eqref{Sig0intrep} can be written as
\[ \Sig(x,y) = 2 \pi \int_\scrM  k(-t, x-z) \,k(t, z-y)\:\gamma^0 \: dt\, dz \:. \]
\end{Lemma}
\Proof Using~\eqref{solCauchy2} in~\eqref{stip} and applying~\eqref{ksymm}, we obtain
\begin{align*}
\bra \psi | \phi \ket &= 4 \pi^2 \int_\scrM dt\, dz \int_0^b dx \int_0^b dy \:\Sl k(t, z-x)\,\gamma^0\, \psi(0,x)
\:|\: k(t, z-y)\,\gamma^0\, \phi(0,y)\Sr \\
&= 4 \pi^2 \int_\scrM dt\, dz \int_0^b dx \int_0^b dy \:\Sl \psi(0,x) \:|\: \gamma^0 \,k(-t, x-z)\: k(t, z-y)\,\gamma^0
\phi(0,y)\Sr \:.
\end{align*}
On the other hand, we know from~\eqref{Sdef} and~\eqref{Sig0intrep} that
\[ \bra \psi | \phi \ket = \Big( \psi \,\Big|\, \int_0^b \Sig(x,y)\, \phi(0,y)\:dy  \Big) \:. \]
Comparing these formulas gives the result.
\QED

The {\em{light-cone coordinates}} $(u,v)$ are defined by
\beq \label{uvdef}
u = t+x \qquad \text{and} \qquad v = t-x \:.
\eeq
Then~$du \,dv = 2 dt \,dx$ and
\begin{align*}
t &= \frac{u+v}{2} \:,& \hspace*{-2cm} x &= \frac{u-v}{2} \\
\pu &= \frac{1}{2} \left( \partial_t + \partial_x \right)
 \:,& \hspace*{-2cm} \pv &= \frac{1}{2} \left( \partial_t - \partial_x \right)
\end{align*}
(to improve the readability we denote the indices~$u$ and~$v$ in roman style).
Setting
\beq \label{guvdef}
\gu = \gamma^0 + \gamma^1 \qquad \text{and} \qquad \gv = \gamma^0 - \gamma^1 \:,
\eeq
we obtain the anti-commutation relations
\beq \label{ac}
\big(\gu\big)^2 = 0 = \big(\gv\big)^2 \:,\qquad
\big\{ \gu, \gv \big\} = 4\:.
\eeq
The Dirac operator~\eqref{Dir} and the causal fundamental solution~\eqref{k0} become
\begin{align}
\D &= i \gu \pu + i \gv \pv \label{Duv} \\
k(u,v) &= \frac{1}{4\pi} \,\Big( \gu \:\delta(u) + \gv\: \delta(v) \Big) \:. \label{k0uv}
\end{align}

Combining Lemma~\ref{lemma0kernel} with the integral representation of Proposition~\ref{prpintrep},
we can compute powers of the operator~$\Sig$. For example,
\begin{align*}
\big(\Sig^2\big)(x,y) &= \int_0^b \Sig(x,z)\, \Sig(z,y)\, dz \\
&= 4 \pi^2 \int_0^b dz \int_\scrM d^2\zeta \int_\scrM d^2 \tilde{\zeta} \;k\big(-\zeta^0, x-\zeta^1\big)\,
k\big(\zeta^0, \zeta^1-z\big) \:\gamma^0 \\
&\hspace*{4.1cm} \times k\big(-\tilde{\zeta}^0, z-\tilde{\zeta}^1\big) \, k\big(\tilde{\zeta}^0, \tilde{\zeta}^1-y\big) 
\: \gamma^0\:.
\end{align*}
Now we can carry out the $z$-integral using the group property~\eqref{semigroup0}. This gives
\begin{align*}
\big(\Sig^2\big)(x,y) &= 2 \pi
\int_\scrM d^2 \zeta \int_\scrM d^2 \tilde{\zeta} \: k\big(-\zeta^0, x-\zeta^1\big)\:
k\big(\zeta-\tilde{\zeta}\big) \: k\big(\tilde{\zeta}^0, \tilde{\zeta}^1-y\big) \: \gamma^0 \:.
\end{align*}
Iterating this method, we obtain
\begin{align*}
\big(\Sig^p\big)(x,y) &= 2 \pi \int_\scrM d^2 \zeta_1 \cdots \int_\scrM d^2 \zeta_p \\
&\quad \times
k\big(-\zeta_1^0, x-\zeta_1^1\big)\, k\big(\zeta_1-\zeta_2\big) \cdots k\big(\zeta_{p-1}-\zeta_p\big)\,
k\big(\zeta_p^0, \zeta_p^1-y\big)\:\gamma^0 \:.
\end{align*}
Taking the trace with the help of Corollary~\ref{cortrace}, one
can again apply~\eqref{semigroup} to obtain
\beq \label{trSp}
\tr \big(\Sig^{2q} \big) = \int_\scrM d^2 \zeta_1 \cdots \int_\scrM d^2 \zeta_{2q}
\;\Tr \Big( k\big(\zeta_1-\zeta_2\big) \cdots k\big(\zeta_{2q-1}-\zeta_{2q} \big)\,
k\big(\zeta_{2q} - \zeta_1 \big) \Big)\:.
\eeq
This formula is generally covariant, showing in particular that the trace of~$\Sig^{2p}$ is indeed
independent of the choice of the Cauchy surface.

The above formulas can be simplified considerably using the form of the
causal fundamental solution in light-cone coordinates~\eqref{k0uv} and~\eqref{ac}. Namely,
\begin{align*}
k_0 & \big(\zeta_1-\zeta_2\big) \cdots k_0\big(\zeta_{p-1}-\zeta_p\big)\,
k_0\big(\zeta_p - \zeta_1 \big) \\
=\,& \frac{1}{(4 \pi)^p}
\gu \:\delta(u_1-u_2) \, \gv \:\delta(v_2-v_3) \cdots
+ \frac{1}{(4 \pi)^p}
\gv \:\delta(v_1-v_2) \, \gu \:\delta(u_3-u_3) \cdots \:.
\end{align*}
Taking the trace and again using~\eqref{ac}, we get zero if~$p$ is odd.
If~$p$ is even, we obtain
\beq \label{Trform}
\begin{split}
\Tr \Big( k_0 & \big(\zeta_1-\zeta_2\big) \cdots k_0\big(\zeta_{p-1}-\zeta_p\big)\,
k_0\big(\zeta_p - \zeta_1 \big) \Big) \\
=\,& \frac{1}{(2 \pi)^p} \: \delta(u_1-u_2) \,\delta(v_2-v_3)\, \cdots \delta(u_{p-1}-u_p) \,\delta(v_p-v_1) \\
&+\frac{1}{(2 \pi)^p} \: \delta(v_1-v_2) \,\delta(u_2-u_3)\, \cdots \delta(v_{p-1}-v_p) \,\delta(u_p-u_1)\:.
\end{split}
\eeq

Setting~$p=2$, we see that the result of Corollary~\ref{corm0}~(iii) also holds
for general domains:
\begin{Prp} \label{prpS2} Let~$\scrM$ be a bounded, globally hyperbolic subset of Minkowski space. Then
the trace of~$\Sig^2$ encodes the space-time volume,
\[ \tr \big( \Sig^2 \big) = \frac{\mu(\scrM)}{4 \pi^2}\:. \]
\end{Prp}
\Proof Using~\eqref{Trform} in~\eqref{trSp} in the case~$p=2$, we obtain
\begin{align*}
\tr \big( \Sig^2 \big) &= \int_\scrM d^2 \zeta_1 \int_\scrM d^2 \zeta_2\: \frac{2}{(2 \pi)^2} \:
\delta(u_1-u_2) \, \delta(v_1-v_2) \\
&= \frac{1}{(2 \pi)^2} \int_\scrM d^2 \zeta_1 \int_\scrM d^2 \zeta_2\:  \: \delta^2(\zeta_1-\zeta_2)
= \frac{\mu(\scrM)}{4 \pi^2} \:,
\end{align*}
giving the result.
\QED

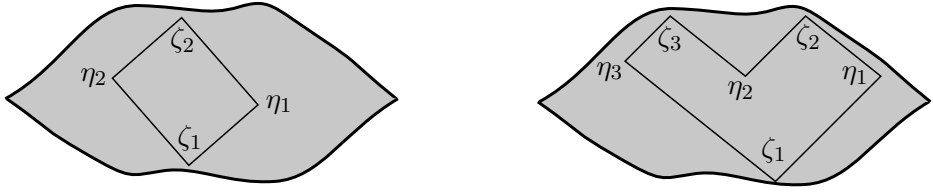
\begin{figure}
\scalebox{1} 
{
\begin{pspicture}(0,-1.2809438)(12.32,1.2809438)
\definecolor{color349g}{rgb}{0.7843137254901961,0.7843137254901961,0.7843137254901961}
\psbezier[linewidth=0.04,fillstyle=gradient,gradlines=2000,gradbegin=color349g,gradend=color349g,gradmidpoint=1.0](12.28,-0.12094381)(11.56,-0.54799604)(11.26,-1.1809438)(10.64,-1.2209438)(10.02,-1.2609438)(9.610406,-1.0117922)(9.16,-1.079395)(8.709594,-1.1469979)(8.74,-1.2073541)(8.12,-0.8452173)(7.5,-0.48308054)(7.78,-0.6181685)(7.08,-0.12094381)
\psbezier[linewidth=0.04,fillstyle=gradient,gradlines=2000,gradbegin=color349g,gradend=color349g,gradmidpoint=1.0](7.1,-0.12094381)(7.9,0.31905618)(8.2,1.1390562)(8.88,1.1190562)(9.56,1.0990562)(9.789594,0.97716856)(10.2,1.0990562)(10.610406,1.2209438)(10.64,1.1206678)(11.26,0.7067973)(11.88,0.29292676)(11.6,0.44731298)(12.3,-0.12094381)
\psbezier[linewidth=0.04,fillstyle=gradient,gradlines=2000,gradbegin=color349g,gradend=color349g,gradmidpoint=1.0](0.0,-0.08094381)(0.8,0.3590562)(1.1,1.1790562)(1.78,1.1590562)(2.46,1.1390562)(2.689594,1.0171685)(3.1,1.1390562)(3.510406,1.2609438)(3.54,1.1606679)(4.16,0.7467973)(4.78,0.33292675)(4.5,0.48731297)(5.2,-0.08094381)
\psbezier[linewidth=0.04,fillstyle=gradient,gradlines=2000,gradbegin=color349g,gradend=color349g,gradmidpoint=1.0](5.22,-0.08094381)(4.5,-0.507996)(4.2,-1.1409438)(3.58,-1.1809438)(2.96,-1.2209438)(2.550406,-0.97179216)(2.1,-1.039395)(1.6495941,-1.1069978)(1.68,-1.167354)(1.06,-0.80521727)(0.44,-0.44308054)(0.72,-0.5781685)(0.02,-0.08094381)
\rput{-48.784664}(0.80445576,1.8051852){\psframe[linewidth=0.02,dimen=outer](3.1734788,0.6365768)(1.6119018,-0.6054354)}
\usefont{T1}{ptm}{m}{n}
\rput(2.46,-0.5959438){$\zeta_1$}
\usefont{T1}{ptm}{m}{n}
\rput(3.64,-0.15594381){$\eta_1$}
\usefont{T1}{ptm}{m}{n}
\rput(2.36,0.7040562){$\zeta_2$}
\usefont{T1}{ptm}{m}{n}
\rput(1.18,0.20405619){$\eta_2$}
\pspolygon[linewidth=0.02](8.84,1.0190561)(8.24,0.41905624)(10.24,-1.1809437)(11.64,0.21905623)(10.64,1.0190562)(9.84,0.21905623)
\usefont{T1}{ptm}{m}{n}
\rput(10.22,-0.75594383){$\zeta_1$}
\usefont{T1}{ptm}{m}{n}
\rput(11.3,0.22405618){$\eta_1$}
\usefont{T1}{ptm}{m}{n}
\rput(10.68,0.74405617){$\zeta_2$}
\usefont{T1}{ptm}{m}{n}
\rput(9.78,0.02405619){$\eta_2$}
\usefont{T1}{ptm}{m}{n}
\rput(8.84,0.7240562){$\zeta_3$}
\usefont{T1}{ptm}{m}{n}
\rput(8.04,0.2840562){$\eta_3$}
\end{pspicture} 
}
\caption{The geometric constraints given by~$\Theta(\zeta_1, \ldots, \zeta_q)$
for~$q=2$ (left) and~$q=3$ (right).}
\label{figconstraint}
\end{figure}

For general~$q$, one can compute the trace of~$\Sig^{2q}$ with the following method. First, by
renaming the variables one sees that the two summands in~\eqref{Trform} give the same
contribution to the trace. Thus
\beq
\begin{split}
\tr \big( \Sig^{2q} \big)
&= \frac{2}{(2 \pi)^{2q}} \int_\scrM d^2 \zeta_1 \int_\scrM d^2 \eta_1 \cdots 
\int_\scrM d^2 \zeta_q \int_\scrM d^2 \eta_q \\
&\qquad \times
\delta(u_1-\tilde{u}_1) \,\delta(\tilde{v}_1-v_2)\, \cdots \delta(u_q-\tilde{u}_q) \,\delta(\tilde{v}_q-v_1)\:,
\end{split} \label{trS2q}
\eeq
where the points~$\eta_j$ have the coordinates~$(\tilde{u}_j, \tilde{v}_j)$.
Carrying out the integrals over $\eta_1, \ldots, \eta_q$, we obtain
\begin{align*}
\tr \big( \Sig^{2q} \big)
&=\frac{2}{(2 \pi)^{2q}}\: \frac{1}{2^q} \int_\scrM d^2 \zeta_1 \cdots \int_\scrM d^2 \zeta_q\:
\Theta(\zeta_1, \ldots, \zeta_q) \:,
\end{align*}
where the function~$\Theta(\zeta_1, \ldots, \zeta_q)$, which takes the values zero and one, 
gives geometric constraints for the position of the points~$\zeta_1, \ldots, \zeta_q$.
More precisely, introducing the points
\begin{align*}
\eta_1 = (\zeta_1^0, \zeta_2^1)\;, \;\;
\eta_2 = (\zeta_2^0, \zeta_3^1)\;, \quad \ldots \quad
\eta_{q-1} = (\zeta_{q-1}^0, \zeta_q^1) \;, \;\;
\eta_q = (\zeta_q^0, \zeta_1^1) \:,
\end{align*}
the function~$\Theta$ is defined by
\[ \Theta(\zeta_1, \ldots, \zeta_q) = \left\{ \begin{array}{ll}
1 & \text{if~$\eta_1, \ldots, \eta_q \in \scrM$} \\
0 & \text{otherwise}\:. \end{array} \right. \]
The geometric constraints are illustrated in Figure~\ref{figconstraint}.
Via these constraints, the trace of~$\Sig^p$ depends on the
geometry of the boundary curves of~$\scrM$. While these constraints are rather complicated in general, 
in the case~$q=2$ they can be easily understood giving the following result.
\begin{Prp} \label{prp71}
Denoting by~$J(\zeta)$ the set of all points which can be joined from~$\zeta$ by a
causal curve,
\beq \label{trS4}
\tr(\Sig^4) = \frac{1}{8 \pi^2} \int_\scrM \mu\big( \scrM \cap J(\zeta) \big) \:d^2\zeta \:.
\eeq
\end{Prp}
\Proof For fixed~$\zeta_1$, the region where~$\Theta(\zeta_1, \zeta_2)=1$
coincides with~$J(\zeta_1)$ (see the left of Figure~\ref{figconstraint}).
This gives~\eqref{trS4}.
\QED

\subsection{Length of Causal Curves and the Largest Eigenvalue} \label{seclarge}
We now turn attention to the length of causal curves.
The {\em{length}}~$\ell(\alpha)$ of a causal curve~$\alpha : (0,1) \rightarrow \scrM$
is defined by
\beq \label{lengthtimelike}
\ell(\alpha) = \int_0^1 \sqrt{ \big( \alpha'(\tau)^0 \big)^2 - \big( \alpha'(\tau)^1 \big)^2 } \, d\tau\:.
\eeq

\begin{Prp} \label{prp81}
Let~$\alpha$ be a causal curve.
Then the length of this curve (as defined by~\eqref{lengthtimelike})
is bounded in terms of the largest eigenvalue~$\lambda$ of~$\Sig$ by
\beq \label{lamlower}
\lambda \geq \frac{\ell}{4 \pi} \:.
\eeq
This inequality is sharp if~$\scrM$ is a causal diamond.
Conversely, if equality holds in~\eqref{lamlower} and~$\scrM$ is connected,
then~$\scrM$ is a causal diamond.
\end{Prp}
\Proof Let~$\alpha$ be a timelike curve with end points~$\zeta$ and~$\zeta'$.
We let~$D$ be the causal diamond whose upper and lower corners are~$\zeta$ and~$\zeta'$,
and again denote the left and right corners by~$\eta$ and~$\eta'$.
We may replace~$\alpha$ by the straight line joining~$\zeta$ and~$\zeta'$, because this
increases the length of~$\alpha$.
By a Lorentz transformation we can arrange that~$\eta=0$ and~$\eta'=(0, \ell)$ with~$\ell>0$.
We choose on~$D$ the orthonormal functions
\[ \psi_L = \frac{1}{\sqrt{2 \pi \ell}} \begin{pmatrix} 1 \\ 0 \end{pmatrix}
\:,\qquad \psi_R = \frac{1}{\sqrt{2 \pi \ell}} \begin{pmatrix} 0 \\ 1 \end{pmatrix} \:. \]
Then
\[ \lambda \geq \big|( \psi_L | \Sig \psi_R ) \big| = \big| \bra \psi_L | \psi_R \ket \big| =
\frac{1}{2 \pi \ell} \int_D dt\, dx = \frac{1}{2 \pi \ell}\: \frac{\ell^2}{2}
= \frac{\ell}{4 \pi} \:, \]
giving the result.

If~$\scrM$ is a causal diamond, the inequality~\eqref{lamlower}
is sharp according to Lemma~\ref{lemmasimple} in case~$K=1$. In order to prove the converse statement,
assume that~$\scrM$ is connected and that equality holds in~\eqref{lamlower}.
Then~$\lambda = |( \psi_L | \Sig \psi_R )|$, and the Rayleigh-Ritz principle
implies that the wave functions~$\psi_L \pm \psi_R$ are eigenvectors of~$\Sig$ with eigenvalues~$\pm \lambda$.
Next, we choose a Cauchy surface~$\scrN$ which goes through the left and right corner
of the above causal diamond~$D$. Then any wave function~$\psi$ on~$\scrN$ which is supported
outside~$D \cap \scrN$ is obviously orthogonal to~$\psi_L$ and~$\psi_R$.
Using that the vectors~$\psi_L \pm \psi_R$ are eigenvectors, it follows that
\[ 0 = ( \psi | \Sig \psi_{L\!/\!R} ) = \bra \psi | \psi_{L\!/\!R} \ket \:. \]
Choosing~$\psi$ as a piecewise constant function, one sees that the causal future and
past of~$\scrN \setminus D$ does not intersect the causal future and past of~$D$.
Since~$\scrM$ is connected, we conclude that~$\scrM \setminus D$ is empty.
This gives the result.
\QED

\subsection{Length of Spacelike Curves and~$\tr(\Sig_+)$} \label{sectrSp}
We now come to the analysis of the length of spacelike curves.
The {\em{length}}~$\ell(\alpha)$ of a spacelike curve~$\alpha : (0,1) \rightarrow \scrM$
is defined by
\[ \ell(\alpha) = \int_0^1 \sqrt{ \big( \alpha'(\tau)^1 \big)^2 - \big( \alpha'(\tau)^0 \big)^2 } \, d\tau\:. \]
We first note that for globally hyperbolic subsets of Minkowski space, the supremum
of the length of spacelike curves is always {\em{larger}} or equal to the supremum of the
lengths of causal curves,
\beq \label{ineqlength}
\sup_{\alpha \text{ causal}} \ell(\alpha) \leq \sup_{\alpha \text{ spacelike}} \ell(\alpha) \:.
\eeq
Namely, if~$\alpha$ is a causal curve with end points~$\zeta$ and~$\zeta'$,
then the corresponding causal diamond~$D(\zeta, \zeta')$ is contained in the space-time~$\scrM$.
But then the straight line joining the left and right corners of the causal diamond is a spacelike curve
whose length is at least as large as that of~$\alpha$.
An example for a space-time where the maximal length of spacelike curves is much larger than
the length of causal curves is shown on the left of Figure~\ref{figcauchy}.

\begin{figure}
\scalebox{1} 
{
\begin{pspicture}(0,-0.48157784)(4.232671,0.48157784)
\rput{-48.784664}(0.21908844,0.205681){\psframe[linewidth=0.02,dimen=outer](0.5341119,0.13514483)(0.13855912,-0.41261348)}
\psbezier[linewidth=0.04](0.012671014,-0.16705288)(1.3126711,-0.067052886)(0.812671,0.032947112)(1.8126711,0.19294712)(2.812671,0.35294712)(3.572671,0.112947114)(4.212671,-0.34705287)
\rput{-48.784664}(0.33077443,0.671979){\psframe[linewidth=0.02,dimen=outer](1.0335554,0.24047744)(0.7791156,-0.2979461)}
\rput{-48.784664}(0.40784535,1.1053042){\psframe[linewidth=0.02,dimen=outer](1.5497818,0.3239513)(1.2955602,-0.11805707)}
\rput{-48.784664}(0.50317603,1.5155344){\psframe[linewidth=0.02,dimen=outer](2.0780044,0.42581826)(1.7673377,-0.019924026)}
\rput{-48.784664}(0.64093024,1.8793198){\psframe[linewidth=0.02,dimen=outer](2.5508049,0.41348436)(2.2345371,0.052409858)}
\rput{-48.784664}(0.83816326,2.2342722){\psframe[linewidth=0.02,dimen=outer](3.103839,0.38474068)(2.661503,0.0011535607)}
\rput{-48.784664}(1.1242645,2.5852034){\psframe[linewidth=0.02,dimen=outer](3.6592064,0.18264006)(3.1661355,-0.07674583)}
\rput{-48.784664}(1.4821786,2.8945014){\psframe[linewidth=0.02,dimen=outer](4.237519,-0.08184862)(3.6278234,-0.29225716)}
\end{pspicture} 
}
\caption{Approximating a spacelike curve by lightlike rectangles.}
\label{figapprox}
\end{figure}
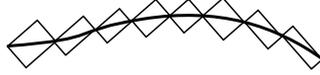

The inequality~\eqref{ineqlength} suggests that for estimating the length of spacelike curves,
it is not sufficient to consider a single eigenvalue. Instead, one should form a 
suitable sum of eigenvalues.
More precisely, we must consider the trace of the operator~$\Sig_+$ defined
as the positive spectral part of~$\Sig$,
\[ \Sig_+ = \chi_{(0, \infty)}(\Sig) \,\Sig \:. \]
\begin{Prp} \label{prp91} Let~$\alpha : [0,1] \rightarrow \scrM$ be a spacelike curve.
Then the length of the curve is bounded from above by the trace of~$\Sig_+$,
\beq \label{lengthestim}
\tr(\Sig_+)  \geq \frac{\ell(\alpha)}{4 \pi} \:.
\eeq
\end{Prp}
\Proof By a Poincar{\'e} transformation we can arrange that the two end points of the curve~$\alpha$
lie on the $x$-axis. We approximate~$\alpha$ by rectangles with lightlike sides
(see Figure~\ref{figapprox}).
By choosing the rectangles sufficiently small (and possibly leaving out small parts at the beginning or
end of the curve), we can arrange that the rectangles all lie in~$\scrM$.
Next, we choose a simple domain such that the rectangles are all part of the rectangles
of the simple domain. Thus we can label the rectangles which approximate the curve by
\[ \Delta^{k_1}_{l_1}, \ldots, \Delta^{k_N}_{l_N} \:. \]
The fact that the curve is spacelike implies that the indices~$k_j$
are all different, and the indices~$l_j$ are also all different.
Choosing again the orthonormal basis~\eqref{ONB}, this implies that the vectors
\[ \psi^{k_1,0}_R, \ldots, \psi^{k_N,0}_R \qquad \text{and} \qquad  \psi^{l_1,0}_L, \ldots, \psi^{l_N,0}_L \]
are orthonormal. Moreover,
\beq \label{lengthid}
\sum_{j=1}^N \bra \psi^{k_j,0}_R | \psi^{l_j,0}_L \ket
= \frac{1}{2 \pi \, \sqrt{2}} \sum_{j=1}^N \sqrt{\mu(\Delta^{k_j}_{l_j})} 
\rightarrow \frac{\ell(\alpha)}{4 \pi} \:,
\eeq
where the last convergence refers to the limit where the size of the rectangles tends to zero
(note that area of a lightlike rectangle is half the square of the Minkowski length of its diagonals).

It remains to show that the trace of~$\Sig_+$ bounds the left side of~\eqref{lengthid}.
Similar as shown in Lemma~\ref{lemmasimple} for simple domains, in the massless case
the fermionic signature operator is block off-diagonal,
\[ \Sig = \begin{pmatrix} 0 & \Sig_R \\ \Sig_L & 0 \end{pmatrix}\:. \]
Exactly as in Corollary~\ref{corm0}, one concludes that
\beq \label{trsigp}
\tr(\Sig_+) = \tr \left( \sqrt{\Sig_L^* \Sig_L} \right) \:.
\eeq
We denote the eigenvalues of~$\Sig_L^* \Sig_L$ (counting with multiplicities)
by~$\nu_n$, with the ordering
\[ \nu_1 \geq \nu_2 \geq \nu_3 \cdots \:. \]
Let~$(\psi_n)$ be a corresponding orthonormalized eigenvector basis.
The computation
\[ ( \Sig_L \psi_n \,|\, \Sig_L \psi_{n'}) = ( \psi_n \,|\, \Sig_L^* \Sig_L \psi_{n'})
= \nu_{n}\: \delta_{n,n'} \]
shows that the~$\psi_n$ are mapped to orthogonal vectors.
Choosing~$\phi_n$ as orthonormal vectors which are collinear to~$\Sig_L \psi_n$,
we conclude that
\beq \label{nuform}
\nu_n = (\psi_n \,|\, \Sig_L^* \Sig_L \psi_n) 
= (\Sig_L \psi_n \,|\,\Sig_L \psi_n) = \big| (\phi_n | \Sig_L \psi_n) \big|^2 \:.
\eeq
Taking the square root and using~\eqref{trsigp}, we obtain
\[ \tr(\Sig_+) = \sum_{n=1}^\infty \big| (\phi_n | \Sig_L \psi_n) \big| \:. \]

Now we can apply the min-max principle in the following way
(for basics see~\cite[Section~XIII.1]{reed+simon4}).
The first~$n$ eigenvalues can be computed by noting that the
orthonormal sets~$\psi_1, \ldots, \psi_n$ and~$\phi_1, \ldots, \phi_n$
maximize the expression in~\eqref{nuform}, in the sense that
\begin{align*}
\nu_1 &= \sup_{\phi_1, \psi_1 \text{with}\|\phi_1\| = \|\psi_1\|=1}
\big| (\phi_1 | \Sig_L \psi_1) \big|^2 \\
\nu_n &= \sup_{\stackrel{\phi_n, \psi_n \text{with}\|\phi_n\| = \|\psi_n\|=1,}
{\phi_n \perp \phi_1,\ldots, \phi_{n-1},\;\; \psi_n \perp \psi_1,\ldots, \psi_{n-1}}}
\big| (\phi_n | \Sig_L \psi_n) \big|^2 \:.
\end{align*}
As a consequence, we can estimate the trace of~$\Sig_+$ from below by adding up the
absolute values of the expectation values for any pair of orthonormal systems~$(\phi_n)$
and~$(\psi_n)$. In particular,
\[ \tr(\Sig_+) \geq \sum_{i=1}^N \big| ( \psi^{k_j,0}_R | \Sig_L \psi^{l_j,0}_L) \big| \:. \]
Noting that the last inner products coincide precisely with the summands in~\eqref{lengthid},
which are all non-negative, the result follows.
\QED
Exactly as for the inequality~\eqref{lamlower}, the inequality~\eqref{lengthestim} is sharp
for a simple domain consisting of one causal diamond (see Lemma~\ref{lemmasimple} in the case~$K=1$).
But it is not sharp general, as can be seen in Example~\ref{excounter}:
The fermionic signature operators~$\Sig$ and~$\tilde{\Sig}$ are isospectral, so that the left
side of~\eqref{lengthestim} coincides.
But the length of the longest spatial curve on the right of~\eqref{lengthestim}
is different for the simple domains corresponding to~$\Sig$ and~$\tilde{\Sig}$.
More precisely, these lengths are given by~$a+b+c = a + b + \delta/ab$ respectively~$d+e+f = 2 + \delta$. 
Analyzing the equations for~$a$ and~$b$ for small~$\delta$, 
one finds that these lengths differ by a term~$\sim \sqrt{\delta}$.

\subsection{A Reconstruction Theorem} \label{secreconstruct}
As shown in Example~\ref{excounter}, the spectrum of~$\Sig$ in general does not
determine the geometry of~$\scrM$. This raises the question which additional
structures must be given in order to encode the geometry completely.
We propose that in order to describe the space-time geometry, one should not consider~$\Sig$
as an operator on an abstract Hilbert space, but instead as an operator on a space of
spinorial functions on a given Cauchy surface. More precisely, we use the
identification
\[ \H_0 = L^2(\scrN, S\scrM) \simeq L^2(\scrN, S\scrN) \oplus L^2(\scrN, S\scrN) \:, \]
where the two direct summands correspond to the left- and right-handed components
of the spinors, respectively (note that the fiber of the intrinsic spinor bundle~$S\scrN$ is
isomorphic to~$\C$).
Knowing~$\Sig$ on the hypersurface~$\scrN$, the geometry of the ambient space-time
is completely determined:

\begin{Thm} \label{thmreconstruct}
Let~$\Sig$ and~$\tilde{\Sig}$ be the fermionic signature operators corresponding
to two bounded, globally hyperbolic sets~$\scrM, \,\,\tilde{\!\!\scrM} \subset \R^{1,1}$.
Moreover, let~$\scrN$ and~$\,\,\tilde{\!\!\scrN}$ be two Cauchy surfaces in~$\scrM$ respectively~$\,\,\tilde{\!\!\scrM}$
of the same length (with respect to the induced Riemannian metrics).
Identifying~$\scrN$ and~$\,\,\tilde{\!\!\scrN}$ via an isometry, and also identifying the corresponding
Hilbert spaces via this isometry,
\[ \H_0 = L^2(\scrN, S\scrN) \oplus L^2(\scrN, S\scrN) \simeq L^2(\,\,\tilde{\!\!\scrN}, S\,\,\tilde{\!\!\scrN}) \oplus L^2(\,\,\tilde{\!\!\scrN}, S\,\,\tilde{\!\!\scrN}) \:, \]
we assume that
\[ \Sig = \tilde{\Sig}\:. \]
Then there is a Poincar{\'e} transformation~$\Lambda \in 
\R^{1,1} \rtimes \SO(1,1)$ which maps the space-time regions and the corresponding
Cauchy surfaces to each other, i.e.
\[ \Lambda \scrM = \,\,\tilde{\!\!\scrM} \qquad \text{and} \qquad \Lambda \scrN = \,\,\tilde{\!\!\scrN}\:. \]
\end{Thm} \noindent
We note for clarity that~$\SO(1,1)$ are the isometries of~$\R^{1,1}$ of determinant one.
These transformations include reflections at the origin (which flip both the spatial orientation
and the time orientation), but they do {\em{not}} include time reversals nor
spatial inversions.

To avoid repetitions, the proof will be given in Section~\ref{secreconstruct2} for general surfaces.

\section{The Massive Case} \label{secmassive}
\subsection{Solution of the Cauchy Problem} \label{secgenmass}
We now generalize the integral representation~\eqref{solCauchy2} as well
as the group property~\eqref{semigroup0} to the massive case.
\begin{Lemma} \label{lemmabessel} A Dirac solution~$\psi \in \H_m$
can be expressed in terms of the initial data at~$t=0$ by
\beq \label{cauchyrep}
\psi(t,x) = 2 \pi \int_0^b k_m(t,x-y)\, \gamma^0\, \psi(0,y)\: dy\:,
\eeq
where~$k_m(t,x)$ is the distribution
\begin{align*}
k_m(t,x) =\;& \frac{1}{4\pi} \,(\gamma^0+\gamma^1) \:\delta(t+x) +
\frac{1}{4\pi} \,(\gamma^0-\gamma^1) \:\delta(t-x) \\
& -\frac{im}{4 \pi}\: J_0 \Big( m \sqrt{t^2-x^2} \Big)\: \epsilon(t)\: \Theta(t^2-x^2) \\
&-\frac{m}{4 \pi} \: \big(t \gamma^0 - x \gamma^1 \big) \: \frac{J_1 \Big( m \sqrt{t^2-x^2} \Big)}{\sqrt{t^2-x^2}}
\:\epsilon(t)\: \Theta(t^2-x^2) \:.
\end{align*}
Moreover,
\beq \label{semigroup}
k_m(\zeta-\tilde{\zeta}) = 2 \pi \int_0^b k_m(\zeta^0, \zeta^1-x)\, \gamma^0\, k_m(-\tilde{\zeta}^0, x-\tilde{\zeta}^1)\:
dx \:.
\eeq
\end{Lemma}
\Proof We recall the general method for constructing the time evolution operator
(for details see for example~\cite[Section~2]{finite} or~\cite[Section~4.2]{intro}).
The unique solvability of the Cauchy problem gives rise to the existence of
advanced and retarded Green's functions~$s^\vee$ and~$\s^\wedge$
(see for example~\cite{baer+ginoux}).
The causal fundamental solution~$k_m$ is defined as the difference of the advanced
and retarded Green's functions; more precisely,
\beq \label{kmdef}
k_m := \frac{1}{2\pi i} \left( s_m^{\vee}-s_m^{\wedge} \right) .
\eeq
Then the Cauchy problem for the Dirac equation with initial data~$\psi_\scrN$ on a Cauchy surface~$\scrN$
with future-directed normal~$\nu$ has the solution (see~\cite[Lemma~2.1]{finite})
\[ \psi(\zeta) = 2 \pi \int_\scrN k_m(\zeta;\zeta')\, \nu^j(\zeta')\, \gamma_j\, \psi_\scrN(\zeta')\: d\mu_\scrN(\zeta')\:. \]
Choosing~$\scrN$ as the Cauchy surface at~$t=0$, this formula simplifies to~\eqref{cauchyrep}.
(Indeed, to see the correspondence, one should keep in mind that in Minkowski space,
the causal Greens function depends only on the difference vector~$\zeta-\zeta'$,
so that in~\eqref{cauchyrep} we can use the notation~$k_m(\zeta) \equiv k_m(\zeta; 0)$).

In Minkowski space, the causal fundamental solution is given as the integral over the mass shell.
More precisely,
\beq \label{kmfourier}
k_m(t,x) = \frac{1}{(2 \pi)^2} \int_{\R^2} (\omega \gamma^0 - k \gamma^1 + m)\:
\delta(\omega^2-k^2-m^2)\: \epsilon(\omega)\: e^{-i \omega t + i k x}\:d\omega \,dk
\eeq
(for detail see~\cite[Section~4.2]{intro} or~\cite[\S2.2]{pfp}).
We now compute the Fourier transform in~\eqref{kmfourier}.
First of all,
\begin{align*}
k_m(t,x) &= \frac{1}{(2 \pi)^2} \:(i \gamma^j \partial_j + m) \int_{\R^2}
\delta(\omega^2-k^2-m^2)\: \epsilon(\omega)\: e^{-i \omega t + i k x}\:d\omega \,dk \\
&= \frac{1}{(2 \pi)^2} \:(i \gamma^j \partial_j + m) \int_{\R \setminus [-m,m]}
\frac{\epsilon(\omega)\: e^{-i \omega t}}{2 \sqrt{\omega^2-m^2}}
\left( e^{i \sqrt{\omega^2-m^2} \,x} + e^{-i \sqrt{\omega^2-m^2} \,x} \right) d\omega \\
&= \frac{1}{(2 \pi)^2} \:(i \gamma^j \partial_j + m) \int_{\R \setminus [-m,m]}
\frac{\epsilon(\omega)\: e^{-i \omega t}}{\sqrt{\omega^2-m^2}}
\:\cos\Big( \sqrt{\omega^2-m^2} \,x \Big) d\omega \\
&= -\frac{i}{2 \pi^2} \:(i \gamma^j \partial_j + m) \int_m^\infty
\frac{\sin(\omega t)}{\sqrt{\omega^2-m^2}}
\:\cos\Big( \sqrt{\omega^2-m^2} \,x \Big) d\omega \:.
\end{align*}
In the case~$m=0$, we obtain
\begin{align*}
k_0(t,x) &= \frac{1}{2 \pi^2} \:\gamma^j \partial_j \int_0^\infty
\frac{\sin(\omega t)}{\omega}
\:\cos (\omega x )\: d\omega \\
&= \frac{1}{2 \pi^2} \int_0^\infty \left( \gamma^0 \cos(\omega t) \:\cos (\omega x ) 
+ \gamma^1 \sin(\omega t) \:\sin (\omega x ) \right) d\omega \\
&= \frac{1}{4 \pi^2} \int_{-\infty}^\infty \left( \gamma^0 \cos(\omega t) \:\cos (\omega x ) 
+ \gamma^1 \sin(\omega t) \:\sin (\omega x ) \right) d\omega \\
&= \frac{\gamma^0}{4\pi} \,\Big( \delta(t+x) + \delta(t-x) \Big)
-\frac{\gamma^1}{4\pi} \,\Big( \delta(t+x) - \delta(t-x) \Big) \\
&= \frac{1}{4\pi} \,(\gamma^0+\gamma^1) \:\delta(t+x) +
\frac{1}{4\pi} \,(\gamma^0-\gamma^1) \:\delta(t-x) \:,
\end{align*}
in agreement with~\eqref{k0}.
In the case~$m > 0$, we obtain additional Lorentz invariant contributions
in timelike directions. In order to compute them, it is easiest to set~$x=0$ and~$t>0$. Then
the Fourier integral can be carried out in terms of Bessel functions of the first kind
(see~\cite[\S10.2]{DLMF})
\begin{align*}
k_m(t,x) &= -\frac{i}{2 \pi^2} \:(i \gamma^0 \partial_t + m) \int_m^\infty
\frac{\sin(\omega t)}{\sqrt{\omega^2-m^2}}\:d\omega \\
&= -\frac{i}{4 \pi}\: (i \gamma^0 \partial_t + m) J_0(mt)
= -\frac{i m}{4 \pi}\: J_0(mt) -\frac{m}{4 \pi}\: \gamma^0\: J_1(mt) \:,
\end{align*}
where in the last step we used~\cite[eq.~(10.6.2)]{DLMF}.
Rewriting these contributions in Lorentz invariant form and using that~$k_m(t,x)^* = k_m(-t,-x)$
gives the desired formula for~$k_m$ (here the star again denotes the adjoint with respect to the
inner product~$\Sl .|. \Sr_x$). This also concludes the proof of~\eqref{cauchyrep}.

Finally, the identity~\eqref{semigroup} follows immediately by using the group property
of the time evolution operator (similar as explained before~\eqref{semigroup0}).
\QED

The result of the previous lemma gives a pointwise estimate for the solution of the
Cauchy problem.
\begin{Lemma} \label{lemmapointwise}
Let~$\psi$ be a solution of the Dirac equation~\eqref{Dir} with~$m \geq 0$
with initial values~$\psi(0,.) \in C^0((0,b), \C^2)$. Then~$\psi$ satisfies 
for all~$(t,x) \in D$ the pointwise bound
\[ |\psi_{L\!/\!R}(t,x)| \leq |\psi_{L\!/\!R}(0,.)|_{C^0} + 2 \,\sqrt{m t} \:|\psi(0,.)|_{C^0} \:. \]
\end{Lemma}
\Proof We first estimate the Bessel functions in Lemma~\ref{lemmabessel} by
\[ |J_0(x)| \leq \frac{1}{(1+x^2)^{\frac{1}{4}}} \:,\qquad  |J_1(x)| \leq \frac{x}{(1+x^2)^\frac{3}{4}} \:. \]
This gives
\begin{align*}
&|\psi_{L\!/\!R}(t,x)| \leq |\psi_{L\!/\!R}(0,.)|_{C^0} \\
&\;\;+ \frac{|\psi(0,.)|_{C^0}}{2} \int_{-t}^t \left( \frac{m}{(1+m^2 (t^2-x^2))^\frac{1}{4}}
+ \frac{m^2 t}{(1+m^2 (t^2-x^2))^\frac{3}{4}} \right) dx \:.
\end{align*}
Carrying out the integral gives the result.
\QED

\subsection{Regularity of the Image of~$\Sig$} \label{secreg}
We now work out estimates which give information on the regularity of the
functions in the image of~$\Sig$. We again consider the Dirac equation~$(\Dir - m) \psi =0$ in
a bounded, globally hyperbolic space-time~$\scrM \subset D \subset \R^{1,1}$, with the Cauchy
surface~$\scrN = \{0\} \times (0,b)$ (see the right of Figure~\ref{figcauchy}).
Moreover, we let~$\psi, \phi \in \H_m \cap C^\infty(D)$ be smooth solutions of the Dirac equation
inside the causal diamond~$D$. We introduce the wave function
\beq \label{thetadef}
\theta = (\Dir + m)\, \gamma^0 \psi \:.
\eeq
The calculation
\[ (\Dir - m) \,\theta = (\Dir^2 - m^2) \,\gamma^0 \,\psi = 
-(\Box + m^2) \,\gamma^0 \,\psi = - \gamma^0 \,(\Box + m^2) \,\psi = 0 \]
shows that~$\theta$ is again a solution of the Dirac equation. Hence
\begin{align}
( \theta | \Sig \phi) &= \int_\scrM \Sl (\Dir + m)\, \gamma^0 \psi \,|\, \phi \Sr \: d\mu \notag \\
&= \int_\scrM \left( \partial_j \Sl i \gamma^j \, \gamma^0 \psi \,|\, \phi \Sr
+ \Sl \gamma^0 \psi \,|\,  (\Dir + m)\, \phi \Sr \: d\mu \right) d\mu \notag \\
&= -i \left( \int_{\partial \scrM_+}- \int_{\partial \scrM_-} \right) \Sl \psi \,|\, \gamma^0 \slashed{\nu} \phi \Sr\:
d\mu_{\partial \scrM} \label{boundary} \\
&\quad+2m \int_\scrM \Sl \psi \,|\, \gamma^0 \phi \Sr \: d\mu \:, \label{inner}
\end{align}
where in the last step we applied the Gauss divergence theorem and used the Dirac equation.
Here~$\partial \scrM_\pm$ are the future and past boundaries of~$\scrM$, and~$\nu$ is the future-directed normal.

If the boundaries~$\partial \scrM_\pm$ of~$\scrM$ are space-like, 
this estimate implies that~$\Sig$ maps to the H\"older continuous functions:
\begin{Prp} \label{prpholder} Assume that the future and past boundaries of~$\scrM$ are space-like. Then there
is a constant~$c = c(\partial \scrM_\pm)$ such that for all~$\psi \in \H_m \cap C^\infty(D)$,
\beq \label{Siges}
\big\| \Sig \,(\Dir + m)\, \gamma^0 \psi \big\| \leq 2 (c + mb)\, \|\psi\| \:.
\eeq
Moreover, the operator~$\Sig$ maps
to the weakly differentiable and H\"older continuous functions,
\[ \Sig \::\: \H_m \rightarrow W^{1,2}(\scrN, \C^2) \hookrightarrow C^{1,\frac{1}{2}}(\scrN, \C^2)\:. \]
The operator~$\Sig : \H_m \rightarrow \H_m$ is compact.
\end{Prp}
\Proof First, the integral~\eqref{inner} can be estimated with the help of Fubini's theorem and the
Schwarz inequality by
\[ \int_\scrM |\Sl \psi \,|\, \gamma^0 \phi \Sr| \: d\mu \leq 2b\, \|\psi\|\, \|\phi\|\:. \]
Moreover, since~$\partial \scrM_\pm$ are space-like curves, we can estimate the
integrand in~\eqref{boundary} in terms of the probability density,
\[ \big| \Sl \psi \,|\, \gamma^0 \slashed{\nu} \phi \Sr \big| \leq c(\partial \scrM_\pm)\:
\sqrt{ \Sl \psi \,|\, \slashed{\nu} \psi \Sr \:\Sl \phi \,|\, \slashed{\nu} \phi \Sr } \:. \]
Estimating the resulting line integrals with the help of the Schwarz inequality, we obtain~\eqref{Siges}.

Using that~$\psi$ solves the Dirac equation, we can use the anti-commutation relations to obtain
\beq \label{phirep}
\begin{split}
(\Dir + m)\, \gamma^0 \,\psi &= (\Dir + m)\, \gamma^0 \,\psi - \gamma^0\, (\Dir - m)\, \psi \\
&= \big[ \Dir, \gamma^0 \big] \psi + 2m \gamma^0 \psi 
= 2 i \gamma^1 \gamma^0 \partial_x \psi + 2m \gamma^0 \psi 
= 2 H \psi \:,
\end{split}
\eeq
where~$H$ is the Dirac Hamiltonian
\beq \label{DirHam}
H = -i \gamma^0 \gamma^1 \partial_x + m \gamma^0 \:.
\eeq
Using this identity together with the fact that~$\Sig$ is symmetric, we can
rewrite~\eqref{Siges} in the ``dual form''
\[ \| H \Sig \phi \| \leq (c + mb)\, \|\phi\| \qquad \text{for all~$\phi \in \H_m$}\:. \]
This shows that~$\Sig$ maps to the~$W^{1,2}$-functions.
We now apply Morrey's embedding into the H\"older continuous functions
(see~\cite[Theorem~5.7.6]{evans}).
The compactness of~$\Sig$ follows from the Arzel{\`a}-Ascoli theorem.
\QED

We point out that the above proposition only applies if the boundaries~$\partial \scrM_\pm$ are space-like.
This assumption is crucial in view of the examples of simple domains (see Lemma~\ref{lemmasimple}),
in which case the eigenfunctions were characteristic functions, which are clearly not H\"older continuous.
We now prove a weaker statement without assuming that the curves~$\partial \scrM_\pm$ are space-like.
Thinking of the characteristic functions in simple domains, one is led to considering the total variation.
In fact, we now show that the vectors in the image of~$\Sig$ always have bounded variation.
As usual, we denote the total variation by~$\TV_{[0,b]}$ and denote the functions of
finite total variation by~$\BV([0,b], \C^2)$ (for basic definitions see for example~\cite{evans+gariepy}).
\begin{Prp} \label{prpBV} The fermionic signature operator maps~$\H_m$ to~$\BV([0,b], \C^2)$ and
\[ \TV_{[0,b]}(\Sig \phi) \leq c\, \|\phi\|\:, \]
where the constant~$c$ depends only on~$m$ and~$b$.
The operator~$\Sig : \H_m \rightarrow \H_m$ is compact.
\end{Prp}

We begin with a preparatory lemma.
\begin{Lemma} For any smooth solutions~$\psi, \phi \in \H_m$ and~$\theta$
according to~\eqref{thetadef}, the following estimate holds:
\beq \label{thetaes}
\big| ( \theta | \Sig \phi) \big| \leq 
8 \sqrt{b}\:\big(1+\sqrt{mb} \big) \|\phi\|\: |\psi|_{C^0} \:.
\eeq
\end{Lemma}
\Proof We want to estimate~$( \theta | \Sig \phi)$ in terms of the Hilbert norm~$\|\phi\|$
and the sup-norm $|\psi|_{C^0}$. To this end, we estimate~\eqref{inner} by
\beq \label{ines}
\int_\scrM \big| \Sl \psi \,|\, \gamma^0 \phi \Sr\big| \: d\mu
\leq 2 b\, \|\psi\|\, \|\phi\| \leq 2 b^\frac{3}{2}\: |\psi|_{C^0}\, \|\phi\| \:.
\eeq
In~\eqref{boundary} we first apply the Schwarz inequality,
\[ \int_{\partial \scrM_+} \big| \Sl \psi \,|\, \gamma^0 \slashed{\nu} \phi \Sr \big| \:
d\mu_{\partial \scrM} \leq \|\phi\|
\left( \int_{\partial \scrM_+} \Sl \gamma^0 \psi \,|\, \slashed{\nu} \gamma^0  \psi \Sr\: d\mu_{\partial \scrM}
\right)^\frac{1}{2} \:. \]
We would like to relate the last line integral to a corresponding integral on the Cauchy surface~$t=0$.
To this end, we first note that the line integral can be recovered as the boundary integral
when applying the Gauss divergence theorem to the vector
field~$\Sl \gamma^0 \psi \,|\, \gamma^j \gamma^0 \phi \Sr$.
However, this vector field is not divergence-free, because
\begin{align*}
\partial_j &\Sl \gamma^0 \psi \,|\, \gamma^j \gamma^0  \phi \Sr =
2 \re \Sl \gamma^0 \psi \,|\, \gamma^j \gamma^0 \partial_j \phi \Sr \\
&= 4 \re \Sl \gamma^0 \psi \,|\, \gamma^0 \gamma^0 \partial_t \phi \Sr - 2 \re \Sl \gamma^0 \psi \,|\, \gamma^0 \gamma^j \partial_j \phi \Sr \\
&= 4 \re \Sl \psi \,|\, \gamma^0 \partial_t \phi \Sr = 2 \partial_t \Sl \psi \,|\, \gamma^0 \phi \Sr \:.
\end{align*}
Hence
\begin{align*}
\int_{\partial \scrM_+} &\Sl \gamma^0 \psi \,|\, \slashed{\nu} \gamma^0  \psi \Sr \: d\mu_{\partial \scrM}
- \int_0^b\Sl \psi \,|\, \gamma^0  \psi \Sr(0,x)\: dx
= \int_{\scrM \cap \{t \geq 0\}} 2 \partial_t \Sl \psi \,|\, \gamma^0 \phi \Sr\: dx\, dt \\
&= 2 \int_0^b\Sl \psi \,|\, \gamma^0 \psi \Sr \big(T(x),x \big)\: dx
-  2 \int_0^b\Sl \psi \,|\, \gamma^0 \psi \Sr(0,x)\: dx\:,
\end{align*}
where we parametrized~$\partial \scrM_+$ as the graph~$\{ (T(x), x) \:|\: x \in [0,b]\}$.
We conclude that
\[ \int_{\partial \scrM_+} \big| \Sl \gamma^0 \psi \,|\, \slashed{\nu} \gamma^0  \psi \Sr \big|\: d\mu_{\partial \scrM}
= 2 \int_0^b\Sl \psi \,|\, \gamma^0 \psi \Sr \big(T(x),x\big) \: dx
- 2 \,\|\psi\|^2 \:. \]
Applying the pointwise estimate of Lemma~\ref{lemmapointwise}, we obtain
\[ \int_{\partial \scrM_+} \big| \Sl \gamma^0 \psi \,|\, \slashed{\nu} \gamma^0  \psi \Sr \big|\: d\mu_{\partial \scrM}
\leq 2 b \,\big(1+2 \,\sqrt{mb} \big)^2 |\psi|^2_{C^0} \:. \]
We conclude that
\[ \int_{\partial \scrM_+} \big| \Sl \psi \,|\, \gamma^0 \slashed{\nu} \phi \Sr \big| \:
d\mu_{\partial \scrM} \leq 2 \sqrt{b}\:\big(1+\sqrt{mb} \big) \|\phi\|\: |\psi|_{C^0} \:. \]
The integral over~$\partial \scrM_-$ can be treated similarly.
Combining these estimates with~\eqref{ines} gives the result.
\QED

\Proof[Proof of Proposition~\ref{prpBV}]
Using~\eqref{phirep}, we can write~\eqref{thetaes} as
\[ \big| ( H \psi | \Sig \phi) \big| \leq 
4 \sqrt{b}\:\big(1+\sqrt{mb} \big) \|\phi\|\: |\psi|_{C^0} \:. \]
Thus for every~$\phi \in \H_m$, we have a bounded linear functional on~$C^0([0,b])$.
The Riesz representation theorem (see~\cite[Chapter~2]{rudin} or~\cite[Theorem~S.5]{reed+simon}
for a proof in one dimension) yields that there is a bounded regular signed Borel measure such that
\beq \label{HSig}
(H \psi |\Sig \phi) = \int_0^b \psi(x)\: d\mu(x) 
\eeq
and
\[ |\mu|([0,b]) \leq 4 \sqrt{b}\:\Big(1+\sqrt{mb} \Big) \|\phi\| \:. \]
Choosing~$\psi \in C^\infty_0((0,b))$ with compact support, we conclude
that function~$\Sig \phi$ is weakly differentiable, and
\beq \label{Hweak}
(H \Sig \phi)\, dx = d\mu \qquad \text{as a measure}\:.
\eeq
Moreover, choosing a function~$\psi$ with~$\psi(0) \neq 0$, the vanishing of the
boundary terms when integrating by parts in~\eqref{HSig}.
Combining these facts, we can compute~$\Sig \phi$ by integration.
Namely, writing~\eqref{Hweak} with the help of~\eqref{DirHam} in the form
\[  \left( \partial_x - i  m \gamma^1 \right) \Sig \phi = -i \gamma^1 \gamma^0 \,d\mu \:, \]
we obtain the explicit solution
\[ \big(\Sig \phi \big)(x) = e^{i m \gamma^1 x} \int_0^x e^{-i m \gamma^1 \tau}\: \big( -i \gamma^1 \gamma^0 \big)
\, d\mu(\tau) \:. \]
Differentiating the last equation, we obtain the estimate
\[ \big| (\Sig \phi)'(x) \big| \leq m\,e^{2 m b} \,|\mu|((0,b)) +  d|\mu|(x) \:, \]
showing that the total variation of the function~$\Sig \phi$ is bounded by a constant~$c=c(m,b)$.

Finally, the compactness of~$\Sig$ follows
from Helly's selection theorem (see for example~\cite[Section~VIII.4]{natanson}).
\QED

\subsection{Asymptotics of the Small Eigenvalues} \label{secsmall}
The analysis of the regularity of the image of the fermionic signature operator
(see Propositions~\ref{prpholder} and~\ref{prpBV}) showed in particular that~$\Sig$
is a compact operator. Thus it has a pure point spectrum and finite-dimensional eigenspaces,
and the eigenvalues can accumulate only at the origin. In particular, we
can count the eigenvalues of~$\Sig$ with multiplicities by~$\lambda_1, \lambda_2, \ldots$
and order them such that
\beq \label{lambdaord}
|\lambda_1| \geq |\lambda_2| \geq \cdots \:.
\eeq

We begin with an example where~$\Sig$ has infinite rank.
\begin{Example} {\bf{(A triangular domain)}}
We let~$\scrM \subset \R^{1,1}$ be the triangular domain shown in Figure~\ref{figtriangle}
and for simplicity the massless Dirac equation. Then the eigenvalues of the fermionic signature
operator (ordered according to~\eqref{lambdaord}) satisfy for~$n \geq 5$ the inequalities
\[ \frac{b}{8 \pi^2}\: \frac{4}{n+3} \leq |\lambda_n| \leq \frac{b}{8 \pi^2}\: \frac{4}{n-4} \:. \]
\Proof
\begin{figure}
\scalebox{1} 
{
\begin{pspicture}(0,-1.195)(4.28,1.165)
\definecolor{color204b}{rgb}{0.8,0.8,0.8}
\usefont{T1}{ptm}{m}{n}
\rput(0.27,-1.01){$0$}
\psline[linewidth=0.02cm,linestyle=dashed,dash=0.16cm 0.16cm](0.26,-0.695)(3.96,-0.695)
\pspolygon[linewidth=0.02,fillstyle=solid,fillcolor=color204b](0.26,-0.695)(3.96,-0.695)(2.11,1.155)
\usefont{T1}{ptm}{m}{n}
\rput(3.97,-1.01){$b$}
\end{pspicture} 
}
\caption{A triangular domain.}
\label{figtriangle}
\end{figure}
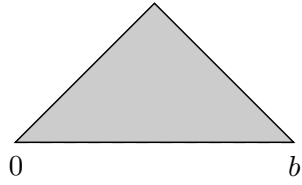
On~$\H_0$ we choose the orthonormal basis of the solution space (cf.~\eqref{psiLR})
\[ \psi^n_L(t,x) = \frac{1}{\sqrt{2 \pi b}} \begin{pmatrix} 1 \\ 0 \end{pmatrix} e^{\frac{2 \pi i}{b}\, n (x+t)} \:,\qquad
\psi^n_R(t,x) = \frac{1}{\sqrt{2 \pi b}} \begin{pmatrix} 0 \\ 1 \end{pmatrix} e^{\frac{2 \pi i}{b}\, n (x-t)} \]
where~$n \in \Z$. Then for any~$n,n' \neq 0$,
\begin{align}
\bra \psi^n_L | \psi^{n'}_R \ket &= \frac{1}{2 \pi b}\: \int_\scrM e^{-\frac{2 \pi i}{b}\, n (x+t)} \,
e^{\frac{2 \pi i}{b}\: n' (x-t)}\, dt\, dx \notag \\
&= \frac{1}{4 \pi b}\: \int_0^b du \int_{-u}^0 dv \; e^{-\frac{2 \pi i}{b}\, nu} \, e^{-\frac{2 \pi i}{b}\, n' v} \notag \\
&= \frac{i}{8 \pi^2\: n'}\: \int_0^b e^{-\frac{2 \pi i}{b}\, nu} \, \big( 1 - e^{\frac{2 \pi i}{b}\, n' u} \big)\:du
= -\frac{ib}{8 \pi^2} \: \frac{\delta_{n, n'}}{n}\:. \label{nasy}
\end{align}
where we again chose the light-cone coordinates~\eqref{uvdef}.
The matrix elements with~$n=0$ or~$n'=0$ are a bit more complicated, and we do not compute them here.
Instead, we only analyze~$\Sig$ on the orthogonal complement of the two-dimensional
subspace~${\mathcal{N}}:= \text{span}(\psi^0_L, \psi^0_R)$. Denoting the orthogonal projection on~${\mathcal{N}}^\perp$
by~$\pi_\perp$, a short computation using~\eqref{nasy} shows that the operator
\[ \frac{8 \pi^2}{b}\: \pi_\perp \,\Sig \,\pi_\perp \quad \text{has the eigenvalues} \quad
\pm 1, \;\pm \frac{1}{2}, \; \pm \frac{1}{3}, \;\ldots \:, \]
each of multiplicity two.

We now estimate the eigenvalues of~$\Sig$ from above and below using the min-max principle.
Since the spectrum is symmetric (see Proposition~\ref{prpsymm}), we know that
\beq \label{minmaxeq}
\big| \lambda_{2\ell+1} \big| = \big| \lambda_{2\ell+2} \big|
= \inf_{J \subset \H_0, \;\dim J=\ell} \;\;\sup_{\psi \perp J,\; \|\psi\|=1}\;\;
( \psi | \Sig \psi ) \:,
\eeq
giving the upper bound
\begin{align*}
\big| \lambda_{2\ell+1} \big|, \big| \lambda_{2\ell+2} \big| &\leq \inf_{J \subset \H_0, \;\dim J=\ell,\; J \supset {\mathcal{N}}} \;\;\sup_{\psi \perp J,\; \|\psi\|=1}\;\;
( \psi | \Sig \psi ) \\
&= \inf_{K \subset {\mathcal{N}}^\perp, \;\dim K=\ell-2} \;\;\sup_{\psi \perp K, \; \|\psi\|=1}\;\;
( \psi \,|\, \pi_\perp\,\Sig\,\pi_\perp \psi ) \leq \frac{b}{8 \pi^2}\: \frac{2}{\ell-1} \:.
\intertext{Similar, we can estimate~\eqref{minmaxeq} from below to obtain}
\big| \lambda_{2\ell+1} \big|, \big| \lambda_{2\ell+2} \big| &\geq \inf_{J \subset \H_0, \;\dim J=\ell} \;\;\sup_{\psi \perp J,\;
\psi \perp {\mathcal{N}},\;\|\psi\|=1}\;\; ( \psi | \Sig \psi ) \\
&= \inf_{J \subset {\mathcal{N}}^\perp, \;\dim J=\ell} \;\;\sup_{\psi \perp J,\;\|\psi\|=1}\;\; ( \psi \,|\, \pi_\perp\,\Sig\,\pi_\perp \psi )
\geq \frac{b}{8 \pi^2}\: \frac{2}{\ell+2} \:.
\end{align*}
This concludes the proof.
\QED \end{Example}

This example shows that in general we cannot expect a decay
of~$|\lambda_n|$ for large~$n$ faster than~$\sim 1/n$.
Indeed, in the next theorem we prove this $1/n$-decay:

\begin{Thm} \label{thmasy}
Representing the boundary of~$\partial \scrM$ as a graph,
\[ \partial \scrM_\pm = \big\{ (T_\pm(x), x) \::\: x \in [0,b] \big\} \:, \]
we introduce the dimensionless constant~$c$ by
\beq \label{cdepend}
c = (1+mb) \bigg( 1 + 4  \sum_\pm \TV_{[0,b]} \,  T'_\pm \bigg) \:.
\eeq
Then
\[ |\lambda_n| \leq \frac{c b}{n}\:. \]
\end{Thm} \noindent
Before coming to the proof, we remark that it is not clear whether the dependence
of the constant~$c$ in~\eqref{cdepend} on the total variation of~$T'_\pm$  is only a technical assumption
for our proof, or whether this assumption is needed for the theorem to hold.

We again work in light-cone coordinates~$u$ and~$v$.
As in~\eqref{Dirm0}, we denote the two components of the spinors by indices~$L$ and~$R$.
Then the Dirac equation can be written as
\beq \label{DirLR}
i \pv \psi_L = \frac{m}{2}\: \psi_R \:,\qquad i \pu \psi_R = \frac{m}{2}\: \psi_L \:.
\eeq
This allows us to rewrite the spatial derivatives (which we denote by a prime) as
\beq \label{psiprime}
\psi'_L = \pu \psi_L + \frac{i m}{2}\: \psi_R \:,\qquad
\psi'_R = -\pv \psi_L - \frac{i m}{2}\: \psi_L\:.
\eeq
Moreover, the space-time inner product becomes
\begin{align*}
\bra \psi | \phi \ket &= \int_\scrM \Sl \psi | \phi \Sr \: dt \,dx
= \int_\scrM \left( \overline{\psi_L} \phi_R + \overline{\psi_R} \phi_L \right) \:dt \,dx \:.
\end{align*}
Combining these relations, we can compute the inner product of the spatial derivatives
of two Dirac solutions:
\begin{Lemma} Let~$\psi, \phi \in \H_m$ be smooth solutions of the Dirac equation. Then
\begin{align}
\bra \psi' | \phi' \ket &=\int_{\R^2} \left( \overline{\psi_L} \:\phi_R
+ \overline{\psi_R} \:\phi_L \right) \puv \chi_\scrM\: dt \,dx \label{t1} \\
&\quad +\frac{im}{2} \int_\scrM \left( \overline{\psi'_R}\: \phi_R -\overline{\psi_R}\: \phi'_R
- \overline{\psi'_L}\: \phi_L +\overline{\psi_L}\: \phi'_L \right) dt \,dx  \label{t2}
\end{align}
(where~$\puv \chi_\scrM$ denotes the distributional derivative of the characteristic function).
\end{Lemma}
\Proof We first rewrite the spatial derivatives using~\eqref{psiprime} in terms of
derivatives with respect to~$u$ and~$v$,
\begin{align*}
\bra \psi' | \phi' \ket &=-\int_\scrM \left( \overline{\pu \psi_L} \:\pv \phi_R
+ \overline{\pv \psi_R} \:\pu \phi_L \right) dt \,dx \\
&\quad -\frac{im}{2} \int_\scrM \left( \overline{\pv \psi_R}\: \phi_R -\overline{\psi_R}\: \pv \phi_R
+ \overline{\pu \psi_L}\: \phi_L -\overline{\psi_L}\: \pu \phi_L \right) dt \,dx \\
&\quad -\frac{m^2}{4} \int_\scrM \left( \overline{\psi_L} \:\phi_R
+ \overline{\psi_R} \:\phi_L \right) dt \,dx \:.
\end{align*}
In the first integral, we integrate both derivatives by parts. Whenever the derivatives
hit the wave functions, we apply the Dirac equation~\eqref{DirLR}.
Combining all the resulting terms, we obtain
\begin{align*}
\bra \psi' | \phi' \ket &=\int_{\R^2} \left( \overline{\psi_L} \:\phi_R
+ \overline{\psi_R} \:\phi_L \right) \puv \chi_\scrM\: dt \,dx \\
&\quad - \frac{im}{2} \int_\scrM \left( \overline{\pv \psi_R}\: \phi_R -\overline{\psi_R}\: \pv \phi_R
+ \overline{\pu \psi_L}\: \phi_L -\overline{\psi_L}\: \pu \phi_L \right) dt \,dx \\
&\quad -\frac{m^2}{2} \int_\scrM \left( \overline{\psi_L} \:\phi_R
+ \overline{\psi_R} \:\phi_L \right) dt \,dx \:.
\end{align*}
Expressing the remaining derivatives of the wave functions with the help of~\eqref{psiprime}
as spatial derivatives, we obtain the result.
\QED

Next, we need to estimate the terms~\eqref{t1} and~\eqref{t2}.
In~\eqref{t2} we can use Fubini and the Schwarz inequality,
\[ |\eqref{t4}| \leq mb \:\big( \|\psi'\|\, \|\phi\| + \|\psi\|\, \|\phi'\| \big)\: . \]
The analysis of the boundary terms~\eqref{t1} is more subtle.
We only analyze the boundary terms on~$\partial \scrM_+$, because the past boundary can be
analyzed similarly. It is again useful to write~$\partial \scrM_+$
as a graph of a function~$T(x)$ over~$[0,b]$. We first consider the case that~$T$ is smooth;
the non-smooth situation will be obtained below by approximation.
The fact that~$\partial \scrM_+$ is non-timelike implies that~$|T'(x)| \leq 1$.
Then
\begin{align*}
\puv \chi_\scrM &= \frac{1}{4}\: (\partial_t^2 - \partial_x^2) \: \Theta(T(x) - t) \\
&= \frac{1}{4}\: \delta' \big( T(x)-t \big) \,\big( 1-T'(x)^2 \big) - \frac{1}{4} \: \delta\big( T(x)-t \big)\: T''(x) \\
&=  -\frac{1}{4}\: \partial_t \delta \big( T(x)-t \big) \,\big( 1-T'(x)^2 \big) - \frac{1}{4} \: \delta\big( T(x)-t \big)\: T''(x) \:.
\end{align*}
Using this relation in~\eqref{t1}, in the term involving~$\partial_t \delta( T(x)-t)$ we may integrate by parts.
We thus obtain
\begin{align*}
\int_{\R^2}& \left( \overline{\psi_L} \:\phi_R
+ \overline{\psi_R} \:\phi_L \right) \puv \chi_\scrM\: dt \,dx \\
&=-\frac{1}{4} 
\int_0^b  T''(x) \left( \overline{\psi_L} \:\phi_R
+ \overline{\psi_R} \:\phi_L \right)|_{t=T(x)} \: dx \\
&\quad +\frac{1}{4} \int_0^L \big( 1-T'(x)^2 \big)\:
\partial_t \left( \overline{\psi_L} \:\phi_R
+ \overline{\psi_R} \:\phi_L \right)|_{t=T(x)}\: dx \:.
\end{align*}
Using the Dirac equation~\eqref{DirLR}, we can rewrite the time derivatives in terms
of spatial derivatives,
\begin{align}
\int_{\R^2}& \left( \overline{\psi_L} \:\phi_R
+ \overline{\psi_R} \:\phi_L \right) \puv \chi_\scrM\: dt \,dx \notag \\
&=-\frac{1}{4} 
\int_0^b  T''(x) \left( \overline{\psi_L} \:\phi_R
+ \overline{\psi_R} \:\phi_L \right)|_{t=T(x)} \: dx \label{t3} \\
&\quad +\frac{1}{4} \int_0^b \Big( 1-T'(x)^2 \Big)
\Big( \overline{\psi'_L} \:\phi_R - \overline{\psi_L} \:\phi'_R
- \overline{\psi'_R} \:\phi_L + \overline{\psi_R} \:\phi'_L \Big) \Big|_{t=T(x)}\: dx \label{t4} \:.
\end{align}
In the integral~\eqref{t3} we estimate the wave functions pointwise with the help of
Lemma~\ref{lemmapointwise},
\[ |\eqref{t3}| \leq (1+2 \sqrt{mb})^2 \:|\psi(0,.)|_{C^0} \, |\phi(0,.)|_{C^0} \; \TV_{[0,b]} \,T'\:. \]
In order to get an idea for how to estimate~\eqref{t4}, we first
rewrite the scalar product as an integral over~$\partial \scrM_+$ with integration measure~$dx$,
\begin{align*}
( \psi | \psi) &= -\int_{\R^2} \Sl \psi | \gamma^j \psi \Sr\: \partial_j \Theta(T(x) - t) \: dt\, dx\\
&= \int_{\R^2} \Sl \psi | (\gamma^0 - T'(x)\, \gamma^1) \phi \Sr\; \delta(T(x) - t))\: dt\, dx \\
&= \int_0^b \Sl \psi | (\gamma^0 - T'(x)\, \gamma^1) \psi \Sr|_{t=T(x)}\: dx \\
&= \int_0^b \Big( \big(1+T'(x) \big)\, |\psi_L|^2 + \big( 1-T'(x) \big) |\psi_R|^2 \Big) \Big|_{t=T(x)}\: dx\:.
\end{align*}
Writing the factor~$(1-T'(x)^2)$ in~\eqref{t4} as~$(1-T')(1+T')$, we can always group the factors~$1+T'$
and~$1-T'$ together with the components~$L$ respectively~$R$. Applying the Schwarz inequality, we obtain
\[ |\eqref{t4}| \leq
\frac{1}{4} \:\big( \|\psi'\|\, \|\phi\| + \|\psi\|\, \|\phi'\| \big) \:. \]
In the above estimates we made use of the fact that~$T$ is twice differentiable.
However, the estimates can be extended by approximation to the situation when~$T$
is differentiable almost everywhere and~$T'$ has bounded total variation.

Combining all the terms gives the following estimate:
\begin{Lemma} \label{lemmaderes}
Suppose that the future and past boundaries~$\partial \scrM_\pm$ are parametrized
by functions~$T_\pm \in C^0((0,b))$. Then
\begin{align*}
\big| \bra \psi' | \phi' \ket \big| &\leq \frac{1}{2}\: (1+mb) \:\big( \|\psi'\|\, \|\phi\| + \|\psi\|\, \|\phi'\| \big) \\
&\quad +(1+2 \sqrt{mb})^2 \:|\psi(0,.)|_{C^0} \, |\phi(0,.)|_{C^0} \; \sum_\pm \TV_{[0,b]} \,  T'_\pm \:.
\end{align*}
\end{Lemma} \noindent

\Proof[Proof of Theorem~\ref{thmasy}]
We apply the min-max principle in the form
\[ |\lambda_{2n+1}| = \inf_{\stackrel{I, J \subset \H_m,}{
\dim I = \dim J = n}} \| \pi_{J^\perp} \,\Sig\, \pi_{I^\perp} \| \:, \]
where~$\| . \|$ denotes the sup-norm. In fact, the infimum is attained if~$I$ and~$J$
are invariant subspaces of~$\Sig$ which together span the spectral subspace corresponding to the
eigenvalues~$\lambda_1, \ldots, \lambda_{2n}$.
We choose an orthonormal basis~$(\e_{k,s})_{k \in \Z, s \in \{\pm\}}$ of~$\H_m$
formed of plane-wave solutions,
\[ \e_{k,\pm}(t,x) = \frac{1}{\sqrt{4 \pi b}\: \omega}\:
(\pm \omega \gamma^0 - k \gamma^1+ m ) \,\chi\, e^{\mp i \omega t + i \frac{k x}{b}} \:, \]
where~$\omega(k):=\sqrt{k^2/b^2+m^2}$, and~$\chi$ is the fixed spinor~$\chi=(1,i)/\sqrt{2}$.
We choose~$I$ and~$J$ as the $(4k+2)$-dimensional subspace
\[ \H_{(k)} := \text{span} \big( \e_{-k,\pm}, \ldots, \e_{k, \pm} \big) \:. \]
Then
\[ |\lambda_{8 k+5}| \leq \| \pi_{\H_{(k)}^\perp} \,\Sig\, \pi_{\H_{(k)}^\perp} \|
= \sup_{\psi \in \H_{(k)}^\perp,\; \|\psi\|=1} \big| \la \psi | \Sig \psi \ra \big| \:, \]
and applying Lemma~\ref{lemmaderes} gives
\beq
|\lambda_{8 k+5}| \leq 
\sup_{\psi \in \H_{(k)}^\perp,\; \|\psi\|=1} \bigg(
(1+mb) \: \| \theta \|
+(1+2 \sqrt{mb})^2 \:|\theta|^2_{C^0} \sum_\pm \TV_{[0,b]} \,  T'_\pm \bigg) \:, \label{inter}
\eeq
where~$\theta$ is a primitive of~$\theta$,
\[ \theta(x) = \int_0^x \psi(0,y)\, dy\:. \]
In order to estimate~$\theta$, we expand~$\psi$ in the basis~$(\e_{k, \pm})$ and integrate,
\begin{align*}
\psi(t,x) &= \sum_{|\ell| > k, \;s} c_{\ell,s} \: \e_{\ell,s}(t,x) \\
\theta(x) &= \sum_{|\ell| > k, \;s} c_{\ell,s} \: \frac{1}{\sqrt{4 \pi b}\: \omega}\:
(\pm \omega \gamma^0 - k \gamma^1+ m ) \,\chi\, \frac{b}{i k} \left(e^{i k x}-1 \right) .
\end{align*}
As a consequence,
\begin{align*}
\|\theta\|^2 &= \sum_{|\ell| > k, \;s} |c_{\ell,s}|^2\: \frac{b^2}{\ell^2} \leq \frac{b^2}{k^2}\: \|\psi\|^2 \\
|\theta|_{C^0} &\leq \sum_{|\ell| > k, \;s} \frac{|c_{\ell,s}|}{\sqrt{4 \pi b}}\: \frac{b}{|\ell|}
\leq \sqrt{\frac{b}{4 \pi}} \bigg( \sum_{|\ell| > k, \;s} |c_{\ell,s}|^2 \bigg)^\frac{1}{2}
\bigg( \sum_{|\ell| > k, \;s} \frac{1}{\ell^2} \bigg)^\frac{1}{2} \\
&= \sqrt{\frac{b}{4 \pi}} \:\|\psi\| \bigg( \sum_{|\ell| > k, \;s} \frac{1}{\ell^2} \bigg)^\frac{1}{2}
\leq \sqrt{\frac{b}{4 \pi}} \:\|\psi\| \ \:\frac{2}{\sqrt{k}}\:.
\end{align*}
Using this inequality in~\eqref{inter}, we conclude that
\[ |\lambda_{8k+5}| \leq \frac{b}{k} \bigg(
(1+mb) 
+\frac{(1+2 \sqrt{mb})^2}{\pi} \sum_\pm \TV_{[0,b]} \,  T'_\pm \bigg) . \]
Simplifying the constant with the Schwarz inequality gives the result.
\QED

\subsection{Representation of~$\Sig$ as an Integral Operator}
We now generalize methods and results of Sections~\ref{sec22} and~\ref{sectrace}
to the massive case.

\begin{Prp} The statements of Proposition~\ref{prpintrep}
and Corollary~\ref{cortrace} also hold in the massive case.
\end{Prp}
\Proof According to Lemma~\ref{lemmabessel}, the solution of
the Cauchy problem can be written as
\begin{align*}
\psi(t,x) = \begin{pmatrix} \psi_L(0, x+t) \\ \psi_R(0, x-t) \end{pmatrix}
+ \int_0^b K(t, x-x') \, \psi(0, x')\: dx'
\end{align*}
with a bounded kernel~$K$,
\[ |K(t,x)| < c \qquad \text{for all~$t, x \in \R$}\:. \]
Using this representation in~\eqref{stip}, multiplying out and estimating each term gives
\begin{align*}
\big| \bra \psi | \phi \ket \big| &\leq
\int_\scrM \left( \|\psi_R(0, x-t)\| \, \|\phi_L(0,x+t)\| + \|\psi_L(0, x+t)\| \, \|\phi_R(0,x-t)\|\right) dt \,dx \\
&\quad + c \left( \int_0^L \|\psi(0,x')\|\, dx' \right) \int_\scrM \left( \| \phi_R(0, x-t)\| + \|\phi_L(0, x+t)\| \right) dt\, dx\\
&\quad + c \left( \int_0^L \|\phi(0,x')\| dx' \right) \int_\scrM \left( \| \psi_R(0, x-t)\| + \|\psi_L(0, x+t)\| \right) dt\, dx \\
&\quad + c^2\, \mu(\scrM) \left(  \int_0^b \|\phi(0,x)\|\, dx \right)  \left(  \int_0^b \|\psi(0,x')\| dx' \right) .
\end{align*}
Now the first integral can be estimated as in the proof of Proposition~\ref{prpintrep}
by the integral over the whole causal diamond.
We conclude that there is a constant~$C$ (which depends only on~$m$ and the geometry of~$\scrM$)
such that for all~$\psi, \phi \in \H_m$,
\[ \big| \bra \psi | \phi \ket \big| \leq C \left( \int_0^b \|\psi(0, x)\| \, dx \right) \left( \int_0^b \|\phi_L(x')\| \, dx' \right) . \]
Now we can proceed exactly as in the proof of Proposition~\ref{prpintrep} and Corollary~\ref{cortrace}.
\QED

Using the solution of the Cauchy problem in Lemma~\ref{lemmabessel}, we
can immediately generalize Lemma~\ref{lemma0kernel}:
\begin{Lemma} \label{lemmakernel} The fermionic signature operator can be written as an integral operator
\[ (\Sig \psi)(x) = \int_0^b \Sig(x,y) \psi(y) \: dy \]
with the distributional kernel
\[ \Sig(x,y) = 2 \pi \int_\scrM  k_m(-t, x-z) \,k_m(t, z-y)\:\gamma^0 \: dt\, dz \:. \]
\end{Lemma}

By iterating this integral representation, one can form composite expressions in the
fermionic signature operator, just as explained in Section~\ref{lemma0kernel}.
In particular, the formula~\eqref{trSp} generalizes to
\begin{align}
\tr \big(\Sig^{2q} \big) &= \int_\scrM d^2 \zeta_1 \cdots \int_\scrM d^2 \zeta_{2q} \notag \\
&\qquad \;\times \Tr \Big( k_m\big(\zeta_1-\zeta_2\big) \cdots k_m\big(\zeta_{2q-1}-\zeta_{2q} \big)\,
k_m\big(\zeta_{2q} - \zeta_1 \big) \Big)\:. \label{trSpm}
\end{align}

\subsection{Symmetry of the Spectrum}
The symmetry argument of Proposition~\ref{prpsymm} no longer applies in the massive
case, because if~$\psi$ solves the Dirac equation for mass~$m$, then~$\Gamma \psi$
is a solution corresponding to the mass~$-m$. But we now given another transformation of the spinors
involving complex conjugation which again shows that the spectrum of~$\Sig$ is
symmetric.

\begin{Prp} \label{prpsymmmass} The spectrum of~$\Sig$ is symmetric with respect to the origin.
\end{Prp}
\Proof We introduce the anti-linear mapping
\beq \label{Adef}
{\mathscr{A}} \::\: \psi \mapsto \Gamma \,\overline{\psi} \:,
\eeq
where the bar denotes complex conjugation and~$\pseudo$ is again the matrix in~\eqref{pseudodef}.
Suppose that~$\psi \in \H_m$ is a solution of the Dirac equation~\eqref{Dir}
Using that the Dirac matrices~\eqref{gamma} have real entries, we obtain
\[ (\D - m) \,\pseudo\, \overline{\psi} = \pseudo \,(-\D - m)\, \overline{\psi} 
= \pseudo \,\overline{ (\D - m) \psi} = 0 \:, \]
showing that~${\mathscr{A}} : \H_m \rightarrow \H_m$ maps solutions to solutions.
Moreover, using~\eqref{ssprod}, \eqref{stip} and~\eqref{print2}, one readily verifies that~${\mathscr{A}}$
preserves the norm but flips the sign of the space-time inner product,
\beq \label{Aact}
({\mathscr{A}} \psi | {\mathscr{A}} \phi) = (\phi | \psi) \:,\qquad
\bra {\mathscr{A}} \psi | {\mathscr{A}} \phi \ket = -\bra \phi | \psi \ket \:.
\eeq

Using the orthogonality of the eigenspaces, the eigenvalue equation~$\Sig \psi = \lambda \psi$
can be written in the equivalent form
\[ (\psi \,|\, \Sig\, \psi) = \lambda\, (\psi | \psi) \qquad \text{and} \qquad
(\phi \,|\, \Sig \,\psi) = 0 \quad \forall\, \phi \perp \psi \:. \]
By definition of the fermionic signature operator~\eqref{Sdef}, this can be written equivalently as
\beq \label{econd}
\bra \psi | \psi \ket = \lambda\, (\psi | \psi) \qquad \text{and} \qquad
\bra \phi | \psi \ket = 0 \quad \forall\, \phi \perp \psi \:.
\eeq
Suppose that~$\psi \in \H_m$ is an eigenvector corresponding to the eigenvalue~$\lambda$.
Then~\eqref{econd} holds. The relations~\eqref{Aact} imply that
\[ \bra {\mathscr{A}} \psi \,|\, {\mathscr{A}} \psi \ket = -\lambda\, ({\mathscr{A}} \psi \,|\, {\mathscr{A}} \psi) \qquad \text{and} \qquad
\bra \phi \,|\, {\mathscr{A}} \psi \ket = 0 \quad \forall\, \phi \perp {\mathscr{A}} \psi \:. \]
Hence~$\Sig {\mathscr{A}} \psi = - \lambda {\mathscr{A}} \psi$, completing the proof.
\QED
In the physics literature, the analog of the transformation~\eqref{Adef} in four space-time dimensions
is referred to as {\em{charge conjugation}} (see for example~\cite[Section~5.2]{bjorken}
or~\cite[Section~3.6]{peskin+schroeder}). The interesting point is that the symmetry under charge conjugations
is broken if external potentials (like an electromagnetic potential) are present. In this case, the
spectrum of the fermionic signature operator will in general no longer be symmetric. The deviation
from charge conjugation symmetry could be detected for example by computing traces of
odd powers of~$\Sig$.

\subsection{Computation of~$\tr(\Sig^2)$} \label{secS2m}
In order to see the effect of the mass on the traces, we now compute the trace of~$\Sig^2$.

\begin{Prp} \label{prpS2massive}
The Hilbert-Schmidt norm of the fermionic signature operator is given by
\[ \tr(\Sig^2) = \frac{\mu(\scrM)}{4 \pi^2} + \frac{m^2}{8 \pi^2} 
\iint_{\scrM \times \scrM} \big(J_0^2+J_1^2 \big)\big(m \sqrt{(\zeta-\zeta')^2} \big)\:
\Theta \big( (\zeta-\zeta')^2 \big) \: d^2 \zeta\, d^2 \zeta' \:. \]
For small~$m$, we have the expansion
\begin{align}
\tr(\Sig^2) &= \frac{\mu(\scrM)}{4 \pi^2} + \frac{m^2}{8 \pi^2} 
\iint_{\scrM \times \scrM} \Theta \big( (\zeta-\zeta')^2 \big) \: d^2 \zeta\, d^2 \zeta' \label{trS2m2} \\
&\quad + \frac{m^4}{32 \pi^2} \iint_{\scrM \times \scrM} (\zeta-\zeta')^2\: \Theta \big( (\zeta-\zeta')^2 \big) \: d^2 \zeta\, d^2 \zeta' + \O(m^6)  \:. \label{trS2m4}
\end{align}
For large~$m$, we have the asymptotics
\beq \label{largem}
\tr(\Sig^2) =  \frac{m}{4 \pi^3}
\iint_{\scrM \times \scrM} \frac{\Theta \big( (\zeta-\zeta')^2 \big)}{\sqrt{(\zeta-\zeta')^2}}\:  \: d^2 \zeta\, d^2 \zeta'
+ \O \Big( \frac{1}{m^0} \Big)\:.
\eeq
\end{Prp}
\Proof We again work in light-cone coordinates~$(u,v)$. Then the causal fundamental solution
of Lemma~\ref{lemmabessel} can be written as
\begin{align*}
k_m(u,v) =\;& \frac{1}{4\pi} \,\Big( \gu \:\delta(u) + \gv\: \delta(v) \Big)
-\frac{im}{4 \pi}\: J_0 \big( m \sqrt{uv} \big)\: \epsilon(u+v)\: \Theta(uv) \\
&-\frac{m}{8 \pi} \: \big(v \gu + u \gv \big) \: \frac{J_1 \big( m \sqrt{uv} \big)}{\sqrt{uv}}
\:\epsilon(u+v)\: \Theta(uv) \:.
\end{align*}
Now the result follows from~\eqref{trSpm} by a straightforward computation
using asymptotic expansion of the Bessel functions.
\QED

Compared to the formula of Proposition~\ref{prpS2}, the dependence on the
mass parameter~$m$ gives additional geometric information:
The term~$\sim m^2$ in~\eqref{trS2m2} has the same structure as the
formula~\eqref{trS4}  for~$\tr(\Sig^4)$ in the massless case.
The term~$\sim m^4$, on the other hand, involves an additional weight factor~$(\zeta-\zeta')^2$.
The formula for large~$m$ in~\eqref{largem} again has a similar structure,
but with yet another weight factor~$1/\sqrt{(\zeta-\zeta')^2}$.
For brevity, we do not work out the geometric meaning of these different integrals.

\section{Lorentzian Surfaces in the Massless Case} \label{secglobhyp}
\subsection{Conformal Embedding into Minkowski Space}
Let~$(\scrM,g)$ be a two-dimen\-sio\-nal time-oriented Lorentzian manifold.
The manifold is {\em{globally hyperbolic}} if it does not contain closed causal
curves and if the causal diamonds~$D(\zeta, \zeta')$
(see~\eqref{causaldiamond}) are compact for all space-time points~$\zeta, \zeta' \in \scrM$
(for details see~\cite{bernal+sanchez3}). It is proven in~\cite{bernal+sanchez} that any
globally hyperbolic space-time admits a smooth foliation~$(\scrN_t)_{t \in \R}$ by spacelike Cauchy
hypersurfaces, defined as follows.

\begin{Def}
A subset of a time-oriented Lorentzian manifold $(\scrM,g)$ is called {\bf Cauchy surface} if it is intersected exactly once by every $C^0$-inextendible future causal curve in $\scrM$. 
\end{Def}

It is a well-known fact that any two-dimensional Lorentzian manifold is locally conformally flat,
in the sense that any point of~$\scrM$ has a neighborhood which is conformal to an open subset
of Minkowski space. It is less well-known that a globally hyperbolic Lorentzian manifolds
admits even a {\em{global}} conformal embedding to Minkowski space:

\begin{Prp}
\label{gh-general}
Let~$(\scrM,g)$ be a globally hyperbolic two-dimensional manifold with a non-compact Cauchy surface $\scrN$. Then there is a conformal map~$\Phi : (\scrM,g) \rightarrow \R^{1,1}$ whose image is open, relatively compact and causally convex (meaning that no future-directed causal curve can leave and reenter~$\Phi(\scrM)$), and such that $\scrN$ is
mapped to $\{ 0 \} \times (0,1)$. 
\end{Prp}
\Proof Since $\scrN$ is non-compact, it is diffeomorphic to $\R$.
We introduce a new metric~$h$ on~$\scrM$ obtained by the conformal change
\[ h = e^{2u}\, g \]
with~$u \in C^\infty(\scrM)$.
A direct computation shows that the scalar curvatures of~$g$ and~$h$ are related by
\[ 2 \Delta_g w + s_g  = e^{2w} s_h . \]
This shows that the equation~$s_h= 0$ is equivalent to the linear
normally hyperbolic equation
\beq \label{weq}
2 \Delta_g w + s_g  = 0 \:.
\eeq
We impose the initial conditions
\beq \label{iv}
w(0, r) = w_0 (r), \qquad \nabla_\nu w (0,r) = w_1(r)
\eeq
(where~$r \in \R$ parametrizes~$\scrN$, and~$\nu$ is again the future-directed normal vector field on~$\scrN$).
The resulting Cauchy problem~\eqref{weq}, \eqref{iv} is globally well-posed
(see for example~\cite{john, taylor1, baer+ginoux}).

We next choose the initial conditions~$w_0$ and~$w_1$ such that~$\scrN$ becomes
a $h$-pregeodesic of length one (a pregeodesic is a geodesic up to reparametrizations):
Specializing the general formulas in~\cite[Theorem~1.159]{besse}, the condition for being a
pregeodesic is
$$0 = \nabla_{\partial_r }^h (e^{-u} \nu) = \nabla^g_{\partial_r} (e^{-u} \nu) + e^{-u}  \frac{\partial u}{\partial_r}  \nu  + e^{-u} \nu(u) \partial_r   \:, $$
which is equivalent to the equation
\[ \nabla_\nu u = - \nabla^g_{\partial_r} \nu = - H^g \:, \]
where $H^g$ is mean curvature. This equation can be satisfied by suitably choosing~$w_1$.
We still have the freedom to choose~$w_0$ arbitrarily. We use this freedom to give~$\scrN$ length one.

Solving the above Cauchy problem, we obtain a flat metric in which~$\scrN$ is totally geodesic.
The proof is completed by applying Lemma~\ref{lemmaisom} below.
\QED
 
\begin{Lemma} \label{lemmaisom}
Let~$(\scrM,h)$ be a two-dimensional, flat Lorentzian manifold which contains a totally geodesic Cauchy surface
of length one. Then~$(\scrM, h)$ is isometric to an open neighborhood
of $\{ 0 \} \times (0,1)$ in $\R^{1,1}$.
\end{Lemma} \noindent
We point out that in the Riemannian case, this proposition does not hold, as there
are examples of flat contractible two-dimensional manifolds which do {\em{not}} admit an isometric
embedding into~$\R^2$: Take any periodic immersed curve $c: \SSS^1 \rightarrow \R^n$ self-intersecting exactly once at $p \in \SSS^1$, like for example the Lemniscate of Bernoulli. 
As it is immersed, it has a normal neighborhood $\scrN$ such that $ c $ extends to a local diffeomorphism $C: \SSS^1 \times \R^{n-1} \rightarrow \scrN$.
We pull back the Euclidean metric to a flat metric $G$ on $\SSS^1 \times \R^{n-1} $ and restrict it to the open subset 
$\scrN:= (\SSS^1 \setminus \{ q \} )  \times \R^{n+1}$
where $q \in \SSS^1 \setminus \{ p \} $. Then the usual rigidity arguments ensure that any other isometric immersion of $(\scrN,G)$ into Euclidean $\R^n$ 
coincides with $C$ up to rigid motions and thus is not an embedding.

\Proof[Proof of Proposition~\ref{lemmaisom}] Parametrizing the Cauchy surface by arc length,
we obtain a $h$-geodesic~$c : (0,1) \rightarrow \scrM$ with~$\scrN=c((0,1))$.

Let us show that, for $T^\perp \scrN:= \{ v \in T_n \scrM \,\vert\, n \in \scrN, v \perp T\scrN \}$, the normal exponential map~$E := {\rm exp}^\scrM \vert_{T^\perp \scrN}$ on~$\scrN$ is {\em{injective}}:
Consider any two timelike geodesics $c_1, c_2$ starting at different points~$x_1, x_2 \in \scrN$,
in the direction of the normal~$\nu$.
These geodesics cannot intersect at a point $p$ as that would be in contradiction to
the Ambrose-Singer theorem (see~\cite[Theorem 10.58]{besse}).
Namely, assume conversely that these geodesics intersect at a point~$p$.
Let~$\Delta$ be the geodesic triangle with vertices~$x_1$, $x_2$ and~$p$.
Since~$\scrN$ is totally geodesic, its normal vector field~$\nu$ is parallel along~$\scrN$.
Moreover, since the two geodesics~$c_1$ and~$c_2$ must intersect transversely,
the parallel transport of~$\nu$ along these geodesics gives two different vectors at~$p$.
Hence the corresponding holonomy of along the triangle~$\Delta$ is non-zero.
On the other hand, the triangle~$\Delta$ clearly is contractible.
But since~$(\scrM,g)$ is flat, the Ambrose-Singer theorem implies that the
the Lie algebra of the connected Lie group~${\rm Hol}_0 (\scrM,g)$ is trivial and
thus~${\rm Hol}_0 (\scrM,g) = \{ 1 \}$.
This implies that the holonomy along any contractible curve in $\scrM$ is the identity.
This is a contradiction.

We next show that~$E$ is also {\em{surjective}}:
For a point~$p \in J^\vee(\scrN)$ in the future of~$\scrN$, we let~$d_p : \scrM \rightarrow [0, \infty)$ be
the distance function from~$p$ (set to zero for spatially separated points).
In globally hyperbolic space-times, this distance function is continuous (see~\cite[Lemma 14.21]{oneillsemi}).
Moreover, the global hyperbolicity of~$\scrM$ implies that the set~$R_p := J^\wedge(p) \cap \scrN $ is compact.
Hence the restriction of~$d_p$ to~$\scrN$ attains a maximum in $R_p$ at a point $q$.
Again due to global hyperbolicity, there is a geodesic curve~$\gamma$ joining~$q$ and~$p$.
The first variational formula implies that~$\gamma$ is perpendicular to $\scrN$ at $q$. Consequently, $p$ has to be in the image of the normal exponential map.

We conclude that we have global Fermi coordinates in $(\scrM,h)$ in which the metric takes the form
\[ h = dt^2 - b(t,x) \,dx^2 \:. \]
A short computation of the curvature tensor shows that~$b$ must not depend on~$t$.
Since~$b(0,x)\equiv 1$ (parametrization of~$c$ by arc length), we find that~$b \equiv 1$ on~$\scrM$.
We conclude that the metric in Fermi coordinates simply is the Minkowski metric.

The above argument can be applied just as well to the past of~$\scrN$. Combining the results for
the past and future of~$\scrN$, we find that~$E$ gives an isometric diffeomorphism from an open subset~$\Omega$
of $\R^2$ endowed with the Minkowski metric to $(\scrM,h)$. It remains to show that~$\Omega$ is
a globally hyperbolic subset of~$R^{1,1}$ with Cauchy surface~$\{0\} \times (0,1)$.
But this follows immediately from the fact that~$\Phi := E^{-1} : (\scrM,h) \rightarrow \Omega \subset \R^{1,1}$ maps
Cauchy surfaces isometrically to Cauchy surfaces.
\QED

\subsection{Conformal Transformation of the Fermionic Signature Operator} \label{secconf}
We again let~$(\scrM,g)$ be a time-oriented, globally hyperbolic Lorentzian surface with
a given Cauchy surface~$\scrN$.
According to Proposition~\ref{gh-general}, we can identify~$\scrM$ with an open subset
of Minkowski space~$\R^{1,1}$, endowed with the conformal metric
\beq \label{gconf}
g = f(t,x)^2 \left( dt^2 - dx^2 \right) \:,\qquad f \in C^\infty(\scrM)\:.
\eeq
Moreover, this proposition allows us to arrange that the Cauchy surface~$\scrN$ is the set~$\{0\} \times (0,1)$.

From now on, we denote all quantities referring to the metric~$g$ for clarity with a tilde,
whereas the quantities without a tilde refer to the flat Minkowski metric.
We consider the massless Dirac equation
\[ \tilde{\Dir} \tilde{\psi} = 0 \:. \]
This equation as well as its solutions can be described most conveniently using
the {\em{conformal invariance of the massless Dirac equation}} (see for example~\cite{hitchin, hijazi}),
which implies that
\beq \label{confDir}
\tilde{\Dir} = f^{-\frac{3}{2}} \,\Dir\, f^{\frac{1}{2}} \:, \qquad \tilde{\psi} = f^{-\frac{1}{2}}\: \psi \:,
\eeq
where~$\Dir$ is again the Dirac operator in Minkowski space~\eqref{Dir},
and~$\psi$ is a solution of the form~\eqref{psiLR}.
The scalar product on the solutions becomes
\begin{align*}
(\tilde{\psi} | \tilde{\phi}) &= \int_\scrN \Sl \tilde{\psi} | \slashed{\nu} \tilde{\phi} \Sr|_x\: d\mu_\scrN(x) \\
&= \int_0^1 \Sl \tilde{\psi} | \gamma^0 \, \tilde{\phi} \Sr|_{(0,x)}\: f(0,x)\: dx
= \int_0^1 \Sl \psi | \gamma^0 \, \phi \Sr|_{(0,x)}\: dx \:,
\end{align*}
showing that the scalar product on the solutions is conformally invariant.
We again denote the corresponding Hilbert space of solutions by~$(\H_0, (.|.))$.
The space-time inner product~\eqref{stip}, however, does involve the conformal factor, because
\begin{align*}
\bra \tilde{\psi} | \tilde{\phi} \ket &= \int_\scrM \Sl \tilde{\psi} | \tilde{\phi} \Sr\: d\mu \\
&= \int_\scrM \Sl \tilde{\psi} | \tilde{\phi} \Sr\: f(t,x)^2\: dt\, dx
= \int_\scrM \Sl \psi | \phi \Sr\: f(t,x)\: dt\, dx \:.
\end{align*}
As a consequence, the fermionic signature operator has a non-trivial dependence
on the conformal factor.

Before we can define the fermionic signature operator again by~\eqref{Sdef}, we need
to make sure that the space-time inner product is bounded~\eqref{bounded}.
To this end, we assume that~$(\scrM,g)$ has {\bf{finite lifetime}} in the
sense that it admits a foliation~$(\scrN_t)_{t \in (t_0, t_1)}$ by Cauchy surfaces with a bounded
time function~$t$ such that the function~$\la \nu, \partial_t \ra$ is bounded on~$\scrM$ (where~$\nu$ denotes
the future-directed normal on~$\scrN_t$ and~$\la \nu, \partial_t \ra \equiv g(\nu, \partial_t)$).
Then~\eqref{bounded} holds for a suitable constant~$c$ (see~\cite[Proposition~3.5]{finite}).
Thus~\eqref{Sdef} defines~$\Sig$ as a bounded symmetric operator on the Hilbert space~$\H_0$.

In the next lemma we again represent the fermionic signature operator as an integral operator.
\begin{Lemma} \label{lemmakernelcurv} The fermionic signature operator can be written as an integral operator
\beq \label{Srepconf}
(\Sig \tilde{\psi})(x) = \int_0^1 \Sig(x,y) \,\tilde{\psi}(y) \: f\big(0,y)\, dy
\eeq
with the distributional kernel
\beq \label{Sigconf}
\Sig(x,y) = 2 \pi \: f(0,x)^{-\frac{1}{2}}\: f(0,y)^{-\frac{1}{2}} 
\int_\scrM  f(t,z)\: k(-t, x-z) \,k(t, z-y)\:\gamma^0 \: dt\, dz \:,
\eeq
where~$k$ is the causal fundamental solution of Minkowski space~\eqref{k0}.
\end{Lemma}
\Proof In view of the transformation of the Dirac operator in~\eqref{confDir}, the advanced Green's
function~$\tilde{s}^\vee_0$ (defined by the
equation~$\tilde{\Dir} \tilde{s}^\vee_0 = \1$) transforms conformally as~$\tilde{s}^\vee_0
= f^{-\frac{1}{2}} s^\vee_0 f^{\frac{3}{2}}$. Writing this operator with an integral kernel
and keeping in mind the transformation of the volume forms, one finds
\[ \tilde{s}^\vee_0(\zeta, \zeta') =  f(\zeta)^{-\frac{1}{2}}\, s^\vee(\zeta,\zeta')\, f(\zeta')^{-\frac{1}{2}} \:. \]
The retarded Green's function transforms in the same way. Thus, introducing the
causal fundamental solution~$k$ similar to~\eqref{kmdef}, we obtain
\[ \tilde{k}(\zeta,\zeta') = \frac1{2\pi i}\left( \tilde{s}^{\vee}-\tilde{s}^{\wedge} \right)(\zeta, \zeta') =
f(\zeta)^{-\frac{1}{2}}\, k(\zeta, \zeta')\, f(\zeta')^{-\frac{1}{2}} \:. \]

The solution formula for the Cauchy problem~\eqref{solCauchy2} and~\eqref{cauchyrep}
generalizes to (see~\cite[Lemma~2.1]{finite})
\begin{align*}
\tilde{\psi}(\zeta) &= \int_\scrN \tilde{k} \big(\zeta, \zeta' \big)\: \slashed{\nu}(\zeta')\: \tilde{\psi}({\zeta'})\:
d\mu_\scrN({\zeta'}) \\
&= f(\zeta)^{-\frac{1}{2}} \int_0^1 k\big( \zeta - (0,x) \big) \: \gamma^0 \: \tilde{\psi}(0,x)\:
f(0,x)^\frac{1}{2}\:dx\:.
\end{align*}
Modifying the proof of Lemma~\ref{lemma0kernel} in an obvious manner, one
obtains the integral representation~\eqref{Srepconf} with
\begin{align*}
&\Sig(x,y) = 2 \pi \int_\scrM \tilde{k}\big((0,x), \zeta \big) \,\tilde{k}\big(\zeta, (0,y) \big) \:
\slashed{\nu} \big( (0,y) \big) \, \: d\mu(\zeta) \\
&= 2 \pi \: f(0,x)^{-\frac{1}{2}} \int_\scrM k\big((0,x)-\zeta \big) \,f(\zeta)^{-1}\,k\big(\zeta-(0,y) \big) \:
\gamma^0 \: f(0,y)^{-\frac{1}{2}}\:  f(\zeta)^2\, d^2\zeta \:.
\end{align*}
This concludes the proof.
\QED

We now compute the kernel more explicitly by transforming to light-cone coordinates~\eqref{uvdef}
and using the form the distribution~$k$ in~\eqref{k0uv}.
We extend the function~$f$ by zero to all of~$\R^{1,1}$ and denote this function for clarity
by~$\chi_\scrM f$.
\begin{Lemma} \label{lemmaSigcurv}
The integral kernel~\eqref{Sigconf} can be written as
\beq \label{Sigcurv}
\begin{split}
\Sig(x,y) &= \frac{1}{16 \pi} \: f(0,x)^{-\frac{1}{2}}\: f(0,y)^{-\frac{1}{2}} \\
&\quad \times \bigg\{ \gu \gv\: (\chi_\scrM \, f)\big( i^+(x,y) \big)
+ \gv \gu\: (\chi_\scrM\, f) \big( i^-(x,y) \big) \bigg\}
\,\gamma^0\:,
\end{split}
\eeq
where~$i^\pm$ are the upper and lower points of the corresponding causal diamond defined by
\beq \label{ipm}
i^+(x,y) = \Big(\frac{x-y}{2}, \frac{x+y}{2} \Big) \:,\qquad
i^-(x,y) = \Big(-\frac{x-y}{2}, \frac{x+y}{2} \Big) \:.
\eeq
\end{Lemma}
\Proof Transforming to light-cone coordinates and using~\eqref{k0uv}, the kernel~\eqref{Sigconf}
can be written as
\begin{align*}
\Sig(x,y) &= \pi \: f(0,x)^{-\frac{1}{2}}\: f(0,y)^{-\frac{1}{2}} \\
&\qquad \times
\int_\scrM  f\Big( \frac{u+v}{2}, \frac{u-v}{2} \Big)\: k(x-u, -x-v) \,k(u-y, v+y)\:\gamma^0 \: du\, dv \:,
\end{align*}
where~$k(u,v)$ is given by~\eqref{k0}. Using the explicit form of this distribution, we can carry out the
$u$ and~$v$-integrations to obtain
\begin{align*}
\Sig(x,y) &= \frac{1}{16 \pi} \: f(0,x)^{-\frac{1}{2}}\: f(0,y)^{-\frac{1}{2}} \int_\scrM  f\Big( \frac{u+v}{2}, \frac{u-v}{2} \Big) \\
&\qquad \times \Big( \gu \gv \:\delta(x-u) \,\delta(v+y) + \gv \gu\: \delta(-x-v)\, \delta(u-y) \Big) 
\,\gamma^0 \: du\, dv \\
&= \frac{1}{16 \pi} \: f(0,x)^{-\frac{1}{2}}\: f(0,y)^{-\frac{1}{2}} \\
&\qquad \times \bigg\{ \gu \gv\: (\chi_\scrM \, f) \Big(\frac{x-y}{2}, \frac{x+y}{2} \Big)
+ \gv \gu\: (\chi_\scrM\, f) \Big(\frac{y-x}{2}, \frac{y+x}{2}\Big) \bigg\}
\,\gamma^0 \:. 
\end{align*}
This gives the result.
\QED

\subsection{Computation of~$\tr(\Sig^2)$ and~$\tr(\Sig^4)$: Volume and Curvature}
Having derived explicit formulas for the integral kernel~$\Sig$,
the spectral properties of the fermionic signature operator can be analyzed
similarly as described in Sections~\ref{sec23}--\ref{sectrSp} for subsets of Minkowski space.
Some results (like Proposition~\ref{prp91}) generalize immediately,
whereas for other results (like Proposition~\ref{prp81}) the generalization is less obvious.
For brevity, we shall not reconsider all the results for the Minkowski drum.
Instead, we restrict attention to generalizing Propositions~\ref{prpS2} and~\ref{prp71} to
curved surfaces. The main point of interest is that the resulting formula for~$\tr (\Sig^4)$
involves scalar curvature (see Proposition~\ref{prpcurv} below).

In preparation, we show that the statement of Corollary~\ref{cortrace} still holds, provided
that the space-time volume is finite.
\begin{Lemma} Let~$(\scrM,g)$ be a globally hyperbolic Lorentzian surface
of finite lifetime. If the total $g$-volume~$\mu(\scrM)$ is finite, then the fermionic signature operator
is Hilbert-Schmidt. Moreover, the
traces of even powers of~$\Sig^{2q}$, $q \in \N$, are given by the integrals
\beq \label{trS2qcurv}
\tr(\Sig^{2q}) = \int_0^1 f(0,x_1)\, dx_1 \ldots \int_0^1 f(0,x_{2q})\, dx_{2q}
\Tr \big(\Sig(x_1, x_2) \cdots \Sig(x_{2q}, x_1) \big) \:.
\eeq
\end{Lemma}
\Proof Following the method in the proof of Corollary~\ref{cortrace}, 
the Hilbert-Schmidt property as well as the formula~\eqref{trS2qcurv} in case~$q=1$
follows immediately once we know that
the kernel of the fermionic signature operator is square integrable in the sense that
\beq \label{HScurv}
\int_0^1 \int_0^1 \|\Sig(x,y)\|^2\: f(0,x)\, dx\, f(0,y)\, dy < \infty \:.
\eeq
Estimating~\eqref{Sigcurv}, we obtain
\begin{align*}
\|\Sig(x,y)\| &\leq \frac{1}{16 \pi} \: f(0,x)^{-\frac{1}{2}}\: f(0,y)^{-\frac{1}{2}}\,
\Big( (\chi_\scrM \, f)\big( i^+(x,y) + (\chi_\scrM\, f) \big( i^-(x,y) \big) \Big) \\
\|\Sig(x,y)\|^2 &\leq \frac{2}{(16 \pi)^2} \: f(0,x)^{-1}\: f(0,y)^{-1}\,
\Big( (\chi_\scrM \, f)\big( i^+(x,y)^2 + (\chi_\scrM\, f) \big( i^-(x,y)^2 \big) \Big)
\end{align*}
and thus
\begin{align*}
\int_0^1 & \int_0^1 \|\Sig(x,y)\|^2\: f(0,x)\, dx\, f(0,y)\, dy \\
&\leq \frac{2}{(16 \pi)^2} \int_0^1 \int_0^1 
\Big( (\chi_\scrM \, f)\big( i^+(x,y)^2 + (\chi_\scrM\, f) \big( i^-(x,y)^2 \big) \Big)\: dx\, dy\:.
\end{align*}
Using~\eqref{ipm}, one can rewrite the integrals as a space-time integral to obtain
\[ \int_0^1 \int_0^1 \|\Sig(x,y)\|^2\: f(0,x)\, dx\, f(0,y)\, dy \\
\leq \frac{8}{(16 \pi)^2} \int_\scrM f(t,z)^2\: dt\, dz = \frac{8}{(16 \pi)^2} \: \mu(\scrM)\:, \]
where in the last step we used that~$d\mu = f^2\, dx\, dy$.
This shows~\eqref{HScurv} and concludes the proof in the case~$q=1$.

In order to treat the case~$q>1$, by iterating~\eqref{Srepconf} and using Fubini's theorem,
one obtains an integral representation of~$\Sig^q$ with a kernel which is again
square integrable.
Again arguing as in the proof of Corollary~\ref{cortrace}, we obtain the result.
\QED

\begin{Prp} Let~$(\scrM,g)$ be a globally hyperbolic Lorentzian surface
of finite lifetime and finite volume. Then
\[ \tr \big(\Sig^2 \big) = \frac{\mu(\scrM)}{4 \pi^2} \:. \]
\end{Prp}
\Proof Evaluating~\eqref{trS2qcurv} for the kernel~\eqref{Sigconf} and~\eqref{Sigcurv}, 
in generalization of~\eqref{trSp} and~\eqref{trS2q} we obtain 
\begin{align}
\tr \big(\Sig^{2q} \big) &= \int_\scrM f(\xi_1)^2 \:d^2 \xi_1 \cdots \int_\scrM f(\xi_{2q})^2 \:d^2 \xi_{2q} \notag \\
&\qquad \times \Tr \Big( k_m\big(\xi_1-\xi_2\big) \cdots k_m\big(\xi_{2q-1}-\xi_{2q} \big)\,
k_m\big(\xi_{2q} - \xi_1 \big) \Big) \notag \\
&= \frac{2}{(2 \pi)^{2q}} \int_\scrM f(\zeta_1)\: d^2 \zeta_1 \int_\scrM f(\eta_1)\: d^2 \eta_1 \cdots 
\int_\scrM f(\zeta_q)\: d^2 \zeta_q \int_\scrM d^2 f(\eta_q)\: \eta_q \notag \\
&\qquad \times \delta(u_1-\tilde{u}_1) \,\delta(\tilde{v}_1-v_2)\, \cdots \delta(u_q-\tilde{u}_q)
\,\delta(\tilde{v}_q-v_1)\:, \label{traceform}
\end{align}
where in the last line we set~$\zeta_j = \xi_{2j-1}$ (having the light-cone coordinates~$(u_j, v_j)$)
and~$\eta_j = \xi_{2j}$ (having the light-cone coordinates~$(\tilde{u}_j, \tilde{v}_j)$). In particular,
\begin{align*}
\tr \big( \Sig^2 \big)
&= \frac{1}{2 \pi^2} \int_\scrM f(\zeta_1)\: d^2 \zeta_1 \int_\scrM f(\eta_1)\: d^2 \eta_1\;
\delta(u_1-\tilde{u}_1) \,\delta(\tilde{v}_1-v_1) \\
&= \frac{1}{4 \pi^2} \int_\scrM f(\zeta_1)^2\: d^2 \zeta_1 = \frac{\mu(\scrM)}{4 \pi^2}\:,
\end{align*}
giving the result.
\QED

\begin{Prp} \label{prpcurv}
Let~$(\scrM,g)$ be a globally hyperbolic Lorentzian surface
of finite lifetime and finite volume. Then
\begin{align*}
\tr \big(\Sig^4\big) &=
\frac{1}{8 \pi^4} \int_\scrM d\mu(\zeta) \int_{J(\zeta)} 
\exp \bigg( \frac{1}{4} \int_{D(\zeta, \zeta')} R\: d\mu \bigg)\, d\mu(\zeta') \:,
\end{align*}
where~$D(\zeta, \zeta')$ is the causal diamond~\eqref{causaldiamond},
and~$R$ denotes scalar curvature.
\end{Prp}
\Proof We evaluate~\eqref{traceform} in the case~$q=2$.
For two causally separated points~$\zeta$ and~$\zeta'$, we again let~$D(\zeta, \zeta')$
be the causal diamond whose upper and lower corners are~$\zeta$ and~$\zeta'$.
The left and right corners of this causal diamond are denoted by~$\eta$ and~$\eta'$, respectively
(similar as in the left of Figure~\ref{figconstraint}). Then
\begin{align}
\tr \big(\Sig^4\big) &=
\frac{1}{8 \pi^4} \int_\scrM f(\zeta_1)\: d^2 \zeta_1 \int_\scrM f(\eta_1)\: d^2 \eta_1
\int_\scrM f(\zeta_2)\: d^2 \zeta_2 \int_\scrM d^2 f(\eta_2)\: \eta_2 \notag \\
&\qquad \times
\delta(u_1-\tilde{u}_1) \,\delta(\tilde{v}_1-v_2)\; \delta(u_2-\tilde{u}_2) \,\delta(\tilde{v}_2-v_1) \notag \\
&= \frac{1}{8 \pi^4} \int_\scrM d^2 \zeta\: \int_{J(\zeta)} d^2  \zeta'\;
f(\zeta)\: f(\eta)\: f(\zeta')\: f(\eta') \:. \label{S4int}
\end{align}
The interesting point is that this {\em{not}} the same as the coordinate invariant quantity
\[ \frac{1}{8 \pi^4} \int_\scrM d\mu(\zeta) \int_{J(\zeta)} d\mu(\zeta') = 
\frac{1}{8 \pi^4} \int_\scrM d^2 \zeta\: \int_{J(\zeta)} d^2  \zeta'\; f(\zeta)^2\: f(\zeta')^2 \:, \]
because the factors~$f$ appear in a different combination. In order to express this difference
geometrically, we first note that scalar curvature is given by
\[ R = 2 K = -\frac{1}{f^2}\: \Box \log(f^2) = -\frac{2}{f^2}\: \Box \log f \]
(see for example~\cite[page~237]{docarmo}, where~$\Box$ denotes the wave operator in Minkowski space).
Integrating this formula for scalar curvature over the causal diamond, introducing light-cone coordinates
and integrating by parts, we obtain
\begin{align*}
\int_{D(\zeta, \zeta')} R\: d\mu
&= -2 \int_{D(\zeta, \zeta')} \Box \log f\: dt\, dx
= -4 \int_{D(\zeta, \zeta')} \puv \log f\: du\, dv \\
&= -4 \big( \log f(\zeta) + \log f(\zeta') - \log f(\eta) - \log f(\eta') \big)\:.
\end{align*}
Hence
\begin{align*}
\exp \bigg( \frac{1}{4} \int_{D(z\eta, \zeta')} R\: d\mu \bigg)
= \frac{f(\eta)\, f(\eta')}{f(\zeta)\, f(\zeta')}
\end{align*}
and thus
\[ f(\zeta)\: f(\eta)\: f(\zeta')\: f(\eta')
= f(\zeta)^2\: f(\zeta')^2 \:
\exp \bigg( \frac{1}{4} \int_{D(\zeta, \zeta')} R\: d\mu \bigg) \:. \]
Using this relation in~\eqref{S4int} concludes the proof.
\QED

\subsection{A Reconstruction Theorem} \label{secreconstruct2}
The goal of this section is to prove Theorem~\ref{thmreconstruct} as well as its generalization
to curved surfaces.
Thus we again let~$(\scrM,g)$ be a time-oriented, globally hyperbolic Lorentzian surface
of finite lifetime together with a Cauchy surface~$\scrN$.
Just as described at the beginning of Section~\ref{secconf}, we
can consider~$\scrM$ as a subset of~$\R^{1,1}$ with the conformally flat metric~\eqref{gconf}.
Moreover, we can arrange that~$\scrN=\{0\} \times (0,1)$.
For an open subset~$I \subset (0,1)$ and a chiral index~$c \in \{L, R\}$ we introduce
\[ \pi_{c,I} := \chi_c\, \chi_I \:, \]
where~$\chi_I$ is the characteristic function, and~$\chi_c$ are the projections on the left- or
right-handed components,
\beq \label{chiLRdef}
\chi_L = \begin{pmatrix} 1 & 0 \\ 0 & 0 \end{pmatrix} \:,\qquad
\chi_R = \begin{pmatrix} 0 & 0 \\ 0 & 1 \end{pmatrix} \:.
\eeq
We consider~$\pi_{c,I} : \H_0 \rightarrow \H_0$ as a multiplication operator on the wave functions~$\tilde{\psi}$
on the Cauchy surface~$\scrN$. Obviously, $\pi_{c,I}$ is a projection operator on~$\H_0$.

Next, for an open subset~$I \subset (0,1)$ we introduce the sets~$K_L(I), K_R(I) \subset \scrM
\subset \R^{1,1}$ obtained
by propagating~$I$ with velocity one to the left respectively right,
\[ K_L(I) = \big\{ (t,x) \in \scrM \,|\, x+t \in I \big\} \:,\qquad K_R(I) = \big\{ (t,x) \in \scrM \,|\, x-t \in I \big\} \:. \]

\begin{Lemma} Assume that for two open subsets~$I, J \subset (0,1)$, the following integral exists,
\[ \int_{K_L(I) \cap K_R(J)} f^2\: d^2 \zeta < \infty \:. \]
Then the operator product~$\pi_{L, I}\, \Sig\, \pi_{R, J}$ is Hilbert-Schmidt, and its Hilbert-Schmidt
norm is given by
\[ \big\| \pi_{L,I} \:\Sig\: \pi_{R, J} \big\|^2_{\text{HS}} = \frac{1}{8 \pi^2}
\int_{K_L(I) \cap K_R(J)} f^2\: d^2 \zeta \:. \]
\end{Lemma}
\Proof We first compute the kernel of the operator~$\pi_{L, I}\, \Sig\, \pi_{R, J}$.
Combining~\eqref{chiLRdef} and~\eqref{guvdef} with~\eqref{gamma}, one sees that
\[ \gu \gv = 4\, \chi_L \qquad \text{and} \qquad \gv \gu = 4\, \chi_R \:. \]
Using Lemma~\ref{lemmaSigcurv}, we obtain
\begin{align}
\pi_{L,I} \:\Sig(x,y)\: \pi_{R, J} &= \frac{1}{4 \pi} \: f(0,x)^{-\frac{1}{2}}\: f(0,y)^{-\frac{1}{2}} \notag \\
&\qquad \times \chi_I(x)\, \chi_J(y) \;(\chi_\scrM \, f)\big( i^+(x,y) \big) \; \chi_L\,\gamma^0 \\
\| \pi_{L,I} \:\Sig(x,y)\: \pi_{R, J} \|^2 &= \frac{1}{16 \pi^2} \:f(0,x)^{-1}\: f(0,y)^{-1} \notag \\
&\qquad \times \chi_I(x)\, \chi_J(y) \;\Big( (\chi_\scrM \, f)\big( i^+(x,y) \big) \Big)^2 . \label{kernel}
\end{align}
Using the Fourier series method in the proof of Corollary~\ref{cortrace}, one concludes that
the operator~$\pi_{L, I}\, \Sig\, \pi_{R, J}$ is Hilbert-Schmidt if and only if the function~\eqref{kernel}
is integrable with respect to the measure~$f(0,x)\, dx\, f(0,y)\, dy$.
In this case, the Hilbert-Schmidt norm satisfies the equality
\[ \big\| \pi_{L,I} \:\Sig\: \pi_{R, J} \big\|^2_{\text{HS}} = \frac{1}{16 \pi^2}
\int_0^1 \int_0^1 \chi_I(x)\, \chi_J(y) \;\Big( (\chi_\scrM \, f)\big( i^+(x,y) \big) \Big)^2\: dx\, dy \:. \]

It remains to interpret the integrand geometrically with the help of the definition of~$i^+$
in~\eqref{ipm}. First of all, due to the factor~$\chi_\scrM$, it suffices to consider the case that~$i^+ \in \scrM$.
Then the condition~$x \in I$ means that the space-time point~$i^+(x,y)$
must lie in~$K_L(I)$. Similarly, the condition~$y \in J$ means that~$i^-(x,y) \in K_R(J)$.
Finally, denoting the components of~$i^+(x,y)$ by~$(t,x)$ and transforming the integration
measure according to~$dx\, dy = 2\, dt\, dx$, the result follows.
\QED

The result of this lemma has a simple geometric interpretation which does not rely on our
embeddings. In order to make this point clear, we now consider~$(\scrM,g)$ as an abstract
oriented, time-oriented, globally hyperbolic manifold of finite lifetime with a given non-compact
Cauchy surface~$\scrN$. We identify the Hilbert space of solutions with the initial values
on the Cauchy surface, i.e.
\[ \H_0 = L^2(\scrN, S\scrM) \simeq L^2(\scrN, S\scrN) \oplus L^2(\scrN, S\scrN) \:, \]
where the two direct summands describe the left- and right-handed components of
the spinors, respectively.
Let~$I$ and~$J$ be open subsets of~$\scrN$. 
Then the multiplication operators~$\chi_{L,I}$ and~$\chi_{R,J}$ are defined on~$\H_0$ in an obvious
way. Moreover, the sets~$K_L(I)$ can be defined as
all points of~$\scrM$ which can be reached from~$I$ by a lightlike geodesic propagating to the left.
Similarly, $K_R(J) \subset \scrM$ is the set of all points which can be reached from~$J$ by a lightlike
geodesic propagating to the right. Moreover, the integrand~$f^2\, d^2 \zeta$ is the same
as the volume measure~$d\mu$ corresponding to the metric~$g$. We thus obtain
the following result.
\begin{Prp} \label{taschenlampe-invariant}
Assume that for two open subsets~$I, J \subset \scrN$, the set~$K_L(I) \cap K_R(J)$ has
finite volume. Then the operator product~$\pi_{L, I}\, \Sig\, \pi_{R, J}$ is Hilbert-Schmidt, and its Hilbert-Schmidt
norm is given by
\[ \big\| \pi_{L,I} \:\Sig\: \pi_{R, J} \big\|^2_{\text{HS}} = \frac{1}{8 \pi^2}\: \mu \Big(
K_L(I) \cap K_R(J) \Big) \:. \]
\end{Prp}

This lemma is very useful because if~$\Sig$ is given as an operator on
the Hilbert space of sections of the spinor bundle on a Cauchy surface, then the volume of the
sets~$K_L(I) \cap K_R(J)$ can be recovered for any open subsets~$I, J \subset \scrN$.
In particular, by choosing the sets~$I$ and~$J$ as small neighborhoods of points~$x,y \in \scrN$,
one may find out whether the null geodesics through~$x$ and~$y$ meet at a space-time point~$i^+(x,y)$.
If they do, one can even determine the volume form at this space-time point.
For subsets of Minkowski space, we thus obtain the statement of Theorem~\ref{thmreconstruct}.

We finally formulate the reconstruction theorem for general surfaces.
For the sake of conceptual clarity, we formulate this result in the language of categories.
Let~$X$ be a locally compact Hausdorff space. By~$C_0(X)$ we
denote the continuous functions on~$X$ which vanish at infinity in the sense that
for every~$\varepsilon>0$ there is a compact subset~$K \subset X$ such that~$|f(x)|< \varepsilon$
for all~$x \in X \setminus K$. Then the celebrated {\em{Gelfand-Naimark theorem}} states that~$X$
can be reconstructed (modulo homeomorphisms) from the single datum of the~$C^*$ algebra~$C_0(X)$.
More specifically, for a commutative $C^*$-algebra~${\mathscr{A}}$ with the property~$\|a^2\|=\|a\|^2$,
the spectrum of~${\mathscr{A}}$ is defined as the set $s({\mathscr{A}})$ of all non-zero $*$-homomorphisms from ${\mathscr{A}}$ to $\C$ with the topology of pointwise convergence. Then the Gelfand-Naimark theorem states that~${\mathscr{A}}$ and $C_0(s({\mathscr{A}}))$
(the latter equipped with the sup-metric) are $*$-isometric by evaluation. Thus~$C_0 \ci s $ is the identity on the family of $C^*$ algebras, modulo $*$-isomorphisms. Moreover, applying~$s$ once again, one finds that~$s \ci C_0 $ is the identity on the family of locally compact Hausdorff topological spaces, modulo homeomorphisms.

Various attempts to extend this approach such as to include geometrical data in the reconstruction have received
much attention in the past decades (see for example~\cite{gracia+varilly} or~\cite{connes13}).
In the same spirit, we define~$\text{\bf{G}}$ as the category of all tuples~$(\scrM,\scrN)$, where $\scrM$ is an oriented
globally hyperbolic surface of finite life-time with a spatially non-compact Cauchy surface~$\scrN$, and the morphisms
given by pair isometries. Next, let~$\text{\bf{H}}$ be the category of isomorphism classes of
triples~$({\mathscr{A}}, \H \oplus \H, \Sig)$ where~$\H$ is a Hilbert space, ${\mathscr{A}}$ is a $C^*$-algebra of bounded linear operators on~$\H$, and~$\Sig$ is a bounded linear operator on~$\H \oplus \H$.
We now construct a functor from~{\bf{G}} to~{\bf{H}}.
Given~$(\scrM,\scrN)$, the fibers~$S_x\scrN$ of the spinor bundle~$S\scrN$ are isomorphic to $\C$.
Moreover, the bundle~$S\scrN$ is canonically isomorphic to the restriction of~$S\scrM$ to~$\scrN$
and projecting to the left- or right-handed component.
We choose~$\H = L^2(\scrN, S\scrN)$. Moreover, we choose~${\mathscr{A}}$ as~$C_0(\scrN)$ acting
by multiplication on~$\H$. Noting that~$\H_0 = L^2(\scrN, S\scrM) \simeq \H \oplus \H$,
we let~$\Sig$ be the fermionic signature operator on~$\H \oplus \H$. This gives rise to the functor
\[ \funk \::\: \text{\bf{G}} \rightarrow \text{\bf{H}} \:. \]

\begin{Thm} \label{thmreconstruct2}
The functor~$\funk$ is injective. 
\end{Thm}
\Proof
First of all, assume that two elements~$(\scrM_1, \scrN_1)$ and~$ (\scrM_2, \scrN_2)$ of~{\bf{G}} are mapped
by~$\funk$ to one and the same element~$({\mathscr{A}}, \H \oplus \H, \Sig)$ of~{\bf{H}}. We need to show that
there is a pair isometry between~$(\scrM_1, \scrN_1)$ and~$(\scrM_2, \scrN_2)$. First, the Gelfand-Naimark theorem
tells us that~${\mathscr{A}}$ is homeomorphic to~$\scrN_1$ and~$\scrN_2$, giving rise to a
homeomorphism~$h : \scrN_1 \rightarrow \scrN_2$. Taking the pull-back of~$h$, we obtain an isomorphism~$\iota$
between the corresponding spaces of~$L^2$-sections~$\H_1$ and~$\H_2$.
Applying Proposition~\ref{taschenlampe-invariant} and choosing~$I=J$ as small neighborhoods
of a point~$x \in \scrN_1 \simeq \scrN_2$, we see that $h$ is actually an isometry (in particular, a diffeomorphism).
Next, we identify~$\scrM_1$ and~$\scrM_2$ via identification of the corresponding sets~$K_L(I) \cap K_R(J)$.
Since this identification obviously preserves the conformal structure, the resulting metrics coincide
up to a conformal factor.
Applying Proposition~\ref{taschenlampe-invariant} once again, we conclude that the volume of the sets~$K_L(I)
\cap K_R(J)$ coincides, proving that the conformal factor is equal to one.
We thus obtain an extension of the above isometry~$\scrN_1 \rightarrow \scrN_2$ to an
isometry~$\scrM_1 \rightarrow \scrM_2$.
This concludes the proof.
\QED

\Thanks {{\em{Acknowledgments:}}
We would like to thank Moritz Reintjes for helpful discussions.


\providecommand{\bysame}{\leavevmode\hbox to3em{\hrulefill}\thinspace}
\providecommand{\MR}{\relax\ifhmode\unskip\space\fi MR }
\providecommand{\MRhref}[2]{%
  \href{http://www.ams.org/mathscinet-getitem?mr=#1}{#2}
}
\providecommand{\href}[2]{#2}

\end{document}